\documentstyle[aps,psfig,epsfig,rmp,multicol]{revtex}
\begin{document}

\title{Statistical Mechanics of Complex Networks}
\author{R\'eka Albert$^{1,2}$ and Albert-L\'aszl\'o Barab\'asi$^2$}
\address{$^1$School of Mathematics, 127 Vincent Hall, University of Minnesota, Minneapolis, Minnesota 55455}
\address{$^2$Department of Physics, 225 Nieuwland Science Hall, University of Notre Dame, Notre Dame, Indiana 46556}
\maketitle
\begin{abstract}Complex networks describe a wide range of systems in nature and society, much quoted examples including the cell, a network of chemicals linked by chemical reactions,
or the Internet, a network of routers and computers connected by physical links. While traditionally these systems were modeled
as random graphs, it is increasingly recognized that the topology and evolution of real
networks is governed by robust organizing principles. Here we review the recent
advances in the field of complex networks, focusing on the statistical
mechanics of network topology and dynamics. After reviewing the
empirical data that motivated the recent interest in networks, we discuss the main models and analytical tools, covering random graphs, small-world and scale-free networks, as well as the interplay between topology and the network's robustness against failures and attacks. 
\end{abstract}
\begin{multicols}{2}

\tableofcontents

\section{INTRODUCTION}
\label{sect_introd}

Complex weblike structures describe a wide variety of systems of high
technological and intellectual importance. For example, the cell is best
described as a complex network of chemicals connected by chemical
reactions; the Internet is a complex network of routers and computers
linked by various physical or wireless links; fads and ideas spread on
the social network whose nodes are human beings and edges represent
various social relationships; the Wold-Wide Web is an enormous virtual
network of webpages connected by hyperlinks. These systems represent
just a few of the many examples that have recently prompted the
scientific community to investigate the mechanisms that determine the
topology of complex networks. The desire to understand such interwoven
systems has brought along significant challenges as well. Physics, a
major beneficiary of reductionism, has developed an arsenal of
successful tools to predict the behavior of a system as a whole from the
properties of its constituents. We now understand how magnetism emerges
from the collective behavior of millions of spins, or how do quantum
particles lead to such spectacular phenomena as Bose-Einstein
condensation or superfluidity. The success of these modeling efforts is
based on the simplicity of the interactions between the elements: there
is no ambiguity as to what interacts with what, and the interaction
strength is uniquely determined by the physical distance. We are at a
loss, however, in describing systems for which physical distance is
irrelevant, or there is ambiguity whether two components interact. While
for many complex systems with nontrivial network topology such ambiguity
is naturally present, in the past few years we increasingly recognize
that the tools of statistical mechanics offer an ideal framework to
describe these interwoven systems as well. These developments have
brought along new and challenging problems for statistical physics and
unexpected links to major topics in condensed matter physics, ranging
from percolation to Bose-Einstein condensation. 

Traditionally the study of complex networks has been the territory of
graph theory. While graph theory initially focused on regular graphs,
since the 1950's large-scale networks with no apparent design principles
were described as random graphs, proposed as the simplest and most
straightforward realization of a complex network. Random graphs were
first studied by the Hungarian mathematicians Paul Erd\H{o}s and
Alfr\'ed R\'enyi. According to the Erd\H{o}s-R\'enyi (ER) model, we
start with $N$ nodes and connect every pair of nodes with probability
$p$, creating a graph with approximately $pN(N-1)/2$ edges distributed
randomly. This model has guided our thinking about complex networks for
decades after its introduction. But the growing interest in complex
systems prompted many scientists to reconsider this modeling paradigm
and ask a simple question: are real networks behind such diverse complex
systems as the cell or the Internet, fundamentally random? Our intuition
clearly indicates that complex systems must display some organizing
principles which should be at some level encoded in their topology as
well. But if the topology of these networks indeed deviates from a
random graph, we need to develop tools and measures to capture in
quantitative terms the underlying organizing principles.

In the past few years we witnessed dramatic advances in this direction,
prompted by several parallel developments. First, the computerization of
data acquisition in all fields led to the emergence of large databases
on the topology of various real networks. Second, the increased
computing power allows us to investigate networks containing millions of
nodes, exploring questions that could not be addressed before. Third,
the slow but noticeable breakdown of boundaries between disciplines
offered researchers access to diverse databases, allowing them to
uncover the generic properties of complex networks. Finally, there is an
increasingly voiced need to move beyond reductionist approaches and try
to understand the behavior of the system as a whole. Along this route,
understanding the topology of the interactions between the components,
i.e. networks, is unavoidable. 

Motivated by these converging developments and circumstances, many
quantities and measures have been proposed and investigated in depth in
the past few years. However, three concepts occupy a prominent place in
contemporary thinking about complex networks. Next we define and briefly
discuss them, a discussion to be expanded in the coming chapters.

{\it Small worlds: } The small world concept in simple terms describes
the fact that despite their often large size, in most networks there is
a relatively short path between any two nodes. The distance between two
nodes is defined as the number of edges along the shortest path
connecting them. The most popular manifestation of "small worlds" is the
"six degrees of separation" concept, uncovered by the social
psychologist Stanley Milgram (1967), who concluded that there was a path
of acquaintances with typical length about six between most pairs of
people in the United States (Kochen 1989). The small world property
appears to characterize most complex networks: the actors in Hollywood
are on average within three costars from each other, or the chemicals in
a cell are separated typically by three reactions. The small world
concept, while intriguing, is not an indication of a particular
organizing principle. Indeed, as Erd\H{o}s and R\'enyi have
demonstrated, the typical distance between any two nodes in a random
graph scales as the logarithm of the number of nodes. Thus random graphs
are small worlds as well.

{\it Clustering: } A common property of social networks is that cliques
form, representing circles of of friends or acquaintances in which every
member knows every other member. This inherent tendency to clustering is
quantified by the clustering coefficient (Watts and Strogatz 1998). Let
us focus first on a selected node $i$ in the network, having $k_i$ edges
which connect it to $k_i$ other nodes. If the first neighbors of the
original node were part of a clique, there would be $k_i(k_i-1)/2$ edges
between them. The ratio between the number $E_i$ of edges that actually
exist between these $k_i$ nodes and the total number $k_i(k_i-1)/2$
gives the value of the clustering coefficient of node $i$
\begin{equation}
\label{eq_clust_coeff}
C_i=\frac{2E_i}{k_i(k_i-1)}.
\end{equation} 
The clustering coefficient of the whole network is the average of all
individual $C_i$'s.
  
In a random graph, since the edges are distributed randomly, the
clustering coefficient is $C=p$ (Sect. \ref{sect_rand_clust}). However,
it was Watts and Strogatz who first pointed out that in most, if not
all, real networks the clustering coefficient is typically much larger
than it is in a random network of equal number of nodes and edges. 

{\it Degree distribution: } Not all nodes in a network have the same
number of edges. The spread in the number of edges a node has, or node
degree, is characterized by a distribution function $P(k)$, which gives
the probability that a randomly selected node has exactly $k$ edges.
Since in a random graph the edges are placed randomly, the majority of
nodes have approximately the same degree, close to the average degree
$\langle k\rangle$ of the network. The degree distribution of a random
graph is a Poisson distribution with a peak at $P(\langle k\rangle)$. On
the other hand recent empirical results show that for most large
networks the degree distribution significantly deviates from a Poisson
distribution. In particular, for a large number of networks, including
the World-Wide Web (Albert, Jeong, Barab\'asi 1999), Internet (Faloutsos
{\it et al.} 1999) or metabolic networks (Jeong {\it el al.} 2000), the
degree distribution has a power-law tail 
\begin{equation}
\label{eq_power}
P(k)\sim k^{-\gamma}.
\end{equation} 
Such networks are called scale-free (Barab\'asi and Albert 1999). While
some networks display an exponential tail, often the functional form of
$P(k)$ still deviates significantly from a Poisson distribution expected
for a random graph. 

\vspace{0.2cm}

These three concepts, small path length, clustering and scale-free
degree distribution have initiated a revival of network modeling in the
past few years, resulting in the introduction and study of three main
classes of modeling paradigms. First, random graphs, which are variants
of the Erd\H{o}s-R\'enyi model, are still widely used in many fields,
and serve as a benchmark for many modeling and empirical studies.
Second, following the discovery of clustering, a class of models,
collectively called small world models, have been proposed. These models
interpolate between the highly clustered regular lattices and random
graphs. Finally, the discovery of the power-law degree distribution has
led to the construction of various scale-free models that, by focusing
on the network dynamics, aim to explain the origin of the power-law
tails and other non-Poisson degree distributions seen in real systems.

The purpose of this article is to review each of these modeling efforts,
focusing on the statistical mechanics of complex networks. Our main goal
is to present the theoretical developments in parallel with the
empirical data that initiated and support the various models and
theoretical tools. To achieve this, we start with a brief description of
the real networks and databases that represent the testground for most
current modeling efforts. 

\section{THE TOPOLOGY OF REAL NETWORKS: EMPIRICAL RESULTS}
\label{sect_real_data}

The study of most complex networks has been initiated by a desire to
understand various real systems, ranging from communication networks to
ecological webs. Thus the databases available for study span several
disciplines. In this section we review briefly those that have been
studied by researchers aiming to uncover the general features of complex
networks. Beyond the description of the databases, we will focus on the
three robust measures of the network topology: average path length,
clustering coefficient and degree distribution. Other quantities, as
discussed in the following chapters, will be again tested on these
databases. The properties of the investigated databases, as well as the
obtained exponents are summarized in Tables \ref{table_cluster} and
\ref{table_gamma}. 

\subsubsection{World-Wide Web}

\begin{figure}[htb]
\centerline{\psfig{figure=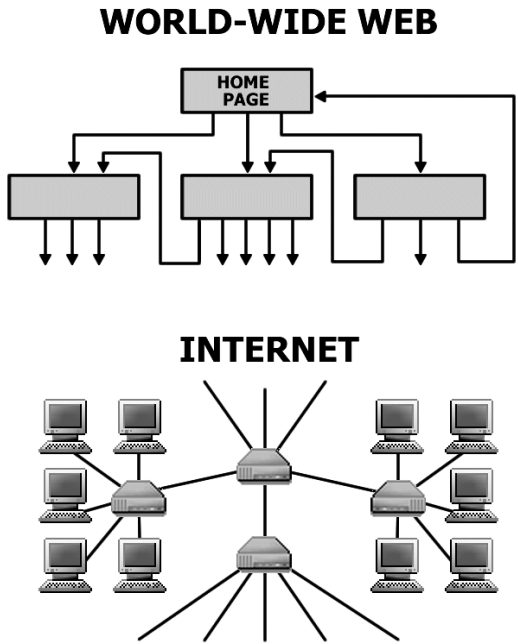,width=2.8in}}
\caption{Network structure of the World-Wide Web and the Internet. Upper panel: the nodes of the World-Wide Web are web documents, connected with directed hyperlinks (URLs). Lower panel: on the Internet the nodes are the routers and computers, the edges are the wires and cables that physically connect them. Figure courtesy of Istv\'an Albert.}
\label{fig_illust}
\end{figure}

The World-Wide Web (WWW) represents the largest network for which
topological information is currently available. The nodes of the network
are the documents (webpages) and the edges are the hyperlinks (URLs)
that point from one document to another (see Fig \ref{fig_illust}). The
size of this network was close to $1$ billion nodes at the end of 1999
(Lawrence and Giles 1998, 1999). The interest in the WWW as a network
has boomed after it has been discovered that the degree distribution of
the webpages follows a power-law over several orders of magnitude
(Albert, Jeong, Barab\'asi 1999, Kumar{\it et al.} 1999). Since the
edges of the WWW are directed, the network is characterized by two
degree distributions: the distribution of outgoing edges, $P_{out}(k)$,
signifies the probability that a document has $k$ outgoing hyperlinks
and the distribution of incoming edges, $P_{in}(k)$, is the probability
that $k$ hyperlinks point to a certain document. Several studies have
established that both $P_{out}(k)$ and $P_{in}(k)$ have power-law tails:
\begin{equation} 
P_{out}(k)\sim k^{-\gamma_{out}} \quad\mbox{and}\quad P_{in}(k)\sim k^{-\gamma_{in}}.
\end{equation} 

\begin{figure}[htb]

\centerline{\hspace{-3cm}\psfig{figure=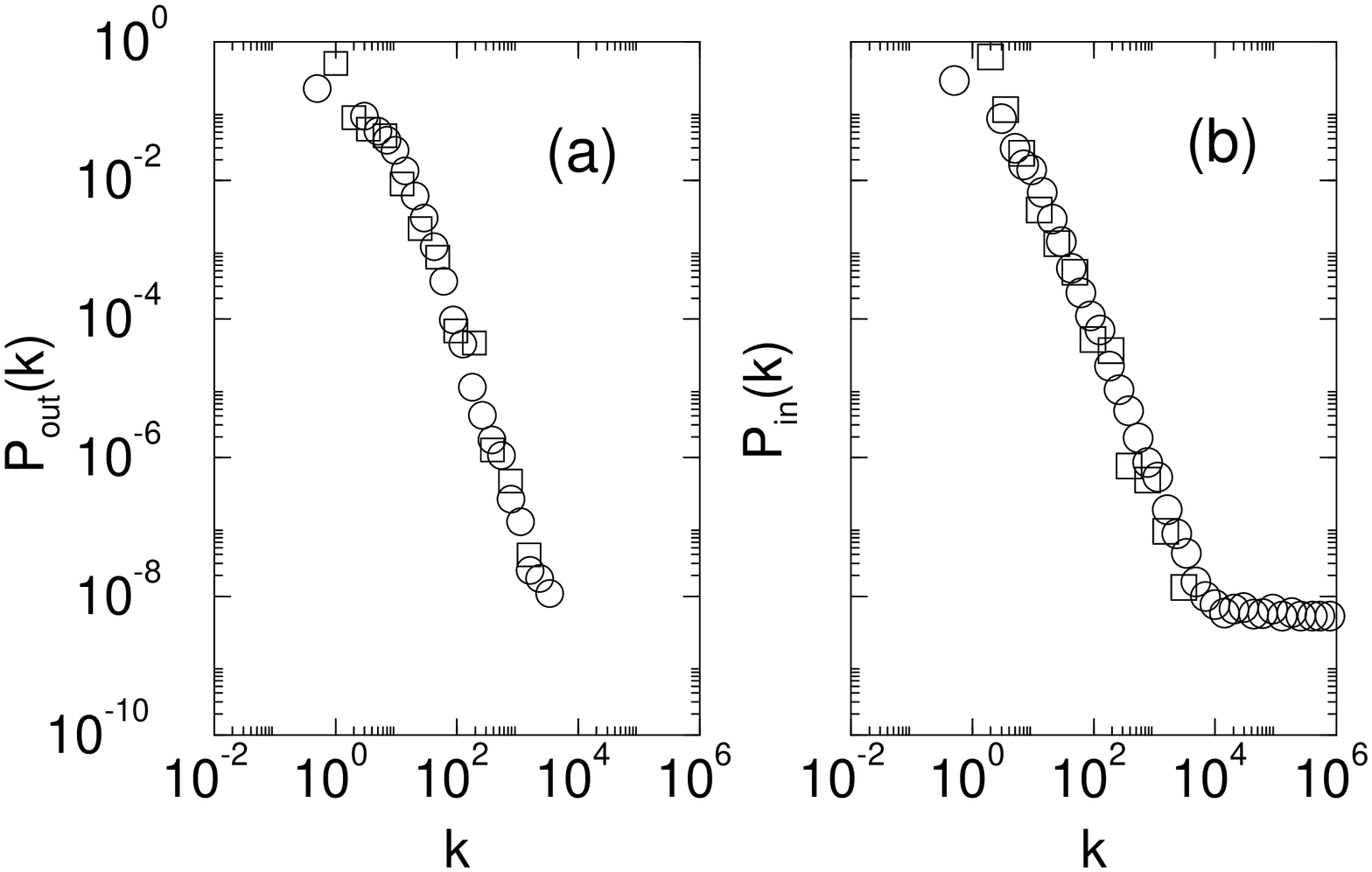,width=2.8in, angle=-90}}
\vspace{-1.5cm}
\caption{Degree distribution of the World-Wide Web from two different 
measurements. Squares correspond to the $325,729$ sample of Albert {\it et al.} 
(1999), and circles represent the measurements of over a $200$ million pages by Broder {\it et al.} (2000) (courtesy of Altavista and  Andrew Tomkins). (a) Degree distribution of the outgoing edges. (b) Degree distribution 
of the incoming edges. The data has been binned logarithmically to reduce noise.}
\label{fig_www}
\end{figure}

Albert, Jeong and Barab\'asi (1999) have studied a subset of the WWW
containing $325,729$ nodes and have found $\gamma_{out}=2.45$ and
$\gamma_{in}=2.1$. Kumar {\it et al.} (1999) used a $40$ million
document crawl by Alexa Inc., obtaining $\gamma_{out}=2.38$ and
$\gamma_{in}=2.1$ (see also Kleinberg {\it et al.} 1999). A later survey
of the WWW topology by Broder {\it et al.} (2000) used two 1999
Altavista crawls containing in total $200$ million documents, obtaining
$\gamma_{out}=2.72$ and $\gamma_{in}=2.1$ with scaling holding close to
five orders of magnitude (Fig. \ref{fig_www}). Adamic and Huberman
(2000) used a somewhat different representation of the WWW, each node
representing a separate domain name and two nodes being connected if any
of the pages in one domain linked to any page in the other. While this
method lumps together often thousands of pages that are on the same
domain, representing a nontrivial aggregation of the nodes, the
distribution of incoming edges still followed a power-law with
$\gamma_{in}^{dom}=1.94$. 

Note that $\gamma_{in}$ is the same for all measurements at the document
level despite the two years time delay between the first and last web
crawl, during which the WWW had grown at least five times larger. On the
other hand, $\gamma_{out}$ has an increasing tendency with the sample
size or time (see Table \ref{table_gamma}). 

Despite the large number of nodes, the WWW displays the small world
property. This was first reported by Albert, Jeong and Barab\'asi
(1999), who found that the average path length for a sample of $325,729$
nodes was $11.2$ and predicted, using finite size scaling, that for the
full WWW of $800$ million nodes that would be around $19$. Subsequent
measurements of Broder {\it et al.} (2000) found that the average path
length between nodes in a $200$ million sample of the WWW is $16$, in
agreement with the finite size prediction for a sample of this size.
Finally, the domain level network displays an average path length of
$3.1$ (Adamic 1999). 

The directed nature of the WWW does not allow us to measure the
clustering coefficient using Eq. (\ref{eq_clust_coeff}). One way to
avoid this difficulty is to make the network undirected, making each
edge bidirectional. This was the path followed by Adamic (1999) who
studied the WWW at the domain level using an 1997 Alexa crawl of $50$
million webpages distributed between $259,794$ sites. Adamic removed the
nodes which have only one edge, focusing on a network of $153,127$
sites. While these modifications are expected to increase somewhat the
clustering coefficient, she found $C=0.1078$, orders of magnitude higher
than $C_{rand}=0.00023$ corresponding to a random graph of the same size
and average degree.

\subsubsection{Internet}

\begin{figure}[htb]
\centerline{\psfig{figure=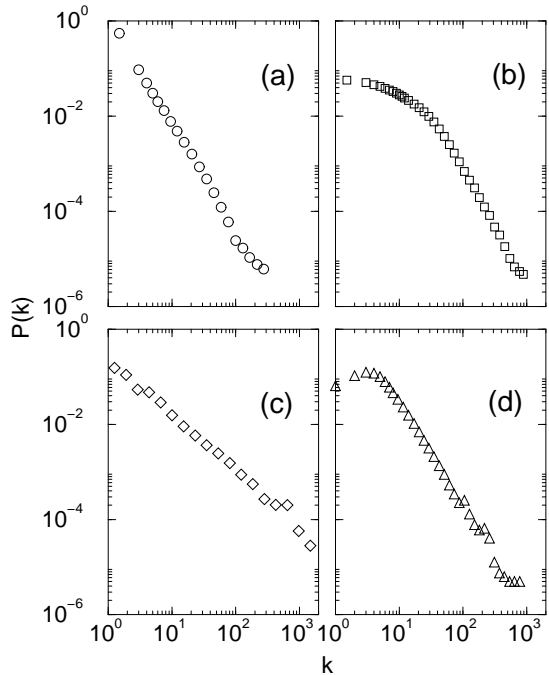,width=2.8in}}
\caption{The degree distribution of several real networks. (a) Internet at the router level. Data courtesy of 
Ramesh Govindan. (b) Movie actor collaboration network (after Barab\'asi and Albert 1999). Note that if TV series are included as well, which aggregate a large number of actors, an 
exponential cutoff emerges for large $k$ (Amaral {\it et al.} 2000). (c) Coauthorship network of high energy physicists (after Newman 2001a,b). (d) Coauthorship network of neuroscientists (after Barab\'asi {\it et al.} 2001).}
\label{fig_comp_pk}
\end{figure}

The Internet is the network of the physical links between computers and
other telecommunication devices (Fig. \ref{fig_illust}). The topology of
the Internet is studied at two different levels. At the router level the
nodes are the routers, and edges are the physical connections between
them. At the interdomain (or autonomous system) level each domain,
composed of hundreds of routers and computers, is represented by a
single node, and an edge is drawn between two domains if there is at
least one route that connects them. Faloutsos {\it et al.} (1999) have
studied the Internet at both levels, concluding that in each case the
degree distribution follows a power-law. The interdomain topology of the
Internet, captured at three different dates between $1997$ and the end
of $1998$, resulted in degree exponents between $\gamma_I^{as}=2.15$ and
$\gamma_I^{as}=2.2$. The $1995$ survey of the Internet topology at the
router level, containing $3,888$ nodes found $\gamma_{I}^r=2.48$
(Faloutsos {\it et al.} 1999). Recently Govindan and Tangmunarunkit
(2000) mapped the connectivity of nearly $150,000$ router interfaces and
nearly $200,000$ router adjacencies, confirming the power-law scaling
with $\gamma_{I}^{r}\simeq 2.3$ (see Fig. \ref{fig_comp_pk}a).

The Internet as a network does display clustering and small path length
as well. Yook {\it et al.} (2001a) and Pastor-Satorras {\it et al.}
(2001), studying the Internet at the domain level between 1997 and 1999
have found that its clustering coefficient ranged between $0.18$ and
$0.3$, to be compared with $C_{rand}\simeq 0.001$ for random networks of
similar parameters. The average path length of the Internet at the
domain level ranged between $3.70$ and $3.77$ (Yook {\it et al.} 2001a,
Pastor-Satorras {\it et al.} 2001), and at the router level it was
around $9$ (Yook {\it et al.} 2001a), indicating its small world
character. 

\subsubsection{Movie actor collaboration network} 

A much studied database is the movie actor collaboration network, based
on the Internet Movie Database that contains all movies and their casts
since the 1890's. In this network the nodes are the actors, and two
nodes have a common edge if the corresponding actors have acted in a
movie together. This is a continuously expanding network, with $225,226$
nodes in 1998 (Watts, Strogatz 1998) which grew to $449,913$ nodes by
May 2000 (Newman, Strogatz and Watts 2000). The average path length of
the actor network is close to that of a random graph with the same size
and average degree, $3.65$ compared to $2.9$, but its clustering
coefficient is more than $100$ times higher than a random graph (Watts
and Strogatz 1998). The degree distribution of the movie actor network
has a power-law tail for large $k$ (see Fig. \ref{fig_comp_pk}b),
following $P(k)\sim k^{-\gamma_{actor}}$, where $\gamma_{actor}=2.3\pm
0.1$ (Barab\'asi and Albert 1999, Amaral {\it et al.} 2000, Albert and
Barab\'asi 2000). 

\subsubsection{Science collaboration graph} 

A collaboration network similar to that of the movie actors can be
constructed for scientists, where the nodes are the scientists and two
nodes are connected if the two scientists have written an article
together. To uncover the topology of this complex graph, Newman
(2001a,b,c) studied four databases spanning physics, biomedical
research, high-energy physics and computer science over a 5 year window
(1995-1999). All these networks show small average path length but high
clustering coefficient, as summarized in Table \ref{table_cluster}. The
degree distribution of the collaboration network of high-energy
physicists is an almost perfect power-law with an exponent of $1.2$
(Fig. \ref{fig_comp_pk}c), while the other databases display power-laws
with a larger exponent in the tail. 

Barab\'asi {\it et al.} 2001 investigated the collaboration graph of
mathematicians and neuroscientists publishing between 1991 and 1998. The
average path length of these networks is around $\ell_{math}=9.5$ and
$\ell_{nsci}=6$, their clustering coefficient being $C_{math}=0.59$ and
$C_{nsci}=0.76$. The degree distributions of these collaboration
networks are consistent with power-laws with degree exponents $2.1$ and
$2.5$, respectively (see Fig. \ref{fig_comp_pk}d). 

\subsubsection{The web of human sexual contacts} 

Many sexually transmitted diseases, including AIDS, spread on a network
of sexual relationships. Liljeros {\it et al.} (2001) have studied the
web constructed from the sexual relations of $2810$ individuals, based
on an extensive survey conducted in Sweden in 1996. Since the edges in
this network are relatively short lived, they analyzed the distribution
of partners over a single year, obtaining both for females and males a
power-law degree distribution with an exponent $\gamma_{f}=3.5\pm 0.2$
and $\gamma_{m}=3.3\pm 0.2$, respectively. 

\subsubsection{Cellular networks} 

Jeong {\it et al.} (2000) studied the metabolism of $43$ organisms
representing all three domains of life, reconstructing them in networks
in which the nodes are the substrates (such as ATP, ADP, $\rm{H_2O}$)
and the edges represent the predominantly directed chemical reactions in
which these substrates can participate. The distribution of the outgoing
and incoming edges have been found to follow power-laws for all
organisms, the degree exponents varying between $2.0$ and $2.4$. While
due to the network's directedness the clustering coefficient has not
been determined, the average path length was found to be approximately
the same in all organisms, with a value of $3.3$. 

The clustering coefficient was studied by Wagner and Fell (2000, see
also Fell and Wagner 2000), focusing on the energy and biosynthesis
metabolism of the {\it Escherichia Coli} bacterium, finding that, in
addition to the power-law degree distribution, the undirected version of
this substrate graph has small average path length and large clustering
coefficient (see Table \ref{table_cluster}). 

Another important network characterizing the cell describes
protein-protein interactions, where the nodes are proteins and they are
connected if it has been experimentally demonstrated that they bind
together. A study of these physical interactions shows that the degree
distribution of the physical protein interaction map for yeast follows a
power-law with an exponential cutoff $P(k)\sim
(k+k_0)^{-\gamma}e^{-(k+k_0)/k_c}$ with $k_0=1$, $k_c=20$ and
$\gamma=2.4$ (Jeong {\it et al.} 2001). 

\subsubsection{Ecological networks} 

Food webs are used regularly by ecologists to quantify the interaction
between various species (Pimm 1991). In a food web the nodes are species
and the edges represent predator-prey relationships between them. In a
recent study, Williams {\it et al.} (2000) investigated the topology of
the seven most documented and largest food webs, such as the Skipwith
Pond, Little Rock Lake, Bridge Brook Lake, Chesapeake Bay, Ythan
Estuary, Coachella Valley and St. Martin Island webs. While these webs
differ widely in the number of species or their average degree, they all
indicate that species in habitats are three or fewer edges from each
other. This result was supported by the independent investigations of
Montoya and Sol\'e (2000) and Camacho {\it et al.} (2001a), who have
shown that food webs are highly clustered as well. The degree
distribution was first addressed by Montoya and Sol\'e (2000), focusing
on the Ythan Estuary, Silwood Park and Little Rock Lake food webs,
considering these networks as being nondirected. Although the size of
these webs is small (the largest of them has $N=186$ nodes), they appear
to share the non-random properties of their larger counterparts. In
particular, Montoya and Sol\'e (2000) concluded that the degree
distribution is consistent with a power-law with an unusually small
exponent of $\gamma\simeq 1.1$. The small size of these webs does give
room, however, for some ambiguity in $P(k)$. Camacho {\it et al.}
(2001a,b) find that for some food webs an exponential fit works equally
well. While the well documented existence of keystone species that play
an important role in the food web topology points towards the existence
of hubs (a common feature of scale-free networks), an unambiguous
determination of the network's topology could benefit from larger
datasets. Due to the inherent difficulty in the data collection process
(Williams {\it et al.} 2000), this is not expected anytime soon. 

\subsubsection{Phone-call network} 

A large directed graph has been constructed from the long distance
telephone call patterns, where nodes are phone numbers and every
completed phone call is an edge, directed from the caller to the
receiver. Abello, Pardalos and Resende (1999) and Aiello, Chung and Lu
(2000) studied the call graph of long distance telephone calls made
during a single day, finding that the degree distribution of the
outgoing and incoming edges follow a power-law with exponent
$\gamma_{out}=\gamma_{in}=2.1$. 

\subsubsection{Citation networks} 

A rather complex network is formed by the citation patterns of
scientific publications, the nodes standing for published articles and a
directed edge representing a reference to a previously published
article. Redner (1998), studying the citation distribution of $783,339$
papers cataloged by the Institute of Scientific Information, and of the
$24,296$ papers published in Physical Review D between $1975$ and
$1994$, has found that the probability that a paper is cited $k$ times
follows a power-law with exponent $\gamma_{cite}=3$, indicating that the
incoming degree distribution of the citation network follows a
power-law. A recent study of V\'azquez (2001) extended these studies to
the outgoing degree distribution as well, obtaining that it has an
exponential tail. 

\subsubsection{Networks in linguistics} 

The complexity of human languages offers several possibilities to define
and study complex networks. Recently Ferrer i Cancho and Sol\'e (2001)
have constructed such a network for the English language, based on the
British National Corpus, words, as nodes, being linked if they appear
next or one word apart from each other in sentences. They have found
that the resulting network of $440,902$ words displays a small average
path length $\ell=2.67$, a high clustering coefficient $C=0.437$, and a
two-regime power-law degree distribution. Words with degree $k\leq 10^3$
decay with a degree exponent $\gamma_{<}=1.5$ while words with
$10^3<k<10^5$ follow a power-law with $\gamma_{>}\simeq 2.7$. 

A different study (Yook, Jeong, Barab\'asi 2001b) linked words based on
their meaning, i.e. two words were connected to each other if they were
known to be synonyms according to the Merriam-Webster Dictionary. The
results indicate the existence of a giant cluster of $22,311$ words from
the total of $23,279$ words which have synonyms, with an average path
length $\ell=4.5$, and a rather high clustering coefficient $C=0.7$
compared to $C_{rand}=0.0006$ for an equivalent random network. In
addition the degree distribution followed had a power-law tail with
$\gamma_{syn}=2.8$. These results indicate that in many respects the
language also forms a complex network with organizing principles not so
different from the examples discussed earlier (see also Steyvers and
Tennenbaum 2001). 

\subsubsection{Power and neural networks} 

The power grid of the western United States is described by a complex
network whose nodes are generators, transformers and substations, and
the edges are high-voltage transmission lines. The number of nodes in
the power grid is $N=4,941$, and $\langle k\rangle =2.67$. In the tiny
($N=282$) neural network of the nematode worm {\it C. elegans}, the
nodes are the neurons, and an edge joins two neurons if they are
connected by either a synapse or a gap junction. Watts and Strogatz
(1998) found that while for both networks the average path length was
approximately equal with that of a random graph with the same size and
average degree, their clustering coefficient was much higher (Table
\ref{table_cluster}). The degree distribution of the power grid is
consistent with an exponential, while for the neural network it has a
peak at an intermediate $k$ after which it decays following an
exponential (Amaral {\it et al.} 2000). 

\subsubsection{Protein folding} 

During folding a protein takes up consecutive conformations.
Representing with a node each distinct state, two conformations are
linked if they can be obtained from each other by an elementary move.
Scala {\it et al.} (2000) studied the network formed by the
conformations of a 2D lattice polymer, obtaining that it has small-world
properties. Specifically, the average path length increases
logarithmically when the size of the polymer (and consequently the size
of the network) increases, similarly to the behavior seen in a random
graph. The clustering coefficient, however, is much larger than
$C_{rand}$, a difference that increases with the network size. The
degree distribution of this conformation network is consistent with a
gaussian (Amaral {\it et al.} 2000). 

\medskip 

The databases discussed above served as motivation and source of
inspiration for uncovering the topological properties of real networks.
We will refer to them frequently to validate various theoretical
predictions, or to understand the limitations of the modeling efforts.
In the remaining of the review we discuss the various theoretical tools
developed to model these complex networks. In this respect, we need to
start with the mother of all network models: the random graph theory of
Erd\H{o}s and R\'enyi. 

\end{multicols}
\newpage
\begin{table}[htb]
\caption{The general characteristics of several real 
networks. For each network we indicated the number of nodes, the average 
degree 
$\langle k\rangle$, the average path length $\ell$ and the clustering coefficient 
$C$. For a comparison we have included the average path length $\ell_{rand}$ 
and 
clustering coefficient $C_{rand}$ of a random graph with the same size and 
average degree. The last column identifies the symbols in Figs. \ref{fig_fit_diam_er} and \ref{fig_fit_ccoef_er}.}
\label{table_cluster}
\begin{tabular}{|c|c|c|c|c|c|c|c|c|}
Network  & Size & $\langle k\rangle$ & $\ell$ & $\ell_{rand}$ & $C$ & 
$C_{rand}$ 
& Reference & Nr.\\
\hline
WWW, site level, undir. & $153,127$ & $35.21$ & $3.1$ & $3.35$ & $0.1078$ & 
$0.00023$ & Adamic 1999 & 1 \\
\hline
Internet, domain level & $3015$ - $6209$ & $3.52$ - $4.11$ & $3.7$ - $3.76$ & 
$6.36$ - $6.18$ & $0.18$ - $0.3$ & $0.001$ & Yook {\it et al.} 2001a, & \\
& & & & & & & Pastor-Satorras {\it et al.} 2001& 2 \\
\hline
Movie actors & $225,226$ & $61$ & $3.65$ & $2.99$ & $0.79$ & $0.00027$ & 
Watts, 
Strogatz 1998 & 3 \\
\hline
LANL coauthorship & $52,909$ & $9.7$ & $5.9$  & $4.79$ & $0.43$ & $1.8\times 
10^{-4}$ & Newman 2001a,b & 4 \\
\hline
MEDLINE coauthorship & $1,520,251$ & $18.1$ & $4.6$  & $4.91$ & $0.066$ & 
$1.1\times 10^{-5}$ & Newman 2001a,b & 5\\
\hline
SPIRES coauthorship & $56,627$ & $173$ & $4.0$  & $2.12 $ & $0.726$ & $0.003$ 
& 
Newman 2001a,b,c & 6\\
\hline
NCSTRL coauthorship & $11,994$ & $3.59$ & $9.7$  & $7.34$ & $0.496$ & $3\times 
10^{-4}$ & Newman 2001a,b & 7\\
\hline
Math coauthorship & $70,975$ & $3.9$ & $9.5$  & $8.2$ & $0.59$ & $5.4\times 
10^{-5}$ & Barab\'asi {\it et al.} 2001 & 8\\
\hline
Neurosci. coauthorship & $209,293$ & $11.5$ & $6$  & $5.01$ & $0.76$ & 
$5.5\times 
10^{-5}$ & Barab\'asi {\it et al.} 2001 & 9\\
\hline
{\it E. coli}, substrate graph & $282$ & $7.35$ & $2.9$ & $3.04$ & $0.32$ & 
$0.026$ & Wagner, Fell 2000 & 10 \\
\hline
{\it E. coli}, reaction graph & $315$ & $28.3$ & $2.62$ & $1.98$ & $0.59$ & 
$0.09$ & Wagner, Fell 2000 & 11\\
\hline
Ythan estuary food web & $134$ & $8.7$ & $2.43$ & $2.26$ & $0.22$ & $0.06$ & 
Montoya, Sol\'e 2000 & 12\\
\hline
Silwood park food web & $154$ & $4.75$ & $3.40$ & $3.23$ & $0.15$ & $0.03$ & 
Montoya, Sol\'e 2000 & 13\\
\hline
Words, cooccurence & $460.902$ & $70.13$ & $2.67$ & $3.03$ & $0.437$ & $0.0001$ & Cancho, Sol\'e 2001 & 14\\
\hline
Words, synonyms & $22,311$ & $13.48$ & $4.5$ & $3.84$ & $0.7$ & $0.0006$ & Yook {\it et al.} 2001 & 15\\
\hline
Power grid & $4,941$ & $2.67$ & $18.7$ & $12.4$ & $0.08$ &$ 0.005$ & Watts, 
Strogatz 1998 & 16 \\
\hline
{\it C. Elegans} & $282$ & $14$  & $2.65$ & $2.25$ & $0.28$ & $0.05$ & Watts, 
Strogatz 1998 & 17 \\
\end{tabular}
\end{table}

\begin{table}[htb]
\caption{The scaling exponents characterizing the degree distribution of several scale-free networks, for which $P(k)$ follows a power-law (\ref{eq_power}). We  
indicate the size of the network, its average degree $\langle k\rangle$ and the cutoff $\kappa$ for the power-law scaling. For directed networks we list separately the indegree ($\gamma_{in}$) and outdegree ($\gamma_{out}$) exponents, while for the 
undirected networks, marked with a star, these values are identical. The columns $l_{real}$, $l_{rand}$ and $l_{pow}$ compare the average path length of real networks with power-law degree distribution and the prediction of random graph theory 
(\ref{path_er}) and that of Newman, Strogatz and Watts (2000) (\ref{path_power}), as discussed in Sect. \ref{sect_sf_graph}. The last column identifies the symbols in Figs. \ref{fig_fit_diam_er} and \ref{fig_fit_ccoef_er}.}
\label{table_gamma}
\begin{tabular}{|c|c|c|c|c|c|c|c|c|c|c|}
Network &  Size & $\langle k\rangle$ & $\kappa$ & $\gamma_{out}$ & $\gamma_{in}$ & $\ell_{real}$ & $\ell_{rand}$ & $\ell_{pow}$ & Reference & Nr.\\
\hline
WWW & $325,729$ & $4.51$ & $900$ & $2.45$ & $2.1$ & $11.2$ & $8.32$ & $4.77$& Albert, Jeong, Barab\'asi 1999 & 
1\\
\hline
WWW & $4\times 10^7$ & $7$ & & $2.38$ & $2.1$ & & & & Kumar {\it et al.} 1999 & 2\\
\hline
WWW & $2\times 10^8$ & $7.5$ & $4,000$ & $2.72$ & $2.1$ &  $16$ & $8.85$ & $7.61$ & Broder {\it et al.} 2000 & 3\\
\hline
WWW, site & $260,000$ & & & & $1.94$ & & & & Huberman, Adamic 2000 & 4\\
\hline
Internet, domain$*$ & $3,015$ - $4,389$ & $3.42$ - $3.76$ & $30-40$ & $2.1$ - $2.2$ & 
$2.1$ 
- $2.2$ & $4$ & $6.3$ & $5.2$ & Faloutsos 1999 & 5\\
\hline
Internet, router$*$ & $3,888$ & $2.57$ & $30$ & $2.48$ & $2.48$ & $12.15$ & $8.75$ & $7.67$ & Faloutsos 1999 & 
6\\
\hline
Internet, router$*$ & $150,000$ & $2.66$ & $60$ & $2.4$ & $2.4$ & $11$ & $12.8$ & $7.47$ & Govindan 2000 & 7\\
\hline
Movie actors$*$ & $212,250$ & $28.78$ & $900$ & $2.3$ &  $2.3$ & $4.54$ & $3.65$ & $4.01$ & Barab\'asi, Albert 
1999 & 8\\
\hline
Coauthors, SPIRES$*$ & $56,627$ & $173$ & $1,100$ & $1.2$ & $1.2$ & $4$ & $2.12$ & $1.95$ & Newman 2001b,c & 9\\
\hline
Coauthors, neuro.$*$ & $209,293$ & $11.54$ & $400$ & $2.1$ & $2.1$ & $6$ & $5.01$ & $3.86$ & Barab\'asi {\it 
et al.} 2001 & 10\\
\hline
Coauthors, math$*$ & $70,975$ & $3.9$ & $120$ & $2.5$ & $2.5$ & $9.5$ & $8.2$ & $6.53$ & Barab\'asi {\it et al.} 
2001 & 11\\
\hline
Sexual contacts$*$ & $2810$ & & & $3.4$ & $3.4$ & & & & Liljeros {\it et al.} 2001 & 12\\
\hline
Metabolic, E. coli & $778$ & $7.4$ & $110$ & $2.2$ & $2.2$ & $3.2$ & $3.32$ & $2.89$ & Jeong {\it et al.} 2000 & 
13\\
\hline 
Protein, S. cerev.$*$ & $1870$ & $2.39$ & & $2.4$ & $2.4$ & & & & Mason {\it et al.} 
2000 
& 14\\
\hline
Ythan estuary$*$ & $134$ & $8.7$ & $35$ & $1.05$ & $1.05$ & $2.43$ & $2.26$ & $1.71$ & Montoya, Sol\'e 2000 & 
14\\
\hline
Silwood park$*$ & $154$ & $4.75$ & $27$ & $1.13$ & $1.13$ & $3.4$ & $3.23$ & $2$ & Montoya, Sol\'e 2000 & 
16\\
\hline
Citation & $783,339$ & $8.57$ & & &  $3$ & & & & Redner 1998 & 17\\
\hline
Phone-call & $53\times 10^6$ & $3.16$ & & $2.1$ & $2.1$ & & & & Aiello {\it et al.} 
2000 
& 18\\
\hline
Words, cooccurence$*$ & $460,902$ & $70.13$ & & $2.7$ & $2.7$ & & & & Cancho, Sol\'e 2001 & 19\\
\hline
Words, synonyms$*$ & $22,311$ & $13.48$ & & $2.8$ & $2.8$ & & & & Yook {\it et al.} 2001 & 20\\
\end{tabular}
\end{table}
\begin{multicols}{2}

\section{RANDOM GRAPH THEORY} \label{sect_random} 

\begin{figure}[htb]
\vspace{-1cm}
\centerline{\psfig{figure=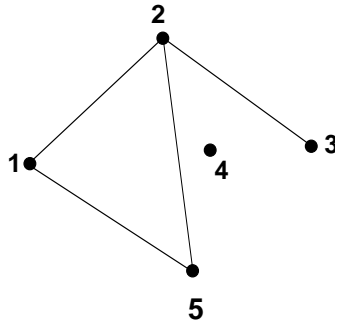,width=1.8in}}
\caption{Illustration of a graph with $N=5$ nodes and $n=4$ edges. The set of 
the 
nodes is $P=\{1, 2, 3, 4, 5\}$ and the edge set is $E=\{\{1,2\}, \{1,5\}, 
\{2,3\}, \{2,5\}\}$.  }
\label{fig_graph}
\end{figure}

In mathematical terms a network is represented by a graph. A graph is a
pair of sets $G=\{P, E\}$, where $P$ is a set of $N$ nodes (or vertices
or points) $P_1$, $P_2$, .. $P_N$ and $E$ is a set of edges (or links or
lines) that connect two elements of $P$. Graphs are usually represented
as a set of dots, each corresponding to a node, two of these dots being
joined by a line if the corresponding nodes are connected (see Fig.
\ref{fig_graph}).

Graph theory has its origins in the $18$th century in the work of
Leonhard Euler, the early work concentrating on small graphs with a high
degree of regularity. In the $20$th century graph theory has become more
statistical and algorithmic. A particularly rich source of ideas has
been the study of random graphs, graphs in which the edges are
distributed randomly. Networks with a complex topology and unknown
organizing principles often appear random, thus random graph theory is
regularly used in the study of complex networks. 

The theory of random graphs was founded by Paul Erd\H{o}s and Alfr\'ed
R\'enyi (1959,1960,1961), after Erd\H{o}s discovered that probabilistic
methods were often useful in tackling problems in graph theory. An
detailed review of the field is available in the classic book of
Bollob\'as (1985), complemented by the review of the parallels between
phase transitions and random graph theory of Cohen (1988), and the guide
of the history of the Erd\H{o}s-R\'enyi approach by Karo\'nski and
Ru\'cinski (1997). In the following we briefly describe the most
important results of random graph theory, focusing on the aspects that
are of direct relevance to complex networks. 

\subsection{The Erd\H{o}s-R\'enyi model} \label{sect_er} 

In their classic first article on random graphs, Erd\H{o}s and R\'enyi
define a random graph as $N$ labeled nodes connected by $n$ edges which
are chosen randomly from the $\frac{N(N-1)}{2}$ possible edges
(Erd\H{o}s and R\'enyi 1959). In total there are
$C_{\frac{N(N-1)}{2}}^{\,n}$ graphs with $N$ nodes and $n$ edges,
forming a probability space in which every realization is equiprobable. 

\begin{figure}[htb]
\centerline{\psfig{figure=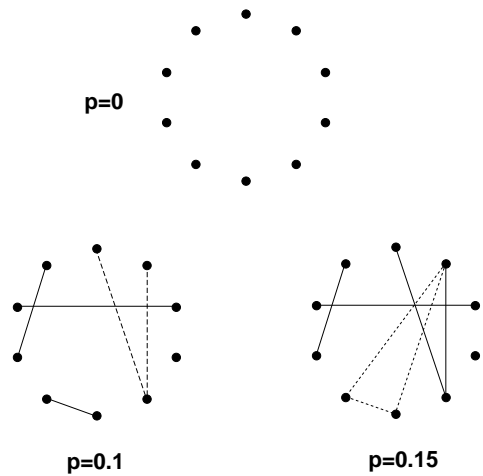,width=2.8in}}
\vspace{-1in}
\caption{Illustration of the graph evolution process for the Erd\H{o}s-R\'enyi 
model. We start with $N=10$ isolated nodes (upper panel), then connect every 
pair 
of nodes with probability $p$. The lower panel of the figure shows two 
different 
stages in the graph's development, corresponding to $p=0.1$ and $p=0.15$. We 
can 
notice the emergence of trees (a tree of order $3$, drawn with dashed lines) and cycles (a cycle of order $3$, drawn with dotted lines) in the 
graph, and a connected cluster which unites half of the nodes at 
$p=0.15=1.5/N$. 
}
\label{fig_graph_evol}
\end{figure}

An alternative and equivalent definition of a random graph is called the
binomial model. Here we start with $N$ nodes, every pair of nodes being
connected with probability $p$ (see Fig. \ref{fig_graph_evol}).
Consequently, the total number of edges is a random variable with the
expectation value $E(n)=p\frac{N(N-1)}{2}$. If $G_0$ is a graph with
nodes $P_1$, $P_2$, .. $P_N$ and $n$ edges, the probability of obtaining
it by this graph construction process is
$P(G_0)=p^n(1-p)^{\frac{N(N-1)}{2}-n}$. 

Random graph theory studies the properties of the probability space
associated with graphs with $N$ nodes as $N\rightarrow \infty$. Many
properties of such random graphs can be determined using probabilistic
arguments. In this respect Erd\H{o}s and R\'enyi used the definition
that almost every graph has a property $Q$ if the probability of having
$Q$ approaches $1$ as $N\rightarrow \infty$. Among the questions
addressed by Erd\H{o}s and R\'enyi some have direct relevance to
understanding complex networks as well, such as: Is a typical graph
connected? Does it contain a triangle of connected nodes? How does its
diameter depend on its size?

The construction of a random graph is often called in the mathematical
literature an evolution: starting with a set of $N$ isolated vertices,
the graph develops by the successive addition of random edges. The
graphs obtained at different stages of this process correspond to larger
and larger connection probabilities $p$, eventually obtaining a fully
connected graph (having the maximum number of edges $n=N(N-1)/2$) for
$p\rightarrow 1$. The main goal of random graph theory is to determine
at what connection probability $p$ will a particular property of a graph
most likely arise. The greatest discovery of Erd\H{o}s and R\'enyi was
that many important properties of random graphs appear quite suddenly.
That is, at a given probability either almost every graph has the
property $Q$ (e.g. every pair of nodes is connected by a path of
consecutive edges) or on the contrary, almost no graph has it. The
transition from a property being very unlikely to being very likely is
usually swift. For many such properties there is a critical probability
$p_c(N)$. If $p(N)$ grows slower than $p_c(N)$ as $N\rightarrow \infty$,
then almost every graph with connection probability $p(N)$ fails to have
$Q$. If $p(N)$ grows somewhat faster than $p_c(N)$, then almost every
graph has the property $Q$. Thus the probability that a graph with $N$
nodes and connection probability $p=p(N)$ has property $Q$ satisfies

\begin{equation}
\label{pc_random}
\lim_{N\rightarrow\infty}P_{N,\,p}(Q)=\left\{\begin{array}{rcl}
0&\mbox {if}& \frac{p(N)}{p_c(N)}\rightarrow 0\\ 1&\mbox {if}&
\frac{p(N)}{p_c(N)}\rightarrow\infty
\end{array}\right..
\end{equation}

An important note is in order here. Physicists trained in critical
phenomena will recognize in $p_c(N)$ the critical probability familiar
in percolation. In the physics literature usually the system is viewed
at a fixed system size $N$ and then the different regimes in
(\ref{pc_random}) reduce to the question whether $p$ is smaller or
larger than $p_c$. The proper value of $p_c$, that is, the limit
$p_c=p_c(N\rightarrow\infty)$ is obtained by finite size scaling. The
basis of this procedure is the assumption that this limit exists,
reflecting the fact that ultimately the percolation threshold is
independent of the system size. This is usually the case in finite
dimensional systems which include most physical systems of interest for
percolation theory and critical phenomena. In contrast, networks are, by
definition, infinite dimensional: the number of neighbors a node can
have increases linearly with the system size. Consequently, in random
graph theory the occupation probability is defined as a function of the
system size: $p$ represents the fraction of the edges which are present
from the possible $N(N-1)/2$. Larger graphs with the same $p$ will
contain more edges, and consequently properties like the appearance of
cycles could occur for smaller $p$ in large graphs than in smaller ones.
This means that for many properties $Q$ in random graphs there is no
unique, $N$-independent threshold, but we have to define a threshold
function which depends on the system size, and
$p_c(N\rightarrow\infty)\rightarrow 0$. On the other hand, we will see
that the average degree of the graph
 
\begin{equation}
\langle k\rangle=2n/N=p(N-1)\simeq pN
\end{equation}
does have a critical value which is independent of the system size. In the 
coming 
subsection we illustrate these ideas by looking at the emergence of various 
subgraphs in random graphs.

\subsection{Subgraphs}
\label{sect_subgraph}

The first property of random graphs studied by Erd\H{o}s and R\'enyi
(1959) was the appearance of subgraphs. A graph $G_1$ consisting of a
set $P_1$ of nodes and a set $E_1$ of edges is a subgraph of a graph
$G=\{P,E\}$ if all nodes in $P_1$ are also nodes of $P$ and all edges in
$E_1$ are also edges of $E$. The simplest examples of subgraphs are
cycles, trees and complete subgraphs (see Fig. \ref{fig_graph_evol}). A
cycle of order $k$ is a closed loop of $k$ edges such that every two
consecutive edges and only those have a common node. That is,
graphically a triangle is a cycle of order 3, while a rectangle is a
cycle of order 4. The average degree of a cycle is equal to $2$, since
every node has two edges. The opposite of closed loops are the trees,
which cannot form closed loops. More precisely, a graph is a tree of
order $k$ if it has $k$ nodes and $k-1$ edges, and none of its subgraphs
is a cycle. The average degree of a tree of order $k$ is $\langle
k\rangle=2-2/k$, approaching $2$ for large trees. Complete subgraphs of
order $k$ contain $k$ nodes and all the possible $k(k-1)/2$ edges, in
other words being completely connected.

Let us consider the evolution process described in Fig.
\ref{fig_graph_evol} for a graph $G=G_{N,p}$. We start from $N$ isolated
nodes, then connect every pair of nodes with probability $p$. For small
connection probabilities the edges are isolated, but as $p$, and with it
the number of edges, increases, two edges can attach at a common node,
forming a tree of order $3$. An interesting problem is to determine the
critical probability $p_c(N)$ at which almost every graph $G$ contains a
tree of order $3$. Most generally we can ask whether there is a critical
probability which marks the appearance of arbitrary subgraphs consisting
of $k$ nodes and $l$ edges. 

In random graph theory there is a rigorously proven answer to this
question (Bollob\'as 1985). Consider a random graph $G=G_{N,p}$. In
addition, consider a small graph $F$ consisting of $k$ nodes and $l$
edges. In principle, the random graph $G$ can contain several such
subgraphs $F$. Our first goal is to determine how many such subgraphs
exist. The $k$ nodes can be chosen from the total number of nodes $N$ in
$C_N^k$ ways and the $l$ edges are formed with probability $p^l$. In
addition, we can permute the $k$ nodes and potentially obtain $k!$ new
graphs (the correct value is $k!/a$, where $a$ is the number of graphs
which are isomorphic to each other). Thus the expected number of
subgraphs $F$ contained in $G$ is \begin{equation} \label{ref_ex}
E(X)=C_N^k\frac{k!}{a}p^l\simeq\frac{N^kp^l}{a}. \end{equation} This
notation suggests that the actual number of such subgraphs, $X$, can be
different from $E(X)$, but in the majority of the cases it will be close
to it. Note that the subgraphs do not have to be isolated, i.e. there
can exist edges with one of their nodes inside the subgraph, but the
other outside of it. 

Equation (\ref{ref_ex}) indicates that if $p(N)$ is such that
$p(N)N^{k/l}\rightarrow 0$ as $N\rightarrow 0$, the expected number of
subgraphs $E(X)\rightarrow 0$, i.e. almost none of the random graphs
contains a subgraph $F$. On the other hand, if $p(N)=cN^{-k/l}$, the
mean number of subgraphs is a finite number, denoted by $\lambda=c^l/a$,
indicating that this function might be the critical probability. The
validity of this finding can be tested by calculating the distribution
of subgraph numbers, $P_p(X=r)$, obtaining (Bollob\'as 1995)
\begin{equation} 
\lim_{N\rightarrow\infty}
P_p(X=r)=e^{-\lambda}\frac{\lambda^r}{r!}.
\end{equation} The probability that
$G$ contains at least one subgraph $F$ is then
\begin{equation}
\label{subgraph_prop}
P_p(G\supset F)=\sum_{r=1}^{\infty}P_p(X=r)=1-e^{-\lambda},
\end{equation}
 which converges to $1$ as $c$ increases. For $p$ values
satisfying $pN^{k/l}\rightarrow\infty$ the probability $P_p(G\supset F)$
converges to $1$, thus, indeed, the critical probability at which
almost every graph contains a subgraph with $k$ nodes and $l$ edges is
$p_c(N)=cN^{-k/l}$.

A few important special cases directly follow from Eq. (\ref{subgraph_prop}):

(a) The critical probability of having a  tree of order $k$ is
$p_c(N)=cN^{-k/(k-1)}$;

(b) The critical probability of having a cycle of order $k$ is
$p_c(N)=cN^{-1}$;

(c) The critical probability of having a complete subgraph of order
$k$ is $p_c(N)=cN^{-2/(k-1)}$.

\subsection{Graph Evolution}
\label{sect_graph_evol}
 
\begin{figure}[htb]
\vspace{-2cm}
\centerline{\hspace{-2.2cm}\psfig{figure=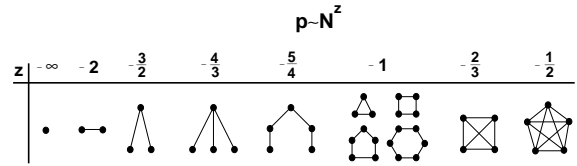,width=3.2in,angle=-90}}
\vspace{-3.7cm}
\caption{The threshold probabilities at which different subgraphs appear in a 
random graph. For $pN^{3/2}\rightarrow 0$ the graph consists of isolated nodes 
and edges. For $p\sim N^{-3/2}$ trees of order $3$ appear, at $p\sim N^{-4/3}$ 
trees of order $4$. At $p\sim N^{-1}$ trees of all orders are present, and in 
the 
same time cycles of all orders appear. The probability $p\sim N^{-2/3}$ marks 
the 
appearance of complete subgraphs of order $4$ and $p\sim N^{-1/2}$ corresponds 
to 
complete subgraphs of order $5$. As $z$ approaches $0$, the graph contains 
complete subgraphs of increasing order.}
\label{fig_transition} 
\end{figure}

It is instructive to view the results discussed above from a different
point of view. Consider a random graph with $N$ nodes and assume that
the connection probability $p(N)$ scales as $N^z$, where $z$ is a
tunable parameter that can take any value between $-\infty$ and $0$
(Fig. \ref{fig_transition}). For $z$ less than $-3/2$ almost all graphs
contain only isolated nodes and edges. When $z$ passes through $-3/2$,
trees of order $3$ suddenly appear. When $z$ reaches $-4/3$, trees of
order $4$ appear, and as $z$ approaches $-1$, the graph contains trees
of larger and larger order. However, as long as $z<-1$, such that the
average degree of the graph $\langle k\rangle=pN\rightarrow 0$ as
$N\rightarrow \infty$, the graph is a union of disjoint trees, and
cycles are absent. Exactly when $z$ passes through $-1$, corresponding
to $\langle k\rangle=$const, even though $z$ is changing smoothly, the
asymptotic probability of cycles of all orders jumps from $0$ to $1$.
Cycles of order $3$ can be also viewed as complete subgraphs of order
$3$. Complete subgraphs of order $4$ appear at $z=-2/3$, and as $z$
continues to increase, complete subgraphs of larger and larger order
continue to emerge. Finally, as $z$ approaches $0$, almost every random
graph approaches the complete graph of $N$ points.
 
Further results can be derived for $z=-1$, i.e. when we have $p\propto
N^{-1}$ and the average degree of the nodes is $\langle k\rangle=$const.
For $p\propto N^{-1}$ a random graph contains trees and cycles of all
order, but so far we have not discussed the size and structure of a
typical graph component. A component of a graph is by definition a
connected, isolated subgraph, also called a cluster in network research
and percolation theory. As Erd\H{o}s and R\'enyi (1960) show, there is
an abrupt change in the cluster structure of a random graph as $\langle
k\rangle$ approaches $1$. 

If $0<\langle k\rangle<1$, almost surely all clusters are either trees
or clusters containing exactly one cycle. Although cycles are present,
almost all nodes belong to trees. The mean number of clusters is of
order $N-n$, where $n$ is the number of edges, i.e. in this range by
adding a new edge the number of clusters decreases by $1$. The largest
cluster is a tree, and its size is proportional to $\ln N$. 

When $\langle k\rangle$ passes the threshold $\langle k\rangle_c=1$, the
structure of the graph changes abruptly. While for $\langle k\rangle<1$
the greatest cluster is a tree, for $\langle k\rangle_c=1$ it has
approximately $N^{2/3}$ nodes and has a rather complex structure.
Moreover for $\langle k\rangle>1$ the greatest (giant) cluster has
$[1-f(\langle k\rangle)]N$ nodes, where $f(x)$ is a function that
decreases exponentially from $1$ to $0$ for $x\rightarrow \infty$. Thus
a finite fraction $S=1-f(\langle k\rangle)$ of the nodes belongs to the
largest cluster. Except for this giant cluster, all other clusters are
relatively small, most of them being trees, the total number of nodes
belonging to trees being $Nf(\langle k\rangle)$. As $\langle k\rangle$
increases, the small clusters coalesce and join the giant cluster, the
smaller clusters having the higher chance of survival. 

Thus at $p_c\simeq 1/N$ the random graph changes its topology abruptly
from a loose collection of small clusters to being dominated by a single
giant cluster. The beginning of the supercritical phase was studied by
Bollob\'as (1984), Kolchin (1986) and Luczak (1990). Their results show
that in this region the largest cluster clearly separates from the rest
of the clusters, its size $S$ increasing proportionally with the
separation from the critical probability, 

\begin{equation}
S\propto (p-p_c).
\end{equation}
As we will see in Sect. \ref{sect_inf_perc}, this dependence is analogous with 
the scaling of the percolation probability in infinite dimensional 
percolation.

\subsection{Degree Distribution}
\label{sect_degree_er}

Erd\H{o}s and R\'enyi (1959) were the first to study the distribution of
the maximum and minimum degree in a random graph, the full degree
distribution being derived later by Bollob\'as (1981). 

In a random graph with connection probability $p$ the degree $k_i$ of a
node $i$ follows a binomial distribution with parameters $N-1$ and $p$
\begin{equation} \label{ind_con} P(k_i=k)=C_{N-1}^{\,k}p^k(1-p)^{N-1-k}.
\end{equation} This probability represents the number of ways in which
$k$ edges can be drawn from a certain node: the probability of $k$ edges
is $p^k$, the probability of the absence of additional edges is
$(1-p)^{N-1-k}$, and there are $C_{N-1}^k$ equivalent ways of selecting
the $k$ endpoints for these edges. Furthermore, if $i$ and $j$ are
different nodes, $P(k_i=k)$ and $P(k_j=k)$ are close to be independent
random variables. To find the degree distribution of the graph, we need
to study the number of nodes with degree $k$, $X_k$. Our main goal is to
determine the probability that $X_k$ takes on a given value, $P(X_k=r)$.

According to (\ref{ind_con}), the expectation value of the number of
nodes with degree $k$ is
\begin{equation}
E(X_k)=NP(k_i=k)=\lambda_k,
\end{equation}
where
\begin{equation}
\label{lambda_k}
\lambda_k=NC_{N-1}^{\,k}p^k(1-p)^{N-1-k}.
\end{equation}

\begin{figure}[htb]
\centerline{\hspace{-0.7in}\psfig{figure=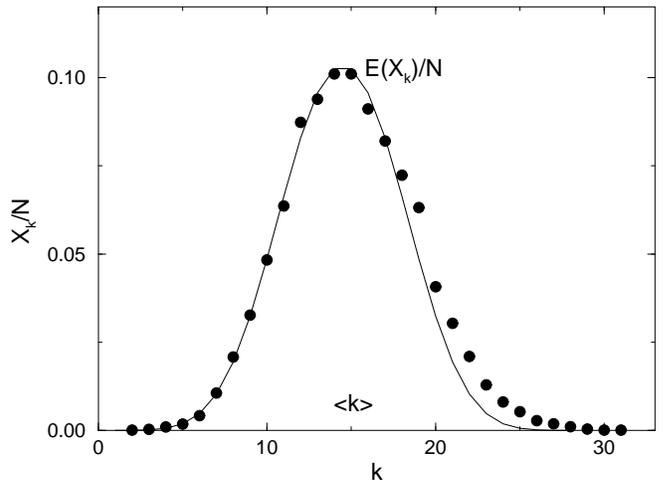,width=2.8in,angle=
-90
}}
\caption{The degree distribution that results from the numerical simulation of 
a 
random graph. We generated a single random graph with $N=10,000$ nodes and 
connection probability $p=0.0015$, and calculated the number of nodes with 
degree 
$k$, $X_k$. The plot compares $X_k/N$ with the expectation value of the 
Poisson 
distribution (\ref{poisson}), $E(X_k)/N=P(k_i=k)$, and we can see that the 
deviation is small. }
\label{fig_pk}
\end{figure}

As in the derivation of the existence
conditions of subgraphs (see Sect. \ref{sect_subgraph}), the distribution of the $X_k$ values, $P(X_k=r)$, approaches a Poisson distribution
\begin{equation}
\label{poisson}
P(X_k=r)=e^{-\lambda_k}\frac{\lambda_k^r}{r!}.
\end{equation} Thus the number of nodes with degree $k$ follows a Poisson
distribution with mean value
$\lambda_k$. Note that the distribution (\ref{poisson}) has as expectation 
value 
the function $\lambda_k$ given by (\ref{lambda_k}) and not a constant. The Poisson distribution decays rapidly for large 
values of $r$, the standard deviation of the distribution being 
$\sigma_k=\sqrt{\lambda_k}$. With a bit of simplification we could say that 
(\ref{poisson}) implies that $X_k$ does not diverge much from the 
approximative 
result $X_k=NP(k_i=k)$, valid only if the nodes are independent (see Fig. 
\ref{fig_pk}). Thus with a good approximation the degree distribution of a 
random 
graph is a binomial distribution
\begin{equation}
P(k)=C_{N-1}^{\,k}p^k(1-p)^{N-1-k},
\end{equation}
which for large $N$ can be replaced by a Poisson distribution
\begin{equation}
\label{poisson2}
P(k)\simeq e^{-pN}\frac{(pN)^k}{k!}=e^{-\langle k\rangle}\frac{\langle 
k\rangle^k}{k!}.
\end{equation}

Since the pioneering paper of Erd\H{o}s and R\'enyi, much work has
concentrated on the existence and uniqueness of the minimum and maximum
degree of a random graph. The results indicate that for a large range of
$p$ values both the maximum and the minimum degrees are determined and
finite. For example, if $p(N)\sim N^{-1-1/k}$ (thus the graph is a set
of isolated trees of order at most $k+1$) almost no graph has nodes with
degree higher than $k$. On the other extreme, if
$p=[\ln(N)+k\ln(\ln(N))+c]/N$, almost every random graph has minimum
degree of at least $k$. Furthermore, for a sufficiently high $p$,
respectively if $pN/\ln(N)\rightarrow\infty$, the maximum degree of
almost all random graphs has the same order of magnitude as the average
degree. Thus despite the fact that the position of the edges is random,
a typical random graph is rather homogeneous, the majority of the nodes
having the same number of edges.
   
\subsection{Connectedness and Diameter}
\label{sect_diam_er} 

The diameter of a graph is the maximal distance between any pair of its
nodes. Strictly speaking, the diameter of a disconnected graph (i.e.
made up of several isolated clusters) is infinite, but it can be defined
as the maximum diameter of its clusters. Random graphs tend to have
small diameters, provided $p$ is not too small. The reason for this is
that a random graph is likely to be spreading: with large probability
the number of nodes at a distance $l$ from a given node is not much
smaller than $\langle k\rangle^l$. Equating $\langle k\rangle^l$ with
$N$ we find that the diameter is proportional with $\ln(N)/\ln(\langle
k\rangle)$, thus it depends only logarithmically on the number of nodes.

The diameter of a random graph has been studied by many authors (see Chung and Lu 2001). A general conclusion
is that for most values of $p$, almost all graphs have precisely the
same diameter. This means that when we consider all graphs with $N$ nodes and connection probability $p$, the range of values 
in 
which the diameters of these graphs can vary is very small, usually 
concentrated around 

\begin{equation}
\label{diam_er}
d=\frac{\ln(N)}{\ln(pN)}=\frac{\ln(N)}{\ln(\langle k\rangle)}.
\end{equation}
 In the following we summarize a few important results:

\begin{itemize}
\item{If $\langle k\rangle=pN<1$ the graph is composed of isolated trees and 
its diameter equals the diameter of a tree.}
\item{If $\langle k\rangle>1$ a giant cluster appears. The diameter of the 
graph equals the diameter of the giant cluster if $\langle k\rangle\geq 3.5$, and is proportional to $\ln(N)/\ln(\langle k\rangle)$.}
\item{If $\langle k\rangle\geq \ln(N)$ the graph is totally connected. Its 
diameter is concentrated on a few values around $\ln(N)/\ln(\langle k\rangle)$.}
\end{itemize}
Another way to characterize the spread of a random graph is to calculate the 
average distance between any pair of nodes, or the average path length. One 
expects that the average path length scales with the number of nodes in the 
same 
way as the diameter 
\begin{equation}
\label{path_er}
\ell_{rand}\sim \frac{\ln(N)}{\ln(\langle k\rangle)}.
\end{equation}

\begin{figure}[htb]
\centerline{\psfig{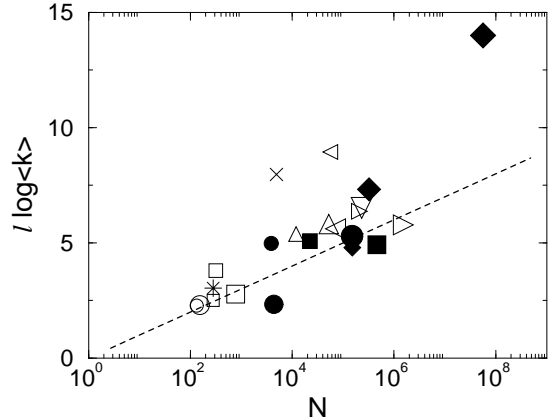}}
\vspace{1cm}
\caption{Comparison between the average path length of real networks and the 
prediction (\ref{path_er}) of random graph theory (dashed line). For 
each 
symbol we indicate the corresponding number in Table \ref{table_cluster} or 
Table 
\ref{table_gamma}: small circle, I.12; large circle, I.13; star, I.17; small 
square, I.10; medium square, I.11; large square, II.13; small filled circle, 
II.6; medium filled circle, I.2; X, I.16; small upwards triangle, I.7; small filled square, I.15; large 
upwards triangle, I.4; small left triangle, I.5; large left triangle, I.6; 
large 
filled circle, II.6; small filled diamond, I.1; small right triangle, I.7, 
downwards triangle, I.3; medium filled diamond, II.1; large filled square, I.14; large right triangle, I.5; 
large filled diamond, II.3.}
\label{fig_fit_diam_er}
\end{figure}

 In Sect. \ref{sect_real_data} we have presented evidence that the average 
path 
length of real networks is close to the average path length of random graphs 
with the same size. Eq. (\ref{path_er}) gives us an opportunity to  better 
compare random graphs and real networks (see Newman 2001a,b). According to Eq. (\ref{path_er}), the 
product $\ell_{rand} \ln(\langle k\rangle)$ is equal to $\ln(N)$, so plotting 
$\ell_{rand} \ln(\langle k\rangle)$ as a function of $\ln(N)$ for random 
graphs 
of different sizes gives a straight line of slope $1$. On Fig. 
$\ref{fig_fit_diam_er}$ we plot this product for several real networks, 
$\ell_{real} \log(\langle k\rangle)$, as a function of the network size, 
comparing it with the prediction of Eq. (\ref{path_er}). We can see that the 
trend of the data is similar with the theoretical prediction, and with several 
exceptions Eq. (\ref{path_er}) gives a reasonable first estimate. 

\vspace{0.5cm}

\subsection{Clustering coefficient}
\label{sect_rand_clust}

As we already mentioned in Sect. \ref{sect_real_data}, complex networks exhibit 
a large degree of clustering. If we consider a node 
in 
a random graph and its first neighbors, the probability that two of these 
neighbors are connected is equal with the probability that two randomly 
selected 
nodes are connected. Consequently the clustering coefficient of a random graph 
is 
\begin{equation}
\label{clust_coeff_er}
C_{rand}=p=\frac{\langle k\rangle}{N}.
\end{equation}

\begin{figure}[htb]
\vspace{0.3cm}
\centerline{\psfig{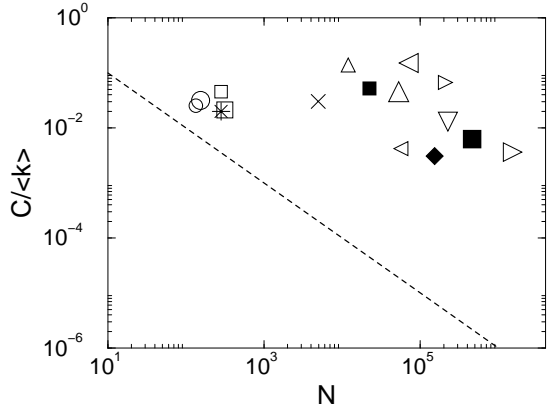}}
\vspace{1cm}
\caption{Comparison between the clustering coefficient of real networks and 
random graphs. All networks 
from 
Table \ref{table_cluster} are included in the figure, the symbols being the 
same 
as in Fig. \ref{fig_fit_diam_er}. The dashed line corresponds to Eq. 
(\ref{clust_coeff_er}).}
\label{fig_fit_ccoef_er}
\end{figure}

According to Eq. (\ref{clust_coeff_er}), if we plot the ratio 
$C_{rand}/\langle 
k\rangle$ as a function of $N$ for random graphs of different sizes, on a 
log-log 
plot they will align along a straight line of slope $-1$. On Fig. 
\ref{fig_fit_ccoef_er} we plot the ratio of the clustering coefficient of real 
networks and their average degree as a function of their size, comparing it 
with the prediction of Eq. (\ref{clust_coeff_er}). The plot convincingly 
indicates that real networks do not follow the prediction of random graphs. 
The 
fraction $C/\langle k\rangle$ does not decrease as $N^{-1}$, instead, it 
appears 
to be independent of $N$. This property is characteristic to large ordered 
lattices, whose clustering coefficient depends only on the coordination number 
of 
the lattice and not their size (Watts and Strogatz 1998).

\subsection{Graph spectra}
\label{sect_spect_er}

 Any graph $G$ with $N$ nodes can be represented by its adjacency matrix $A(G)$ 
with $N\times N$ elements $A_{ij}$, whose value is $A_{ij}=A_{ji}=1$ if nodes 
$i$ 
and $j$ are connected, and $0$ otherwise. The spectrum of graph $G$ is the set 
of 
eigenvalues of its adjacency matrix $A(G)$. A graph with $N$ nodes has $N$ eigenvalues $\lambda_j$, and it is useful to define its spectral 
density as
\begin{equation}
\rho(\lambda)=\frac{1}{N}\sum_{j=1}^{N}\delta(\lambda-\lambda_j),
\end{equation}
which approaches a continuous function if $N\rightarrow \infty$. The interest in spectral properties is related to the fact that the spectral 
density can be directly related to the graph's topological 
features, 
since its $k$th moment can be written as

\begin{equation}
\frac{1}{N}\sum_{j=1}^{N}(\lambda_j)^k=\frac{1}{N}\sum_{i_1,i_2,..\i_k}A_{i_1,
i_2
}A_{i_2i_3}..A_{i_ki_1},
\end{equation} i.e. the number of paths returning to the same node in the graph. Note that these 
paths can contain nodes which were already visited.

\begin{figure}[htb]
\vspace{-2.5cm}
\centerline{\psfig{figure=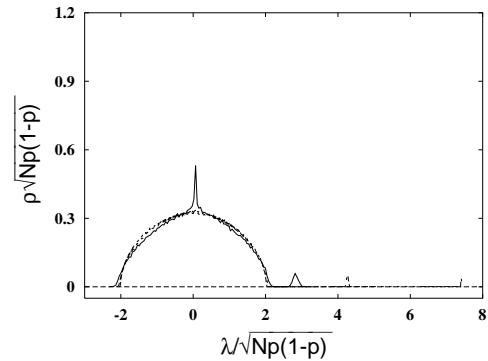,width=2.8in}}
\vspace{-1.6cm}
\caption{Rescaled spectral density of three random graphs having $p=0.05$ and 
size $N=100$ (continuous line), $N=300$ (dashed line) and $N=1000$ 
(short-dashed 
line). The isolated peak corresponds to the principal eigenvalue. After Farkas {\it et al.} 2001.}
\label{fig_spect_er}
\end{figure}

Let us consider a random graph $G_{N,p}$ satisfying $p(N)=cN^{-z}$. For 
$z< 1$ there is an infinite cluster in the graph (see Sect. 
\ref{sect_graph_evol}), and as $N\rightarrow \infty$, any node belongs almost 
surely to the infinite cluster. In this case the spectral density of the 
random 
graph converges to a semicircular distribution (Fig. \ref{fig_spect_er}) 
\begin{equation}
\rho(\lambda)=\left\{\begin{array}{ll}
\frac{\sqrt{4Np(1-p)-\lambda^2}}{2\pi Np(1-p)} &\mbox{if}\quad 
|\lambda|<2\sqrt{Np(1-p)}\\
0 &\mbox{otherwise}
\end{array}\right..
\label{semicirc}
\end{equation}
Known as Wigner's (see Wigner 
1955, 1957, 1958), or the semicircle law, (\ref{semicirc}) has many 
applications in quantum, statistical and solid state physics (Mehta 1991, Crisanti {\it 
et 
al.} 1993, Guhr {\it et al.} 1998). The largest (principal) eigenvalue, $\lambda_1$, is isolated from 
the 
bulk of the spectrum, and it increases with the network size as $pN$.

When $z>1$ the spectral density deviates from the semi-circle 
law. 
The most striking feature of $\rho(\lambda)$ is that its odd moments are equal 
to 
zero, indicating that the only way that a path comes back to the original node 
is 
if it returns following exactly the same nodes. This is a salient feature of a 
tree structure, and, indeed, in Sect. \ref{sect_subgraph} we have seen that in 
this case the random graph is composed of trees.

\section{PERCOLATION THEORY}
\label{sect_perc}

One of the most interesting findings of random graph theory is the existence 
of a 
critical probability at which a giant cluster forms. Translated into network 
language, it indicates the existence of a critical probability $p_c$ such that below $p_c$ the network is composed of isolated clusters but 
above $p_c$ a giant cluster spans the entire network. This phenomenon is 
markedly 
similar to a percolation transition, a topic much studied both in mathematics 
and 
in statistical mechanics (Stauffer and Aharony 1992, Bunde and Havlin 1994, 1996, Grimmett 1999, ben Avraham and Havlin 2000). Indeed, the percolation transition and the emergence 
of 
the giant cluster are the same phenomenon expressed in different languages. 
Percolation theory, however, does not simply reproduce the predictions of 
random 
network theory. Asking questions from  a different perspective, it addresses 
several issues that are crucial for understanding real networks, but are not 
discussed by random graph theory. Consequently, it is important to review the 
predictions of percolation theory relevant to networks, as they are 
crucial 
to understand important aspects of the network topology.

\subsection{Quantities of interest in percolation theory}
\label{sect_def}

\begin{figure}[htb]
\vspace{-1in}
\centerline{\psfig{figure=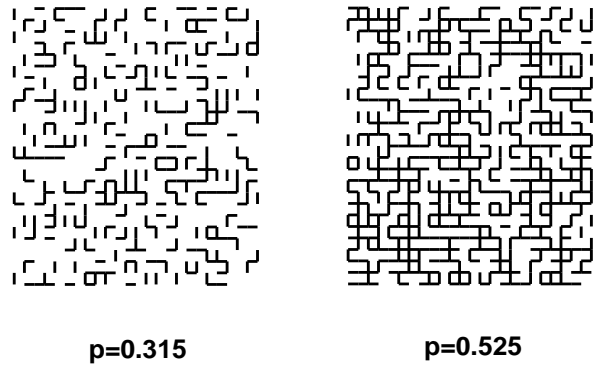,height=4in,width=5.1in,angle=-90}}
\vspace{-2.0in}
\caption{Illustration of bond percolation in 2D. The nodes are placed on a 
$25\times 25$ square lattice, and two nodes are connected by an edge with 
probability $p$. For $p=0.315$ (left), which is below the percolation threshold  $p_c=0.5$, the connected nodes form isolated clusters. For $p=0.525$ (right), which is above 
the percolation threshold, the largest cluster percolates.}
\label{fig_perc}
\end{figure}

Consider a regular $d$-dimensional lattice whose edges are present
with probability $p$ and absent with probability $1-p$. Percolation
theory studies the emergence of paths that percolate the lattice
(starting at one side and ending at the opposite side). For small $p$ only a 
few 
edges are present, thus only small clusters of nodes connected by edges can 
form, 
but at a
critical probability $p_c$, called the percolation threshold, a
percolating cluster of nodes connected by edges appears (see 
Fig.$\,$\ref{fig_perc}). This cluster is also called the infinite cluster, 
because its size diverges as the size of the lattice increases. There are 
several 
much studied versions of percolation, the one presented above being ``bond 
percolation''. The most known alternative is site percolation, in which all 
bonds 
are present and the nodes of the lattice are occupied with probability $p$. In 
a 
similar way as bond percolation, for small $p$ only finite clusters of 
occupied 
nodes are present, but for $p>p_c$ an infinite cluster appears.

 The main quantities of interest in percolation are:

{\it The percolation probability}, $P$, denoting the probability
that a given node belongs to the infinite
cluster:
\begin{equation}
\label{percprob}
P=P_p(|C|=\infty)=1-\sum_{s<\infty} P_p(|C|=s),
\end{equation}
where $P_p(|C|=s)$ denotes the probability that the cluster at the origin has 
size $s$. Obviously
\begin{equation}
P=\left\{\begin{array}{rcl} 0 & \mbox {if}& p<p_c\\ >0 & \mbox {if}&
p>p_c
\end{array}\right..
\end{equation}
 
{\it The average cluster size}, $\langle s\rangle$, defined as
\begin{equation}
\label{clustsize}
\langle s\rangle=E_p(|C|)=\sum_{s=1}^\infty s P_p(|C|=s),
\end{equation}
giving the expectation value of cluster sizes. $\langle s\rangle$ is infinite 
when
$P>0$, thus in this case it is useful to work with the average size of
the finite clusters by taking away from the system the infinite ($|C|=\infty$) 
cluster
\begin{equation}
\langle s\rangle^f=E_p(|C|, |C|<\infty)=\sum_{s<\infty} s P_p(|C|=s).
\end{equation}

{\it The cluster size distribution}, $n_s$, defined as the probability
of a node being the left hand end of a cluster of size $s$,
\begin{equation}
\label{ns}
n_s=\frac{1}{s}P_p(|C|=s).
\end{equation}Note that $n_s$ does not coincide with the probability 
that a node is part of a cluster of size $s$. By fixing the position of the 
node 
in the cluster (asking it to be the left hand end of the cluster), we are 
choosing one of the $s$ possible nodes of the cluster, reflected in the 
division 
of $P_p(|C|=s)$ by $s$, and counting every cluster only once.

These quantities are of interest in random networks as well. There is, 
however, 
an important difference between percolation theory and random networks: 
percolation theory is defined on a regular $d$-dimensional lattice. In a 
random 
network (or graph) we can define a non-metric distance along the edges, but 
since 
any node can be connected by an edge to any other node in the network, there 
is 
no regular small-dimensional lattice a network can be embedded into. However, 
as 
we discuss below, random networks and percolation theory meet exactly in the 
infinite dimensional limit ($d\rightarrow\infty$) of percolation. Fortunately, 
many results in percolation theory can be generalized to infinite dimensions. Consequently, the 
results 
obtained within the context of percolation apply directly to random networks 
as 
well.

\subsection{General results}
\label{sect_gen_perc} 

{\it The subcritical phase ($p<p_c$):} When $p<p_c$, only small clusters of nodes connected by edges are
present in the system. The questions asked in this phase are (i) what is 
the probability that there exists a path $x\leftrightarrow y$ joining two 
randomly chosen nodes $x$ and $y$ and (ii)
what is the rate of decay of $P_p(|C|=s)$ when
$s\rightarrow\infty$. The first result of this type was obtained by Hammersley (1957) who showed that the probability of a path
joining the origin with a node on the surface, $\partial B(r)$, of a box 
centered 
at the origin and with side-length $2r$
decays exponentially if $P<\infty$. We can define a
correlation length $\xi$ as the characteristic
length of the exponential decay

\begin{equation}
\label{subxi}
P_p(0\leftrightarrow \partial B(r))\sim e^{-\frac{r}{\xi}},
\end{equation}
where $0\leftrightarrow \partial B(r)$ means that there is a path from the 
origin 
to an arbitrary node on $\partial B(r)$. Equation (\ref{subxi}) indicates that 
the radius of the finite clusters in the
subcritical region has an exponentially decaying tail, and the correlation 
length 
represents the mean radius of a finite cluster. It was shown (see Grimmett 
1999) 
that $\xi$ is equal to $0$ for
$p=0$ and goes to infinity as $p\rightarrow p_c$.

 The exponential decay of cluster radii implies that the probability that a 
cluster has size $s$, $P_p(|C|=s)$, also decays
 exponentially for large $s$:
\begin{equation}
\label{size_sub}
P_p(|C|=s)\sim e^{-\alpha(p)s}\quad \mbox{as}\quad s\rightarrow\infty,
\end{equation}
where $\alpha(p)\rightarrow\infty$ as $p\rightarrow 0$ and
$\alpha(p_c)=0$.

{\it The supercritical phase ($p>p_c$):} For $P>0$ there is
exactly one infinite cluster (Burton and Keane 1989). In this supercritical phase the previously 
studied 
quantities are dominated by the contribution of the infinite
cluster, thus it is useful to study the
corresponding probabilities in terms of the finite
clusters. The probability that there is a path 
from the origin to the surface of a box of edge length $2r$ which is not part 
of 
the infinite cluster decays exponentially as 
\begin{equation}
\label{superxi}
P_p\left(0\leftrightarrow\partial B(r),|C|<\infty\right) \sim
e^{-\frac{r}{\xi}}.
\end{equation}
Unlike the subcritical phase, though, the decay of the cluster sizes, $P_p(|C|=s<\infty)$, follows a stretch-exponential,  $e^{-\beta(p)s^{\frac{d-1}{d}}}$, offering the first important quantity that depends on the dimensionality of the lattice, but even this dependence vanishes as $d\rightarrow 
\infty$, and the cluster size distribution decays exponentially as in the 
subcritical phase. 

\subsection{Exact solutions: percolation on a Cayley tree}
\label{sect_cayley}

\begin{figure}[htb]
\centerline{\psfig{figure=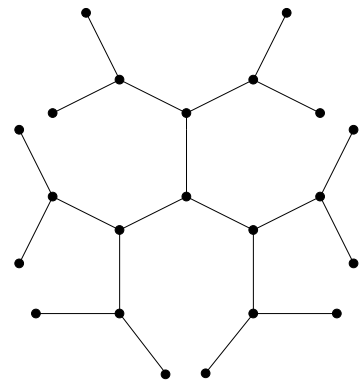,width=2.8in}}
\vspace{-3cm}
\caption{Example of a Cayley tree with coordination number $z=3$. All of the 
nodes have $3$ edges, with the exception of those on the surface, which have 
only one edge. The ratio between the number of 
nodes on the surface and the total number of nodes approaches a constant, $\frac{z-2}{z-1}$, a  property valid only for infinite dimensional objects. 
The average degree approaches $\langle k \rangle=2$ as the size of the tree  goes to infinity, a property common with random trees (see Sect. 
\ref{sect_subgraph}). }
\label{fig_cayley}
\end{figure}

The Cayley tree (or Bethe lattice) is a loopless structure (see Fig.$\,$\ref{fig_cayley}) in which every node has $z$ neighbors, with 
the 
exception of the nodes at the surface. While the 
surface and volume of a regular $d$-dimensional object obey the scaling 
relation
$\mbox{surface}\propto\mbox{volume}^{1-1/d}$, and only in the limit $d\rightarrow\infty$ is 
the 
surface proportional with the volume, for the Cayley tree the number of nodes on the surface is proportional to the total number of nodes (i.e. the volume of the tree). Thus in this respect the Cayley 
tree 
represents an infinite-dimensional object. Another argument for the 
infinite-dimensionality of the Cayley tree is that it has no loops (cycles in 
graph theoretic language). Thus, despite its regular topology, the 
Cayley 
tree represents a reasonable approximation for the topology of a random 
network 
in the subcritical phase, where all the clusters are trees. This is no longer 
true in the supercritical phase, because at the critical probability $p_c(N)$, 
cycles of all order appear in the graph (see Sect. \ref{sect_graph_evol}).

To investigate percolation on a Cayley tree, we assume that each edge is present with probability $p$. Next we discuss the main quantities of interest for this system.

{\it Percolation threshold:} The condition for the 
existence of an infinite path starting from the origin is that at least one of the $z-1$ possible outgoing edges of a node is present, i.e. $(z-1)p\geq 1$. Therefore the percolation threshold is
\begin{equation}
p_c=\frac{1}{z-1}.
\end{equation}

{\it Percolation probability:} For a Cayley tree with $z=3$, for which 
$p_c=1/2$, the percolation probability is given by (Stauffer and Aharony 1992)
\begin{equation}
P=\left\{\begin{array}{ccl} 0 & \mbox {if}& p<p_c=\frac{1}{2}\\ (2p-1)/p^2 & 
\mbox {if}&
p>p_c=\frac{1}{2}
\end{array}\right.
\end{equation}

The Taylor series expansion around $p_c=\frac{1}{2}$ gives $P\simeq 
8(p-\frac{1}{2})$, thus the percolation probability is proportional to the 
deviation from the percolation threshold
\begin{equation}
P\propto (p-p_c)\quad\mbox{as}\,\, p\rightarrow p_c.
\end{equation}

{\it Mean cluster size:} The average cluster size is given by  
\begin{equation}
\langle s\rangle=\sum_{n=1}^\infty 3\times 
2^{n-1}p^n=\frac{3}{2}\frac{1}{1-2p}=\frac{3}{4}(p_c-p)^{-1}.
\end{equation} 
Note that $\langle s\rangle$ diverges as $p\rightarrow p_c$, and it depends on $P$ as a power of the distance $p_c-p$ from the percolation threshold. This behavior is an example of critical phenomena: an order parameter goes to zero 
following a power-law in the vicinity of the critical point (Stanley 1971, Ma 1976).

{\it Cluster size distribution:} The probability of having a cluster of size $s$ is  (Durett 1985) 
\begin{equation}
P_p(|C|=s)=\frac{1}{s}C_{2s}^{s-1}p^{s-1}(1-p)^{s+1}.
\end{equation}
Here the number of edges surrounding the $s$ nodes is $2s$, from which the 
$s-1$ 
inside edges have to be present, and the $s+1$ external ones absent. The 
factor 
$C_{2s}^{s-1}$ takes into account the different cases that can be obtained 
when 
permuting the edges, and the $\frac{1}{s}$ is a normalization factor. Since $n_s=\frac{1}{s} 
P_p(|C|=s)$, after using Stirling's formula we obtain
\begin{equation}
n_s\propto s^{-5/2}p^{s-1}(1-p)^{s+1}.
\end{equation}In the vicinity of the percolation threshold this expression can be 
approximated 
with 
\begin{equation}
\label{cayleyns}
n_s\sim s^{-5/2}e^{-cs}\quad\mbox{with}\,\,c\propto (p-p_c)^2
\end{equation}  

Thus the cluster size distribution follows a power-law with an exponential 
cutoff: only clusters with size $s<s_\xi=1/c\propto (p-p_c)^{-2}$ contribute 
significantly to cluster averages. For these clusters, $n_s$ is effectively 
equal 
to $n_s(p_c)\propto s^{-5/2}$. Clusters with $s\gg s_\xi$ are exponentially 
rare, 
and their properties are no longer dominated by the behavior at $p_c$. The 
notation $s_\xi$ illustrates that as the correlation length $\xi$ is the 
characteristic lengthscale for the cluster diameters, $s_\xi$ is an intrinsic 
characteristic of cluster sizes. The correlation length of a tree 
is 
not well defined, but we will see in the more general cases that $s_\xi$ and 
$\xi$ are related by a simple power-law.
      
\subsection{Scaling in the critical region}
\label{sect_scale_perc}

The principal ansatz of percolation theory is that even the most general 
percolation problem in any dimension obeys a scaling relation similar to Eq. 
(\ref{cayleyns}) near the percolation threshold. Thus in general the cluster 
size 
distribution can be written as
\begin{equation}
\label{gen_scale}
n_s(p)\sim \left\{\begin{array}{rcl}
s^{-\tau}f_{-}(|p-p_c|^{1/\sigma}s)& \mbox{as}& p\leq p_c\\
s^{-\tau}f_{+}(|p-p_c|^{1/\sigma}s)& \mbox{as}& p\geq p_c
\end{array}\right..
\end{equation}

Here $\tau$ and $\sigma$ are critical exponents whose numerical value needs to 
be 
determined,  $f_{-}$ and $f_{+}$ are smooth functions on $[0,\infty)$, and 
$f_{-}(0)=f_{+}(0)$. The results of Sect. \ref{sect_gen_perc} suggest that 
$f_{-}(x)\simeq e^{-Ax}$ and $f_{+}(x)\simeq e^{-Bx^{(d-1)/d}}$ for $x>>1$. 
This 
ansatz indicates that the role of $s_\xi\propto|p-p_c|^{-1/\sigma}$ as a 
cutoff 
is the same as in the Cayley tree. The general form (\ref{gen_scale}) contains 
as 
special case the Cayley tree (\ref{cayleyns}) with $\tau=5/2$, $\sigma=1/2$ and 
$f_{\pm}(x)=e^{-x}$.
 
Another element of the scaling hypothesis is that the correlation length 
diverges 
near the percolation threshold following a power-law:
\begin{equation}
\xi(p)\sim |p-p_c|^{-\nu}\quad\mbox{as}\,\,p\rightarrow p_c.
\end{equation}
This ansatz introduces the correlation exponent, $\nu$, and indicates that 
$\xi$ 
and $s_{\xi}$ are related by a power-law $s_\xi=\xi^{1/\sigma\nu}$. From these two hypotheses we find that the percolation probability (\ref{percprob}) is given by
\begin{equation}
P\sim (p-p_c)^\beta\quad\mbox{with}\,\,\beta=\frac{\tau-2}{\sigma},
\end{equation} 
which scales as a positive power of $p-p_c$ for $p\geq 
p_c$, thus it is $0$ for $p=p_c$ and increases when $p>p_c$. The average size 
of 
finite clusters, $\langle s\rangle^f$, which can be calculated on both sides 
of 
the percolation threshold, obeys 
\begin{equation}
\langle s\rangle^f\sim 
|p-p_c|^{-\gamma}\quad\mbox{with}\,\,\gamma=\frac{3-\tau}{\sigma},
\end{equation} diverging for $p\rightarrow p_c$. The exponents $\beta$ and 
$\gamma$ are called the critical exponents of the percolation probability and 
average cluster size, respectively.

\subsection{Cluster structure}
\label{sect_clust_perc}

Until now we discussed the cluster sizes and radii, ignoring their 
internal 
structure. Let us first focus on the perimeter of a cluster, $t$, denoting the 
number of nodes situated on the most external edges (the leaf nodes). The perimeter $t_s$ of a very large but finite cluster of size $s$ scales 
as (Leath 1976)
\begin{equation}
t_s=s\frac{1-p}{p}+As^\zeta\quad\mbox{as}\,\,s\rightarrow\infty,
\end{equation} 
where $\zeta=1$ for $p<p_c$ and $\zeta=1-1/d$ for $p>p_c$. Thus below $p_c$ the perimeter of the clusters is proportional with their 
volume, a highly irregular property, which is nevertheless true for trees, including the Cayley tree. 

Another way of understanding the unusual structure of finite clusters is by 
looking at the relation between their radii and volume. The correlation length 
$\xi$ is a measure of the mean cluster radius, and we know that $\xi$ scales 
with 
the cutoff cluster size $s_\xi$ as $\xi\propto s_\xi^{1/\nu\sigma}$. Thus the 
finite clusters are fractals (see Mandelbrot 1982), because their size does 
not 
scale as their radius to the $d$th power, but as
\begin{equation}
\label{fract_dim}
s(r)\sim r^{d_f},
\end{equation} where $d_f=1/\sigma\nu$. It can be also shown that at the 
percolation threshold the infinite cluster is still a fractal, but for $p>p_c$ 
it 
becomes a normal $d$-dimensional object.

While the cluster radii and the correlation length $\xi$ are defined using the 
Euclidian distances on the lattice, the chemical 
distance is defined as the length of the shortest path between two 
arbitrary 
sites on a cluster (Havlin and Nossal 1984). Thus  the chemical distance is the equivalent of the distance 
on 
random graphs. The number of nodes within chemical distance $\ell$ scales as 
\begin{equation}
\label{diam_perc}
s(\ell)\sim \ell^{d_{\ell}},
\end{equation}
where $d_{\ell}$ is called the graph dimension of the cluster. While the 
fractal dimension $d_f$ of the Euclidian distances has been related to the other critical exponents, no such relation has been found yet for the graph dimension 
$d_{\ell}$.

\subsection{Infinite dimensional percolation}
\label{sect_inf_perc}
 
Percolation is known to have a critical 
dimension 
$d_c$, below which some exponents depend on $d$, but for any dimension above 
$d_c$ the exponents are the same. While it is generally believed that the 
critical dimension of percolation is $d_c=6$, the dimension independence of 
the 
critical exponents is proven rigorously only for $d\geq 19$ (see Hara and 
Slade 
1990). Thus for $d>d_c$ the results of the infinite dimensional percolation 
theory apply, which predict that
\begin{itemize}
\item{$P\sim (p-p_c)\quad\mbox{as}\,\,p\rightarrow p_c;$}
\item{$\langle s\rangle\sim (p_c-p)^{-1}\quad\mbox{as}\,\,p\rightarrow p_c;$}
\item{$n_s\sim s^{-5/2}e^{-|p-p_c|^{2}s}\quad\mbox{as}\,\,p\rightarrow 
p_c;$}
\item{$\xi\sim |p-p_c|^{-1/2}\quad\mbox{as}\,\,p\rightarrow p_c.$}
\end{itemize}

Consequently, the critical exponents of the infinite dimensional percolation 
are 
$\tau_{\infty}=5/2$, $\sigma_{\infty}=1/2$ and $\nu_{\infty}=1/2$. The fractal dimension of the infinite cluster at the percolation threshold is $d_f=4$, while   graph dimension is $d_{\ell}=2$ (Bunde, Havlin 1996). Thus the characteristic 
chemical distance on a finite cluster or infinite cluster at the percolation threshold scales with its size as
\begin{equation}
\label{diam_clust_inf}
\ell\sim s^{2/d_f}=s^{1/2}.
\end{equation}
  
\subsection{Parallels between random graph theory and percolation}
\label{sect_parall}

In random graph theory we study a graph of $N$ nodes, each pair of nodes 
being connected with probability $p$. This corresponds to percolation in at most $N$ 
dimensions, such that each two connected nodes are neighbors, and the edges 
between graph nodes are the edges in the percolation problem. Since random 
graph 
theory investigates the $N\rightarrow\infty$ regime, it is analogous with 
infinite dimensional percolation.   

We have seen in Sect. \ref{sect_cayley} that infinite dimensional percolation 
is 
similar to percolation on a Cayley tree. The percolation threshold of the 
Cayley tree is $p_c=1/(z-1)$, where $z$ is the coordination number of the 
tree. In a random graph of $N$ nodes 
the coordination number is $N-1$, thus the ``percolation threshold'', denoting the 
connection probability at which a giant cluster appears, should be 
$p_c\simeq1/N$. Indeed, this is exactly the probability at which the phase 
transition leading to the giant component appears in random graphs, as 
Erd\H{o}s 
and R\'enyi has showed (see Sect. \ref{sect_graph_evol}).

In the following we highlight some of the predictions of random graph theory 
and 
infinite dimensional percolation which reflect a complete analogy:

1. For $p<p_c=1/N$
\begin{itemize}
\item{The probability of the giant cluster in a graph, and of the infinite 
cluster in percolation, is equal to $0$.}
\item{The clusters of a random graph are trees, while the clusters in 
percolation 
have a fractal structure and a perimeter proportional with their volume}
\item{The largest cluster in a random graph is a tree with $\ln(N)$ nodes, 
while 
in general for percolation $P_p(|C|=s)\sim e^{-s/s_\xi}$ [ see Eq. 
(\ref{size_sub}) in Sect. \ref{sect_gen_perc}] suggesting that the size of the 
largest cluster scales as $\ln(N)$.}
\end{itemize}

2. For $p=p_c=1/N$
\begin{itemize}
\item{A giant cluster, respectively an infinite cluster appears.}
\item{The size of the giant cluster is $N^{2/3}$, while for infinite 
dimensional 
percolation $P_p(|C|=s)\sim s^{-3/2}$, thus the size of the largest cluster 
scales as $N^{2/3}$.}
\end{itemize}

3. For $p>p_c=1/N$
\begin{itemize}
\item{The size of the giant cluster is $G(c)N$, where 
$\lim_{c\rightarrow\infty}G(c)=1$. The size of the infinite cluster is 
$PN\propto 
(p-p_c)N$.}
\item{The giant cluster has a complex structure, containing cycles. In the 
same 
time the infinite cluster is not fractal anymore, but compact.}
\end{itemize}  

All these correspondences indicate that the phase transition in random graphs 
belongs in the same universality class as mean field percolation. Numerical 
simulations of random graphs (see for example Christensen {\it et al.}, 1998)
have confirmed that the critical exponents of the phase transition are equal 
to 
the critical exponents of the infinite dimensional percolation. The 
equivalence 
of these two theories is very important because it offers us different 
perspectives on the same problem. For example, it is often of interest to look 
at 
the cluster size distribution of a random network with a fixed number of 
nodes. 
This question is answered in a simpler way in percolation theory. However, 
random 
graph theory answers questions of major importance for networks, such as the 
appearance of trees and cycles, which are largely ignored by percolation 
theory.

In some cases there is an apparent discrepancy between the prediction of 
random 
graph theory and percolation theory. For example, percolation theory predicts that the chemical 
distance 
between two nodes in the infinite cluster scales as a power of the size of the cluster 
[see Eq. (\ref{diam_clust_inf})]. On the other hand, random graph theory  
predicts [Eq. (\ref{diam_er})] that the diameter of the infinite cluster 
scales 
logarithmically with its size (see Chung and Lu 2000). The origin of the 
apparent 
discrepancy is that these two predictions refer to different regimes. While 
Eq. 
(\ref{diam_clust_inf}) is valid only in the case where the infinite cluster is 
barely formed (i.e. $p=p_c$, and $\langle k\rangle =1$), and is still a 
fractal, 
the prediction of the random graph theory is valid only well beyond the 
percolation transition, when $\langle k\rangle\gg 1$. Consequently, using 
these two limits we can address the evolution of the chemical distance on the  
infinite cluster (see Cohen {\it et al.} 2001). Thus for a full 
characterization of random networks we need to be aware of both of these 
complementary approaches.

\section{GENERALIZED RANDOM GRAPHS}
\label{sect_sf_graph}
   
In Sect. \ref{sect_real_data} we have seen that real networks differ from 
random 
graphs in that often their degree distribution  
follows a power-law $P(k)\sim k^{-\gamma}$. Since power-laws are free of a 
characteristic scale, these networks are called 'scale-free 
networks' (Barab\'asi, Albert 1999, Barab\'asi, Albert and Jeong 1999). As random graphs do not capture the scale-free character of real 
networks, we need a different model to describe these systems. One approach is 
to 
generalize random graphs by constructing a model which has the degree 
distribution as an input, but is random in all other respects. In other words, 
the edges connect randomly selected nodes, with the constraint that the 
degree distribution is restricted to a power-law. The theory of such 
semi-random 
graphs should answer similar questions as were asked by Erd\H{o}s and 
R\'enyi 
and percolation theory (see Sections \ref{sect_random} and \ref{sect_perc}): 
Is 
there a threshold at which the giant cluster appears? How does the size and 
topology of the clusters evolve? When does the graph become connected? In 
addition, we need to determine the average path length and clustering 
coefficient 
of such graphs. 

The first step in developing such a theory is to identify the relevant 
parameter which, together with the network size, gives a statistically 
complete 
characterization of the network. In the case of random graphs this parameter is 
the connection probability (see Sect. \ref{sect_er}), for
percolation theory it is the bond occupation probability (see Sect. 
\ref{sect_perc}). Since the only restriction for these graphs is that their 
degree distribution needs to follow a power-law, the exponent $\gamma$ of the 
degree distribution could play the role of the control parameter. Accordingly, 
we study scale-free random networks by systematically varying $\gamma$ and 
see if there is a threshold value of $\gamma$ at which the networks' important 
properties abruptly change.

We start by sketching a few intuitive expectations. Consider a large network 
with 
degree distribution $P(k)\sim k^{-\gamma}$, and consider that $\gamma$ 
decreases 
from $\infty$ to $0$. The average degree of the network, or equivalently, the number of edges, 
increases as $\gamma$ decreases, since $\langle k\rangle \sim 
k_{max}^{-\gamma+2}$, where $k_{max}<N$ is the maximum degree of the graph. 
This 
is very similar to the graph evolution process described by Erd\H{o}s and 
R\'enyi 
(see Sect. \ref{sect_graph_evol}). Consequently, we expect that while at large 
$\gamma$ the network consists of isolated small clusters, there is a critical 
value of $\gamma$ at which a giant cluster forms, and at an even smaller 
$\gamma$ 
the network becomes completely connected. 

The theory of random graphs with given degree sequence is relatively recent. 
One 
of the first results is due to Luczak (1992), who showed that almost all 
random 
graphs with a fixed degree distribution and no nodes of degree 
smaller 
than $2$ have a unique giant cluster. Molloy and Reed (1995, 1998) have proven that for 
a random graph with degree distribution $P(k)$ the infinite cluster emerges 
almost surely when
\begin{equation}
\label{eq_qu}
Q\equiv\sum_{k\geq 1}k(k-2)P(k)>0,
\end{equation}
provided that the maximum degree is less than $N^{1/4}$. The method of Molloy 
and 
Reed was applied to random graphs with power-law degree distributions by 
Aiello, 
Chung and Lu (2000). As we show next, their results are in excellent agreement 
with the expectations outlined above. 

\subsection{Thresholds in a scale-free random graph}

Aiello, Chung and Lu (2000)introduce a two-parameter random graph model $P(\alpha, 
\gamma)$ defined as follows: Let $N_k$ be the number of nodes with degree 
$k$. 
$P(\alpha, \gamma)$ assigns uniform probability to all graphs with 
$N_k=e^\alpha 
k^{-\gamma}$. Thus in this model it is not the total number of nodes which is 
specified - along with the exponent $\gamma$ - from the beginning, but the 
number 
of nodes with degree $1$. Nevertheless, the number of nodes and edges in the 
graph can be deduced, noting that the maximum degree 
of the graph is $e^{\alpha/\gamma}$. To find the condition for the appearance of the giant cluster in this model, we insert $P(\alpha, \gamma)$ into Eq. (\ref{eq_qu}), finding as a solution $\gamma_0=3.47875..$. 
Thus when $\gamma>\gamma_0$ the random graph almost surely 
has 
no infinite cluster. On the other hand, when $\gamma< \gamma_0$ there is 
almost 
surely an infinite cluster. 

An important question is whether the graph is connected or not. Certainly for 
$\gamma>\gamma_0$ the graph is disconnected as it is made of independent finite clusters. In the $0<\gamma<\gamma_0$ regime Aiello, Chung and Lu (2000) study the 
size of the second largest cluster, obtaining that for $2\leq \gamma\leq \gamma_0$ the second largest cluster almost 
surely 
has size of order of $\log N$, thus it is relatively small. On the other hand, for $1<\gamma<2$ almost surely every node with degree greater than 
$\log(N)$ belongs to the infinite cluster. The second largest cluster has a 
size 
of order $1$, i.e. its size does not increase as the size of the graph goes to 
infinity. This means that the fraction of nodes in the infinite cluster 
approaches $1$ as the system size increases, thus the graph becomes totally 
connected in the limit of infinite system size. Finally, for $0<\gamma<1$ the graph is almost surely connected.

\subsection{Generating function formalism}
\label{sect_gen_funct} 
  
 A general approach to random graphs with given degree distribution was 
developed by Newman, Strogatz and Watts (2000) using a generating function 
formalism (Wilf 1990). The generating function of the degree distribution, 

\begin{equation}
\label{gen0}
G_0(x)=\sum_{k=0}^\infty P(k)x^k,
\end{equation}
encapsulates all the information 
contained 
in $P(k)$, since 
\begin{equation}
P(k)=\frac{1}{k!}\frac{d^k G_0}{dx^k}\bigg|_{x=0}.
\end{equation}  An important quantity for studying the cluster 
structure is the generating function for the degree distribution of the first 
neighbors of a randomly selected node. This can be obtained in the following 
way: 
a randomly selected edge reaches a node with degree $k$ with probability 
proportional to $kP(k)$ (i.e. it is easier to find a well connected node). If 
we 
start from a randomly chosen node and follow each of the edges starting from 
it, 
then the nodes we visit have their degree distribution generated by $kP(k)$. 
In 
addition, the generating function will contain a term $x^{k-1}$ (instead of 
$x^k$ 
as in Eq. (\ref{gen0})) because we have to discount the edge through which we reached the node. 
Thus the distribution of outgoing edges is generated by the function
\begin{equation}
\label{gen1}
G_1(x)=\frac{\sum_k kP(k)x^{k-1}}{\sum_k kP(k)}=\frac{1}{\langle 
k\rangle}G_0'(x).
\end{equation}
The average number of first neighbors is equal to the average degree of the graph,
\begin{equation}
\label{first}
z_1=\langle k\rangle=\sum_k kP(k)=G_0'(1).
\end{equation}

\subsubsection{Component sizes and phase transitions}

When we identify a cluster using a burning (breadth-first-search) algorithm, we start from an arbitrary node, and follow its edges until we 
reach 
its first neighbors. We record these nodes as part of the cluster, then 
follow 
their outside edges (avoiding the already recorded nodes) and record the nodes we arrive to as second neighbors of the starting node. This process is repeated until no new nodes are found, the set of 
identified nodes forming an isolated cluster. This algorithm is implicitly incorporated into the generating function method. The generating function, $H_1(x)$, for the size distribution of the 
clusters reached by following a random edge satisfies the 
iterative equation 
\begin{equation}
\label{cluster_gen0}
H_1(x)=\frac{\sum_k kP(k)[H_1(x)]^k}{\sum_k kP(k)}=x G_1(H_1(x)). 
\end{equation}
Here $kP(k)$ stands for the probability that a random edge arrives to a node 
with 
degree $k$, and $[H_1(x)]^k$ represents the $k$ ways in which the cluster can 
be 
continued recursively (i.e. by finding the first neighbors of a previously 
found 
node). If we start at a randomly chosen node then we have one such cluster at 
the 
end of each edge leaving that node, and hence the generating function for the 
size of the whole cluster is

\begin{equation}
\label{cluster_gen1}
H_0(x)=x\sum_k P(k)[H_1(x)]^k =xG_0(H_1(x)).
\end{equation}

The average cluster size is given by
\begin{equation}
\langle s\rangle =H_0'(1)=1+\frac{G_0'(1)}{1-G_1'(1)},
\end{equation}
which diverges when $G_1'(1)=1$, a signature 
of the appearance of a giant cluster. Substituting the definition of $G_0(x)$ we can write the condition of the emergence of the giant cluster as
\begin{equation}
\sum_k k(k-2)P(k)=0,
\label{perc_cond}
\end{equation}
identical to Eq. (\ref{eq_qu}) derived by Molloy and Reed (1995). Equation (\ref{perc_cond}) gives an implicit relation for the 
critical 
degree distribution of a random graph: For any degree distribution for which 
the 
sum on the l.h.s. is negative, no giant cluster is present in the graph, but degree 
distributions which give a positive sum lead to the appearance of a giant 
cluster.

When a giant cluster is present, $H_0(x)$ generates the probability 
distribution 
of the finite clusters. This means that $H_0(1)$ is no longer unity but 
instead 
takes on the value $1-S$, where $S$ is the fraction of nodes in the giant 
cluster. We can use this to calculate the size of the giant
 cluster $S$ as (Molloy and Reed 1998)
\begin{equation}
S=1-G_0(u),
\end{equation}
where $u$ is the smallest non-negative real solution of the equation
$
u=G_1(u).
$  

 Since we are dealing with random graphs (although with an 
arbitrary degree distribution), percolation theory (see Sect. \ref{sect_perc}) indicates that close to the 
phase transition the tail of cluster size distribution, $n_s$, behaves as

\begin{equation}
n_s\sim s^{-\tau}e^{-s/s_\xi}.
\end{equation} 
The characteristic cluster size $s_\xi$ can 
be related to the first singularity of $H_0(x)$, $x^*$, and at the phase transition  $x^*=1$ and $s_\xi\rightarrow \infty$. Using a Taylor 
expansion around the critical point we obtain that $H_0(x)$ scales as
\begin{equation}
H_0(x)\sim (1-x)^\alpha \quad\quad \mbox{as}\quad x\rightarrow 1,
\end{equation}
with $\alpha=\frac 1 2$. This exponent can be related to the exponent $\tau$ by 
using the connection between $n_s$ and $H_0(x)$, obtaining 
$
\tau=\alpha+2=\frac 5 2,
$ regardless of degree distribution. Thus close to the critical 
point 
the cluster size distribution follows $n_s^c\sim s^{-5/2}$, as predicted by 
infinite dimensional percolation (Sect. \ref{sect_inf_perc}), but now extended 
to 
a large family of random graphs with arbitrary degree distribution.

\subsubsection{Average path length}

 Extending the method of calculating the 
average 
number of first neighbors, the average number of $m^{\rm th}$ neighbors is
\begin{equation}
z_m=[G_1'(1)]^{m-1}G_0'(1)=\left [\frac{z_2}{z_1}\right ]^{m-1} z_1,
\end{equation}
where $z_1$ and $z_2$ is the number of first and second neighbors. Assuming 
that all nodes in the graph can be reached within $\ell$ steps, we have
\begin{equation}
1+\sum_{m=1}^\ell z_m =N.
\end{equation}
As for most graphs $N>> z_1$ and $z_2>>z_1$, we obtain
\begin{equation}
\label{path}
\ell=\frac{\ln(N/z_1)}{\ln(z_2/z_1)} +1.
\end{equation}

This result reflects several general properties of the average path length:

1. $\ell$ scales logarithmically with $N$ for all random 
graphs, 
irrespective of the degree distribution.

2. $\ell$, a global measure, can be calculated from local 
quantities as the average number of neighbors a node has.

3. Even among the number of neighbors only the average number of first and 
second 
nearest neighbors are important to the calculation of $\ell$, thus 
two graphs with different degree distributions but the same values of $z_1$ and 
$z_2$ 
will have the same average path length.

\subsection{Random graphs with power-law degree distribution}

As an application of the generating function formalism Newman, Watts and 
Strogatz (2000) 
consider the case of a degree distribution of type
\begin{equation}
\label{cutoff_power}
P(k)=Ck^{-\gamma}e^{-k/\kappa} \quad\quad \mbox{for}\quad k\geq 1,
\end{equation} where $C$, $\gamma$ and $\kappa$ are constants. The exponential cutoff, 
present 
in some social and biological networks (see Amaral et al. 2000, Newman 2000a, 
Jeong {\it et al.} 2001), has the technical advantage of making the 
distribution 
normalizable for all $\gamma$, not just $\gamma\geq 2$, as in the case for a 
pure 
power-law. The constant $C$ is fixed by normalization, giving $C=[{\rm 
Li}_{\gamma}(e^{-1/\kappa})]^{-1}$, where ${\rm Li}_n(x)$ is the $n^{\rm th}$ 
polylogarithm of $x$. Thus the degree distribution is characterized by two 
independent parameters, the exponent $\gamma$ and the cutoff $\kappa$.
Following the formalism described above, we find that the size of the infinite cluster is 
\begin{equation}
\label{gclust2}
S=1-\frac{{\rm Li}_{\gamma}(ue^{-1/\kappa})}{{\rm 
Li}_{\gamma}(e^{-1/\kappa})},
\end{equation}
where $u$ is the smallest nonnegative real solution of the equation
$
u={\rm Li}_{\gamma-1}(ue^{-1/\kappa})/[u{\rm 
Li}_{\gamma-1}(e^{-1/\kappa})].
$
For graphs with purely power-law distribution ($\kappa \rightarrow \infty$), 
the above equation becomes 
$
u={\rm Li}_{\gamma-1}(u)/[u\zeta (\gamma-1)]
$, where $\zeta (x)$ is the Riemann $\zeta$-function. For all $\gamma\leq 2$ this 
gives $u=0$, and hence $S=1$, implying that a randomly chosen node belongs to 
the 
giant cluster with probability converging to $1$ as $\kappa \rightarrow\infty$. For 
graphs with $\gamma>2$ this is never the case, even for infinite $\kappa$, 
indicating that such a graph contains finite clusters, i.e. it is not connected, in agreement with the conclusions of Aiello, Chung and Lu (2000).

The average path length is 
\begin{equation}
\label{path_power}
\ell=\frac{\ln N+\ln[{\rm Li}_{\gamma}(e^{-1/\kappa})/{\rm 
Li}_{\gamma-1}(e^{-1/\kappa})]}{\ln [{\rm Li}_{\gamma-2}(e^{-1/\kappa})/{\rm 
Li}_{\gamma-1}(e^{-1/\kappa})-1]}+1,
\end{equation}
which in the limit $\kappa\rightarrow\infty$ becomes
\begin{equation}
\ell=\frac{\log 
N+\ln[\zeta(\gamma)/\zeta(\gamma-1)]}{\ln[\zeta(\gamma-2)/\zeta(\gamma-1)-1]}+
1.
\end{equation}
Note that this expression does not have a finite positive real value for any 
$\gamma<3$, indicating that one must specify a finite cutoff $\kappa$ for the 
degree distribution to get a well-defined average path length. Equations 
(\ref{path}) and (\ref{path_power}) reproduce the result of finite size scaling simulations of the World-Wide Web indicating 
that its average path length scales logarithmically with its size (Albert, 
Jeong, Barab\'asi 1999). But do they offer a good estimate for the average 
path 
length of real networks? In Sect. \ref{sect_real_data} we have seen that the 
prediction of random graph theory is in qualitative agreement with the average 
path length of real networks, but that there also are significant deviations 
from 
it. It is thus important to see if taking into account the correct degree 
distribution gives a better fit.

\begin{figure}[htb]
\vspace{0.5cm}
\centerline{\psfig{figure=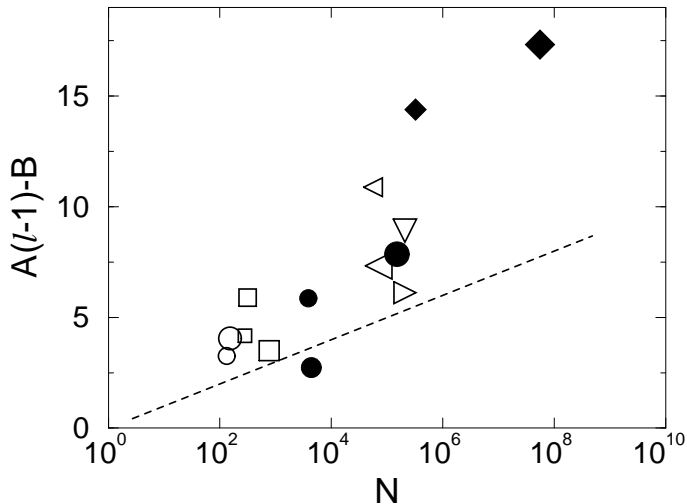,width=2.8in}}
\vspace{1cm}
\caption{Comparison between the average path length of real scale-free 
networks 
and the prediction (\ref{path_power}) of scale-free random graphs (dashed 
line). 
For each network we have plotted $A(\ell-1)-B$ as a function of $N$, where $A$ 
and $B$ are given in the text. The networks included in the figure, indicated 
by 
their number in Table \ref{table_cluster} or Table \ref{table_gamma}, are: 
small 
circle, I.12; large circle, I.13; small square, I.10; medium square, I.11; 
large 
square, II.13; small filled circle, II.6; medium filled circle, I.2; small 
left 
triangle, I.6, large left triangle, I.8; large filled circle, II.7; down 
triangle, I.9; right triangle, I.3, medium filled diamond, II.1, large filled 
diamond, II.3.}
\label{fig_fit_diam_nws}
\end{figure}

In Fig. \ref{fig_fit_diam_nws} we compare the prediction of Eq. 
(\ref{path_power}) with the average path length of real networks by plotting 
$A(\ell-1)-B$ 
in function of the network size $N$, where 
$A=\log[Li_{\gamma-2}(e^{-1/\kappa})/Li_{\gamma-1}(e^{-1/\kappa})-1]$ and
$B=\log[Li_{\gamma}(e^{-1/\kappa})/Li_{\gamma-1}(e^{-1/\kappa})]$, and we use 
the 
cutoff length $\kappa$ as obtained from the empirical degree distributions. For the directed networks we used the $\gamma_{out}$ values. For random networks with the same $N$, $\gamma$ and $\kappa$ as the real networks 
the 
$A\ell -B$ values would align along a straight line with slope $1$ in a 
log-linear plot, given by the dashed line on the figure. The actual values for 
the real networks obey the trend, but they seem to be systematically larger 
than 
the prediction of Eq. (\ref{path_power}), indicating that the average path 
length 
of real networks is larger than that of random graphs with power-law degree 
distribution. This conclusion is further supported by the last three columns of Table 
\ref{table_gamma} which directly compares the average path length of 
real 
networks with power-law degree distribution, $\ell_{real}$, and the estimates 
of 
random graph theory, $\ell_{rand}$, and scale-free random graph theory, 
$\ell_{pow}$. We can see that the general trend is that $\ell_{real}$ is 
larger 
than both $\ell_{pow}$ and $\ell_{rand}$, an indication of the non-random 
aspects of the topology of real networks.

\subsection{Bipartite graphs and the clustering coefficient}
\label{sect_ccoeff_nws}

The clustering coefficient of scale-free random graphs has not been calculated 
yet in the literature, but we can find out its general characteristic if we 
take 
into account that scale-free random graphs are similar to Erd\H{o}s-R\'enyi 
random graphs in the sense that their edges are distributed randomly. 
Consequently, the clustering coefficient of scale-free random graphs converges to $0$ as the network size increases.

\begin{figure}[htb]
\centerline{\psfig{figure=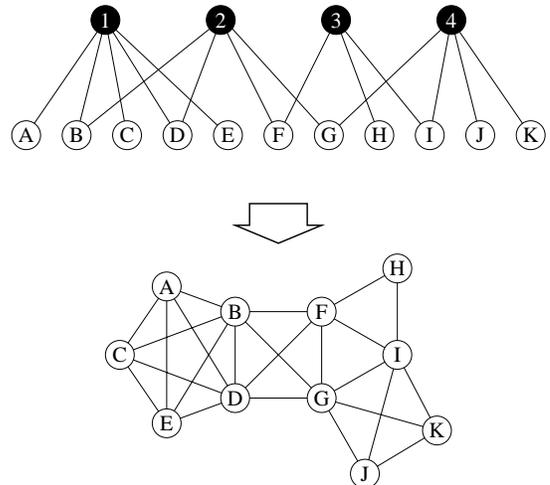,width=2.8in}}
\caption{A schematic representation of a bipartite graph, such as the graph of 
movies and the actors who have appeared in them.  In this small graph we have 
four movies, labeled $1$ to $4$, and eleven actors, labeled $A$ to $K$, with 
edges joining each movie to the actors in its cast. The bottom figure shows the one-mode projection of the graph for the eleven actors. 
After 
Newman {\it et al.} 2000.}
\label{fig_bipartite}
\end{figure}

It is worth noting, however, that some of the real-world networks presented in 
Sect. \ref{sect_real_data}, for example the collaboration networks, can be more 
completely described by bipartite graphs (Newman, Strogatz and Watts 2000). In 
a 
bipartite graph there are two kinds of nodes, and edges connect only nodes of 
different kind. For example, the collaboration network of movie actors is in 
fact 
a projection of a bipartite actor-movie graph, in which the two types of nodes 
are the actors and movies, and an edge connects each movie with the actors 
playing in it (see Fig. \ref{fig_bipartite}). The same argument stands for the 
collaboration between scientists (where scientists and papers are the two types of nodes) and metabolic networks (where nodes can be the substrates or reactions). The generating 
function 
method can be generalized to bipartite graphs (see Newman, Strogatz and Watts 
2000), and it results in a nonvanishing clustering coefficient inherent to the 
bipartite structure 

\begin{equation}
\label{clust_bi}
C=\frac{1}{1+\frac{(\mu_2-\mu_1)(\nu_2-\nu_1)^2}{\mu_1\nu_1(2\nu_1-3\nu_2+\nu_
3)}
},
\end{equation}
where $\mu_n=\sum_k k^nP_a(k)$  and $\nu_n=\sum_k k^nP_m(k)$, and in the 
actor-movie framework, $P_a(k)$ represents the fraction of actors which 
appeared 
in $k$ movies, while $P_m(k)$ means the fraction of movies in which $k$ actors 
have appeared.

The prediction of Eq. (\ref{clust_bi}) has been tested for several 
collaboration 
graphs (Newman, Strogatz, Watts 2000), and in some cases there is an excellent agreement, but in others it deviates by a factor of $2$ from 
the clustering coefficient of the real network. Consequently we can conclude that the order present in real networks is not due solely to the definition of the network, but to a yet unknown organizing principle.

\section{SMALL-WORLD NETWORKS}
\label{sect_small_world}

In Sects. \ref{sect_real_data} and \ref{sect_er} we have seen (see Table 
\ref{table_cluster}, Figs. \ref{fig_fit_diam_er} and \ref{fig_fit_ccoef_er}) 
that 
real-world networks have a small-world character like random graphs, but they 
have unusually large clustering coefficients. Furthermore, as Fig. 
\ref{fig_fit_ccoef_er} demonstrates, the clustering coefficient appears to be 
independent of the network size. This latter property is characteristic to ordered 
lattices, whose clustering coefficient is size independent and depends only on 
the coordination number. For example, a one-dimensional lattice with periodic 
boundary conditions (i.e. a ring of nodes), in which each node is connected to 
the $K$ nodes closest to it (see Fig. \ref{fig_ws}), most of the immediate 
neighbors of any site are also neighbors of one another, i.e. the lattice is 
clustered. For such a lattice the clustering coefficient is
\begin{equation}
C=\frac{3(K-2)}{4(K-1)},
\end{equation}
which converges to $3/4$ in the limit of large $K$. Such low-dimensional 
regular 
lattices, however, do not have short path lengths: for a $d$ dimensional 
hypercubic lattice the average node-node distance scales as $N^{1/d}$, which 
increases much faster with $N$ than the logarithmic increase observed for 
random 
and real graphs. The first successful attempt to generate graphs with high 
clustering coefficients and small $\ell$ is due to Watts and Strogatz (1998). 

\subsection{The Watts-Strogatz (WS) model}
\label{sect_ws}

\begin{figure}[htb]
\centerline{\psfig{figure=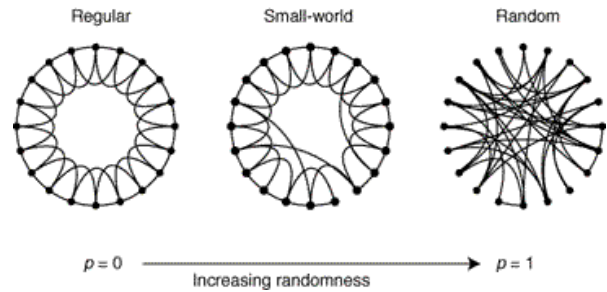,width=3.1in}}
\vspace{1cm}
\caption{The random rewiring procedure of the WS model which interpolates 
between 
a regular ring lattice and a random network without altering the number of 
nodes 
or edges. We start with $N=20$ nodes, each connected to its $4$ 
nearest 
neighbors. For $p=0$ the original ring is unchanged; as $p$ 
increases the network becomes increasingly disordered until for $p=1$ all 
edges 
are rewired randomly. After Watts and Strogatz (1998).}
\label{fig_ws}
\end{figure}

 Watts and Strogatz (1998) proposed a one-parameter model which interpolates 
between an ordered finite dimensional lattice and a random graph. The 
algorithm 
behind the model is the following (Fig. \ref{fig_ws}):  

(1) {\it Start with order}: Start with a ring lattice with $N$ nodes in 
which 
every  node is connected to its first $K$ neighbors ($K/2$ on either side). In 
order to have a sparse but connected network at all times, consider $N\gg K\gg 
\ln(N)\gg 1$. 

(2) {\it Randomize}: Randomly rewire each edge of the lattice with 
probability 
$p$ such that self-connections and duplicate edges are excluded. This process 
introduces $pNK/2$ long-range edges which connect nodes that 
otherwise would be part of different neighborhoods. Varying $p$ the transition 
between order ($p=0$) and  randomness ($p=1$) can be  closely monitored.

This model has its roots in social systems where most people 
are 
friends with their immediate neighbors - neighbors on the same street, 
colleagues, people that their friends introduce them to. On the other hand, everybody has one or 
two friends who are a long way away - people in other countries,  old 
acquaintances, which are represented by the 
long-range edges obtained by rewiring in the WS model.

\begin{figure}[htb]
\centerline{\psfig{figure=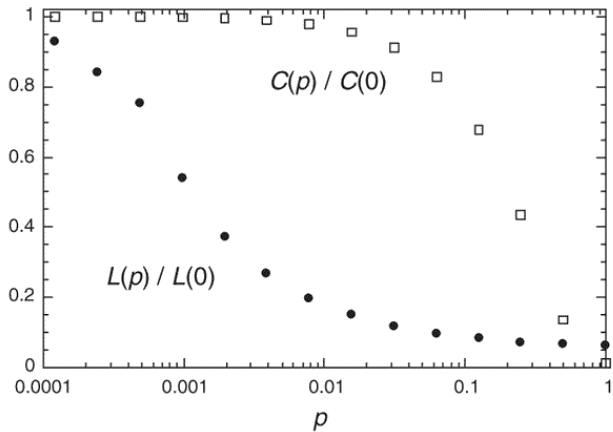,width=3.3in}}
\vspace{1cm}
\caption{Characteristic path length $\ell (p)$ and clustering coefficient 
$C(p)$ 
for the WS model. The data is normalized by the values $\ell (0)$ and 
$C(0)$ for a regular lattice. A logarithmic horizontal scale 
resolves the rapid drop in $\ell (p)$, corresponding to the onset of the 
small-world phenomenon. During this drop $C(p)$ remains almost constant, 
indicating that the transition to a small world is almost undetectable at the 
local level. After Watts and Strogatz (1998).}
\label{fig_wslc}
\end{figure}

To understand the coexistence of small path length and clustering, we study 
the 
behavior of the clustering coefficient $C(p)$ and the average path length 
$\ell(p)$ as a function of the rewiring probability $p$. For a ring lattice  
$\ell(0)\simeq N/2K\gg 1$ and $C(0)\simeq 3/4$, thus $\ell$ scales linearly 
with 
the system size, and the clustering coefficient is large. On  the other hand, 
for 
$p\rightarrow 1$ the model converges to a random graph for which $\ell(1)\sim 
\ln(N)/\ln(K)$  and  $C(1)\sim K/N$, thus $\ell$ scales logarithmically with 
$N$ and the clustering coefficient decreases with $N$. These limiting cases 
might 
suggest that large $C$  is always associated with large $\ell$, and small $C$ 
with small $\ell$. On the  contrary, Watts and Strogatz (1998) found that there is a 
broad interval of $p$ over which $\ell(p)$ is close to $\ell(1)$ yet $C(p)\gg 
C(1)$ (Fig. \ref{fig_wslc}). This regime originates in a rapid drop of 
$\ell(p)$ for small $p$ values, while $C(p)$ stays almost unchanged, resulting in networks that are  clustered but have a small characteristic path  
length.  This coexistence of 
small 
$\ell$ and large $C$ is in excellent agreement with the characteristics of 
real 
networks discussed in Sect. \ref{sect_real_data}, prompting many to call such systems small world networks.

\subsection{Properties of small-world networks}
\label{sect_small_prop}

The pioneering article of Watts and Strogatz started an avalanche of research 
on 
the properties of small-world networks and  the Watts-Strogatz  (WS) model. A 
much studied variant  of the WS model was proposed by Newman and Watts (1999a, 
1999b) in which edges are added between randomly chosen pairs of sites, but no 
edges are removed from the regular lattice. This model is somewhat 
easier 
to analyze than the original Watts-Strogatz model because it does not lead to 
the 
formation of isolated clusters, whereas this can happen in the original model. 
For sufficiently small $p$ and large $N$ this model is equivalent with the WS 
model. In the following  we will summarize the main results regarding the 
properties of small-world models.

\subsubsection{Average path length}
\label{sect_path_length}

As we discussed above, in the WS  model there is a change in the scaling  of 
the 
characteristic path length $\ell$ as the fraction $p$ of the rewired edges is 
increased. For small $p$, $\ell$ scales linearly with the system size, while 
for 
large $p$ the scaling is logarithmic. As discussed by Watts (1999), and Pandit and Amritkar (1999), the origin of the rapid drop in $\ell$ is the 
appearance of shortcuts between nodes. Every  shortcut, created at  random, is 
likely to connect widely separated  parts of  the graph,  and thus has a 
significant impact on the characteristic path  length of the entire graph. 
Even a 
relatively low fraction of shortcuts is sufficient to drastically decrease the 
average path length, yet locally the network remains highly ordered.

\begin{figure}[htb]
\label{fig_scal}
\centerline{\epsfig{figure=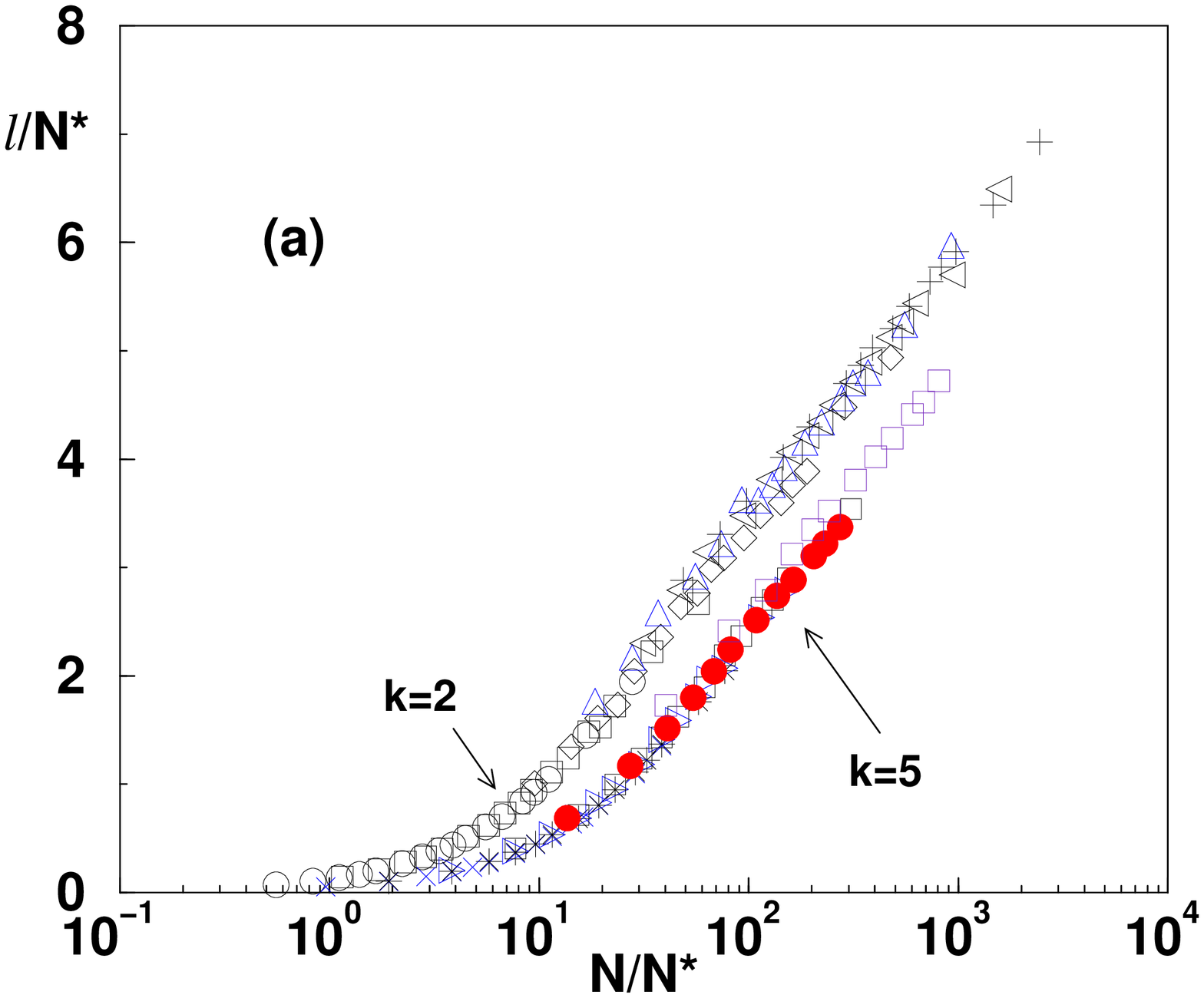,width=1.55in,angle=-90}
	     \epsfig{figure=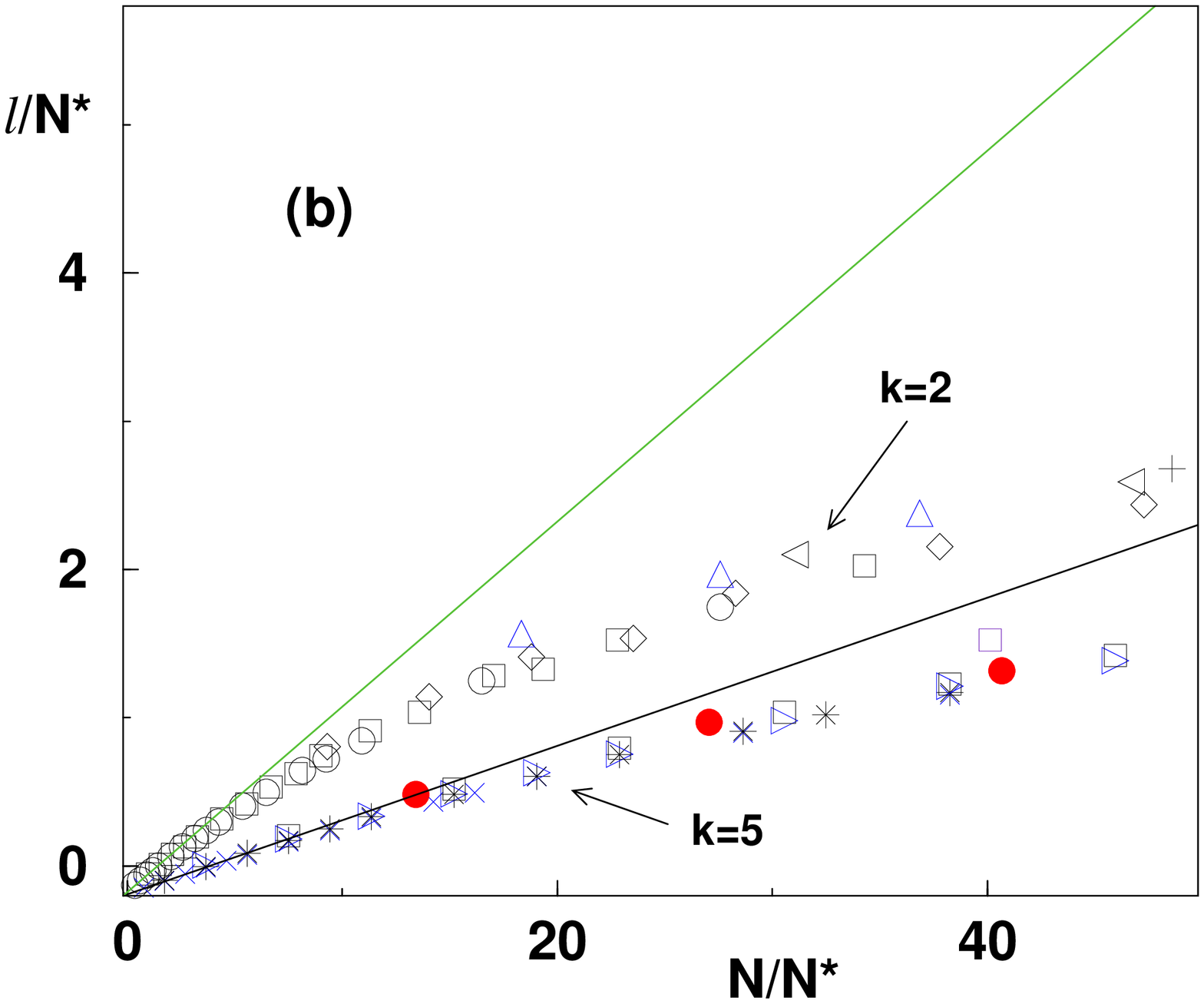,width=1.55in,angle=-90}}
\caption{ Data collapse $\ell (N,p)/N^*(p)$ versus $N/N^*(p)$ for two 
different 
values of $K$. (a) Log-linear scale showing the logarithmic behavior at large 
$N/N^*$; (b) Linear scale showing the linear behavior $\l (N,p)\sim N/(4K)$ at 
small $N/N^*$. After Barrat and Weigt (2000).}
\end{figure}

 An important question regarding the average path length is whether the onset of small-world behavior is dependent on the system size. It was Watts (1999) who first noticed that $\ell$ does not begin to decrease 
until $p\geq 2/NK$, guaranteeing the existence of at least one shortcut. This 
implies that the transition $p$ depends on the system size, or conversely, there exists a $p$-dependent crossover length $N^*$ such that if $N<N^*$, $\ell\sim 
N$, but if $N>N^*$, $\ell\sim \ln(N)$. The concept of the crossover length was 
introduced by Barth\'el\'emy and  Amaral (1999a), who conjectured that the 
characteristic path length scales as (see Fig. \ref{fig_scal}):

\begin{equation}
\label{swscaling}
\ell (N,p)\sim N^* F\left(\frac{N}{N^*}\right),
\end{equation}
where 
\begin{equation}
F(u)=\left\{\begin{array}{lcl}
u& \mbox{if}& u\ll 1\\
\ln(u)& \mbox{if}& u\gg 1
\end{array}\right..
\end{equation}

\begin{figure}[htb]
\centerline{\epsfig{figure=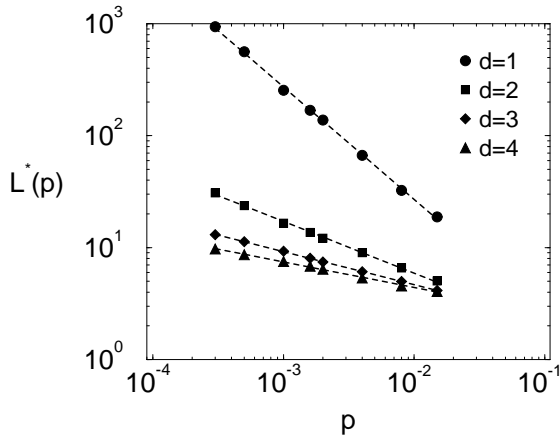,width=2.8in}}
\vspace{1cm}
\caption{ The dependence of the crossover length $N^*$ on the rewiring 
probability in one to four dimensions. The dashed lines represent the scaling 
relation $N^*\sim p^{-1/d}$. After Argollo de Menezes {\it et al.} (2000).}
\label{fig_Nstar}
\end{figure}

Numerical simulations and analytical arguments (Barth\'el\'emy and Amaral 
1999a, 
Barrat 1999, Newman and Watts 1999a,  Bart\'el\'emy and Amaral 1999b, Argollo 
de 
Menezes {\it et al.} 2000, Barrat and  Weigt 2000)  concluded that the 
crossover 
length  $N^*$ scales with $p$ as $N^*\sim p^{-\tau}$ , where $\tau=1/d$ and 
$d$ 
is the dimension of the original lattice to which the random edges are added  
(Fig. \ref{fig_Nstar}). Thus for the original WS model, defined on a circle 
($d=1$), we have $\tau=1$, the onset of  small-world behavior taking place at 
the 
rewiring probability  $p^*\sim 1/N$.

 It is now widely accepted that the  characteristic  
path length  obeys  the general scaling form
\begin{equation}
\label{swscaling_univ}
\ell (N,p)\sim \frac{N}{K} f(pKN^d),
\end{equation}
where $f(u)$ is a universal scaling function that obeys
\begin{equation}
f(u)=\left\{\begin{array}{lcl}
\mbox{constant}& \mbox{if}& u\ll 1\\
\ln(u)/u& \mbox{if}& u\gg 1
\end{array}\right..
\end{equation}
Newman {\it et al.} (1999) have calculated the form of the scaling function 
$f(u)$ for the one-dimensional small-world model using a mean-field method 
which 
is exact for small or large values of $u$, but not in the regime where 
$u\simeq 1$, obtaining  

\begin{equation}
f(u)=\frac{4}{\sqrt{u^2+4u}}\tanh^{-1}\frac{u}{{\sqrt{u^2+4u}}}.
\end{equation}
They also solved for the complete distribution of path lengths within this 
mean-field approximation.

The scaling relation (\ref{swscaling_univ}) has been  confirmed by extensive 
numerical simulations (Newman and Watts 1999a, Argollo de Menezes {\it et al.} 
2000), renormalization group techniques (Newman  and Watts 1999a) and series 
expansions (Newman and Watts 1999b). Equation (\ref{swscaling_univ}) tells us 
that although the average path length in a small-world model appears at 
first 
glance to depend on three parameters  - $p$, $K$ and $N$ - it is in fact 
entirely 
determined by a single scalar function $f(u)$ of a single scalar variable. Note that both the  scaling function $f(u)$ and the scaling 
variable 
$u=pKN^d$ have simple physical interpretations. The variable $u$  is two times 
the average number of random links (shortcuts) on the  graph for a given 
$p$, and $f(u)$ is the average of the fraction by which the distance  between 
two 
nodes is  reduced for a given $u$.  

Several attempts have been made to calculate exactly the distribution of path 
lengths and the average path length $\ell$. Dorogovtsev  and Mendes 2000a 
studied 
a simpler model which contains a ring lattice with directed edges of length 
$1$ 
and  a  central node which is connected with probability  $p$ to  the  nodes 
of 
the lattice by undirected edges  of length $0.5$. They calculated exactly the 
distribution  of path lengths for this model, showing that $\ell/N$ depends 
only  
on the scaling variable $pN$ and the functional form of this dependence is 
similar to the numerically obtained $\ell(p)$ in the WS model.  Kulkarni {\it  
et 
al.} 1999 calculated the probability $P(m|n)$ that two nodes separated by an 
Euclidian distance $n$ have a path  length $m$. They have shown that the 
average 
path length $\ell$ is simply related to  the mean $\langle s\rangle$ and the 
mean 
square $\langle s^2\rangle$ of the shortest distance between two diametrically 
opposite nodes (i.e. separated by the largest   Euclidian distance), according 
to
\begin{equation}
\frac{\ell}{N}=\frac{\langle s\rangle}{N-1}-\frac{\langle s^2\rangle}{L(N-1)}.
\end{equation}
Unfortunately calculating  the shortest distance between opposite nodes is 
just 
as difficult as determining $\ell$ directly.

\subsubsection{Clustering coefficient}
\label{sect_clust_coef_ws}

In addition to a short average path length, small-world networks have a   
relatively high clustering  coefficient. 
The WS model displays this  duality for a wide range of the rewiring  probabilities $p$. In a regular lattice ($p=0$) the clustering coefficient 
does not depend on the size of the lattice but only its topology. As the edges of the network are randomized, the clustering coefficient remains  close to $C(0)$ up to  relatively large values  
of $p$. 

The dependence of  $C(p)$ on $p$ can be derived using a slightly different but equivalent definition of $C$, introduced by Barrat and Weigt (2000). According to this definition, $C'(p)$ is the fraction between the mean number of edges between the neighbors of a node and the mean number of possible edges between those neighbors. In a more graphic formulation (Newman, Strogatz and Watts 2000),
\begin{equation}
C'=\frac{3\times\mbox{number of triangles}}{\mbox{number of connected triples}}.
\end{equation}
Here triangles are trios of nodes in which each node is connected to both of the others, and connected triples are trios in which at least one is connected to both others, the factor $3$ accounting to the fact that each triangle contributes to $3$ connected triples. This definition corresponds to the concept of "fraction of transitive triples" used in sociology (see Wasserman and Faust 1994).

To calculate $C'(p)$ for the WS model, let us start with a regular  lattice  with a clustering coefficient $C(0)$. For 
$p>0$ two neighbors of a node $i$ that were connected at $p=0$ are still 
neighbors of $i$ and connected by and  edge with probability $(1-p)^3$, since 
there are three edges which need to remain intact. Consequently
$
C'(p)\simeq C(0)(1-p)^3.
$ Barrat and Weigt (2000) have verified that the deviation  of $C(p)$ from this  
expression is  small  and goes to  zero as $N\rightarrow\infty$. The corresponding expression for the Newman-Watts model is (Newman, 2001e)
\begin{equation}
C'(p)=\frac{3K(K-1)}{2K(2K-1)+8pK^2+4p^2K^2}.
\end{equation}

\subsubsection{Degree distribution}
\label{sect_degree_ws}

 In the WS model for $p=0$ each  node  has the same degree $K$, thus the degree 
distribution is a delta function centered at $K$. A nonzero $p$ introduces 
disorder in the network, broadening the degree distribution while maintaining 
the 
average degree equal to $K$. Since only a single 
end 
of every edge is rewired ($pNK/2$ edges in total), each node has at least 
$K/2$ 
edges after the rewiring process. Consequently for $K>2$ 
there are no isolated nodes and the network is usually connected, unlike a 
random 
graph which consists of isolated clusters for a wide range of connection 
probabilities. 

For $p>0$ the degree $k_i$ of a vertex $i$ can be written as (Barrat 
and 
Weigt 2000) $k_i=K/2+c_i$ 
where 
$c_i$ can be divided in two parts: $c_i^1\leq K/2$ edges have been left in 
place 
(with probability $1-p$), while $c_i^2=c_i-c_i^1$ edges have been rewired 
towards 
$i$, each with probability $1/N$. The probability distributions of $c_i^1$ and 
$c_i^2$ are
\begin{equation}
P_1(c_i^1)=C_{K/2}^{c_i^1}(1-p)^{c_i^1}p^{K/2-c_i^1}
\end{equation}
and 
\begin{eqnarray}
P_2(c_i^2)&=&C_{pNK/2}^{c_i^2}\left(\frac{1}{N}\right)^{c_i^2}\left(1-\frac{1}
{N}
\right)^{pNK/2-c_i^2}\nonumber\\
&\simeq&\frac{(pK/2)^{c_i^2}}{c_i^2!}e^{-pK/2}
\end{eqnarray}
for  large $N$. Combining these two factors together  the degree distribution 
follows
\begin{equation}
P(k)=\sum_{n=0}^{f(k,K)}C_{K/2}^{n}(1-p)^np^{K/2-n}\frac{(pK/2)^{k-K/2
-n}
}{(k-K/2-n)!}e^{-pK/2}
\label{ws_degree_dist}
\end{equation}
for $k\geq K/2$, where $f(k,K)=min(k-K/2,K/2)$.

\begin{figure}[htb]
\centerline{\epsfig{figure=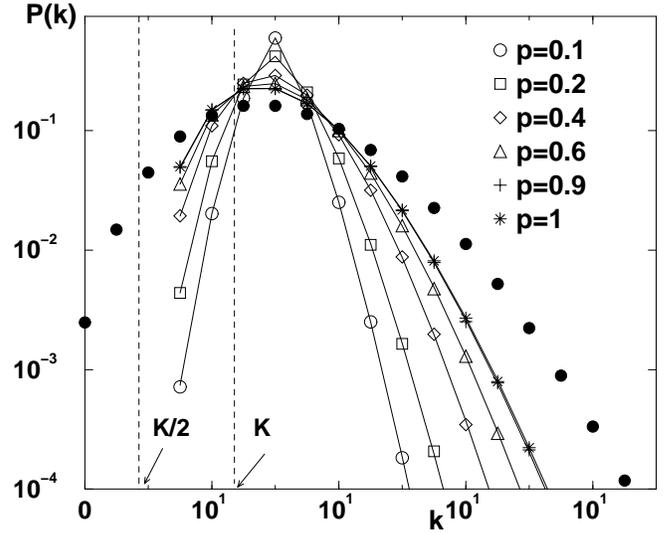,width=2.8in,angle=-90}}
\caption{Degree distribution of the WS model for $K=3$ and various $p$. 
We can see that only $k\geq K/2$ values are present, and the mean degree is 
$\langle k\rangle=K$. The symbols are obtained from numerical simulations of 
the 
WS model with $N=1000$, and the lines correspond to Eq. 
(\ref{ws_degree_dist}). 
As a comparison, the degree distribution of a random graph with the same 
parameters is plotted with filled symbols. After Barrat and Weigt (2000).}
\label{fig_pkws}
\end{figure}

The shape of the degree distribution is similar to that of a random graph: it 
has 
a pronounced peak at $\langle  k\rangle=K$ and decays exponentially for large 
$k$ 
(Fig. \ref{fig_pkws}). Thus the topology of the network is relatively 
homogeneous, all nodes having approximately the same  number of edges.
 
\subsubsection{Spectral properties}
\label{sect_swspectra}

\begin{figure}[htb]
\vspace{-2cm}
\centerline{\psfig{figure=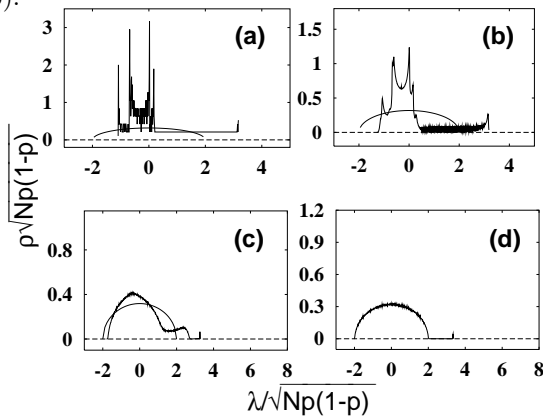,width=2.8in}}
\vspace{-2cm}
\caption{Spectral density of small-world networks, compared to the semi-circle 
law corresponding to random graphs (solid line). The rewiring probabilities  
are $p=0$ (a), $p=0.01$ (b), $p=0.3$ (c) and $p=1$ (d). After Farkas 
{\it 
et al.} (2001)}
\label{fig_spect_sw}
\end{figure}

As discussed in Sect. \ref{sect_spect_er}, the spectral density 
$\rho(\lambda)$ of a graph reveals important information about its topology. 
Specifically, we have seen that for large random graphs $\rho(\lambda)$ 
converges 
to a semi-circle. It comes as no surprise that the spectrum of the WS model depends on 
the 
rewiring probability $p$ (Farkas {\it et al.} 2001). For $p=0$ the network is regular and periodical, 
consequently $\rho(\lambda)$ contains numerous singularities (Fig. 
\ref{fig_spect_sw}a). For intermediate values of $p$ these singularities 
become 
blurred, but $\rho(\lambda)$ retains a strong skewness (Fig. 
\ref{fig_spect_sw}b,c). Finally, as $p\rightarrow 1$, $\rho(\lambda)$ 
approaches 
the semi-circle law characterizing random graphs (Fig. \ref{fig_spect_sw}d). 
While the details of the spectral density change considerably with $p$, the 
third 
moment of $\rho(\lambda)$ is consistently high, indicating a high number of 
triangles in the network. Thus the results summarized in Fig. 
\ref{fig_spect_sw} 
allow us to conclude that the high number of triangles is a 
basic 
property of the WS model (see also Gleis {\it et al.} 2000). The high regularity of small-world models for a broad range of $p$ is underlined by the results concerning the spectral properties of the Laplacian operator, which tells us about the time evolution of a diffusive field on the graph (Monasson 2000).

\section{THE SCALE-FREE MODEL}

The empirical results discussed in Sect. \ref{sect_real_data} demonstrate that 
many large networks are scale-free, that is, their degree distribution 
follows a power-law for large $k$. Furthermore, even for those networks for 
which 
$P(k)$ has an exponential tail, the degree distribution significantly deviates 
from a Poisson. We have seen in Sects. \ref{sect_degree_er} and 
\ref{sect_degree_ws} that random graph theory and the WS model cannot 
reproduce this feature. While it 
is straightforward to construct random graphs which have power-law degree 
distribution (Sect. \ref{sect_sf_graph}), these constructions only postpone an important question: what is 
the mechanism responsible for the emergence of scale-free networks? We will see in this section that 
this quest will require a shift from modeling network topology to modeling the 
network assembly and evolution. While at this point these two approaches do 
not 
appear to be particularly distinct, we will find that there is a fundamental 
difference between the modeling approach we took in random graphs and the 
small-world models, and the one required to reproduce the power-law degree 
distribution. While the goal of the former models is to construct a graph with correct topological features, modeling scale-free networks will 
put 
the emphasis on capturing the network dynamics. That is, the underlying 
assumption behind evolving or dynamic networks is that if we capture correctly 
the processes that assembled the networks that we see today, then we will obtain 
their topology correctly as well. Dynamics takes the driving role, topology 
being 
only a byproduct of this modeling philosophy.  

\subsection{Definition of the scale-free (SF) model}
\label{sect_sf_mod}

The origin of the power-law degree distribution in networks was first 
addressed 
by Barab\'asi and Albert (1999), who argued that the scale-free nature of real 
networks is rooted in two generic mechanisms common in many real networks. The 
network models discussed thus far assume that we start with a fixed number $N$ 
of 
vertices that are then randomly connected or 
rewired, without modifying $N$. In contrast, most real world 
networks describe open systems which {\it grow} by the continuous addition of 
new 
nodes. Starting from a small nucleus of nodes, the number of 
nodes 
increases throughout the lifetime of the network by the subsequent addition of 
new nodes. For example the WWW grows exponentially in time by the addition of 
new 
web pages or the research literature constantly grows by the publication of 
new 
papers.

Second, network models discussed so far assume that the probability that two nodes 
are 
connected (or their connection is rewired) is independent of the nodes' 
degree, 
i.e. new edges are placed  randomly. Most real networks, however, exhibit {\it 
preferential attachment}, such that the likelihood of connecting to a node 
depends on the node's degree. For example, a webpage will more likely include 
hyperlinks to popular documents with already high degree, because such highly 
connected documents are easy to find and thus well known, or a new manuscript 
is 
more likely to cite a well known and thus much cited publications than less 
cited 
and consequently less known papers.

These two ingredients, growth and preferential attachment, inspired the 
introduction of the scale-free (SF) 
model that has a power-law degree distribution. The algorithm of the SF model 
is 
the following: 

(1) {\it Growth}:
Starting with a small number ($m_0$) of nodes, at every timestep we add a new 
node with $m$($\leq m_0$) edges that link the new node to $m$ different nodes 
already present in the system. 

(2) {\it Preferential attachment}:
 When choosing the nodes to which the new node connects, we assume that the 
probability $\Pi$ that a new node will be connected to node $i$ depends on the 
degree $k_i$ of node $i$, such that 
\begin{equation}
\label{pi}
\Pi(k_i)=\frac{k_i}{\sum_j k_j}.
\end{equation}

After $t$ timesteps this algorithm results in a network with $N=t+m_0$ nodes 
and 
$mt$ edges. Numerical simulations indicated that this network evolves into a 
scale-invariant state with the probability that a node has $k$ edges following 
a 
power-law with an exponent $\gamma_{SF}=3$ (see Fig. \ref{fig_sfres}). The 
scaling exponent is independent of $m$, the only 
parameter in the model.

\subsection{Theoretical approaches}
\label{sect_cont_theor}

The dynamical properties of the scale-free model can be addressed using various analytic approaches. The continuum theory proposed by Barab\'asi and Albert (1999), focuses on the dynamics of node degrees, followed by the master equation approach of Dorogovtsev, Mendes and Samukhin (2000a) and the rate equation approach introduced by Krapivsky, Redner and Leyvraz (2000). As these methods are often used interchangeably in the subsequent section, we review each of them next.

{\it Continuum theory:} The continuum approach introduced in  Barab\'asi and Albert (1999) and Barab\'asi, Albert and 
Jeong (1999) calculates the time dependence of the degree $k_i$ of 
a 
given node $i$. This degree will increase every time a new node enters the 
system 
and links to node $i$, the probability of this process being $\Pi(k_i)$. 
Assuming 
that $k_i$ is a continuous real variable, the rate at which $k_i$ changes is 
expected to be proportional to $\Pi(k_i)$. Consequently, $k_i$ satisfies the 
dynamical equation
\begin{equation}
\frac{\partial k_i}{\partial t} = m \Pi(k_i)=m 
\frac{k_i}{\sum_{j=1}^{N-1}k_j}.
\end{equation}
The sum in the denominator goes over all nodes in the system except the newly 
introduced one, thus its value is  $\sum_j k_j=2mt-m$, leading to
\begin{equation}
\frac{\partial k_i}{\partial t}=\frac{k_i}{2t}. 
\end{equation}
The solution of this equation, with the initial condition 
that every node $i$ at its introduction has $k_i(t_i)=m$, is 
\begin{equation}
\label{connect}
k_i(t)=m\left(\frac {t}{t_i}\right)^\beta, \quad\quad\mbox{with}\quad 
\beta=\frac{1}{2}. 
\end{equation}
Equation (\ref{connect}) indicates that the degree of all nodes evolves the 
same 
way, following a power-law, the only difference being the intercept of the 
power-law. 
 
Using (\ref{connect}), the probability that a node  
has a degree $k_i(t)$ smaller than $k$, $P(k_i(t)<k)$, 
can be written as 
\begin{equation}
\label{prob}
P(k_i(t)<k)=P(t_i>{{m^{1/\beta}t}\over{k^{1/\beta}}}). 
\end{equation}
Assuming that we add the nodes at equal time intervals to the network, the 
$t_i$ 
values have a constant probability density 
\begin{equation}
P(t_i) = \frac{1}{m_0+t}. 
\end{equation}
Substituting this into Eq. (\ref{prob}) we obtain that 
\begin{equation}
P\left(t_i>{{m^{1/\beta} 
t}\over{k^{1/\beta}}}\right)=1-{{m^{1/\beta}t}\over{k^{1/\beta}(t+m_0)}}.
\end{equation} 
The degree distribution $P(k)$ can be obtained using
\begin{equation}
P(k)=\frac{\partial P(k_i(t)<k)}{\partial k}=\frac{2m^{1/\beta}t}{m_0+t}\, 
\frac{1}{k^{1/\beta+1}}
\label{cube},
\end{equation} 
predicting that asymptotically ($t\rightarrow \infty$) 
\begin{equation}
\label{beta_gamma}
P(k)\sim 2m^{1/\beta} k^{-\gamma}, \quad \mbox{with}\quad\gamma=\frac{1}{\beta}+1=3
\end{equation}
being independent of $m$, in agreement with the numerical results.

\begin{figure}[htb]
\centerline{\hspace{-2cm}\psfig{figure=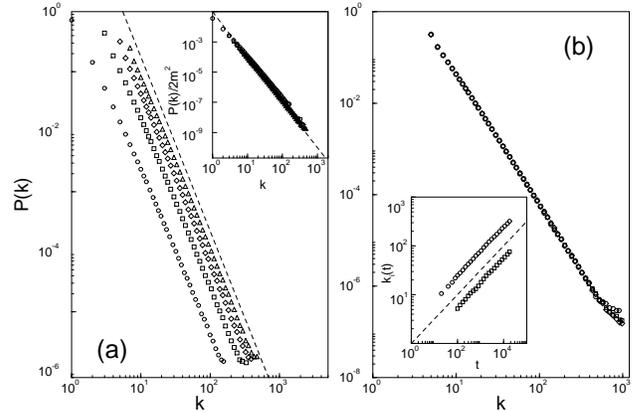,width=2.8in,angle=-90}}
\vspace{-1cm}
\caption{(a) Degree distribution of the scale-free model, with 
$N=m_0+t=300,000$ 
and $m_0=m=1$ (circles), $m_0=m=3$ (squares), $m_0=m=5$ (diamonds) and $
m_0=m=7$ (triangles). The slope of the dashed line is $\gamma=2.9$. The inset  
shows the rescaled distribution (see text) $P(k)/2m^2$ for the same values of 
$m$, the slope of the dashed line being $\gamma=3$. (b) $P(k)$ for $m_0=m=5$ 
and 
system sizes $N=100,000$ (circles), $N=150,000$ (squares) and $N=200,000$ 
(diamonds). The inset shows the time-evolution for the degree of two vertices, 
added to the system at $t_1=5$ and $t_2=95$. Here $m_0=m=5$, and the dashed 
line 
has slope $0.5$, as predicted by Eq.$\,$(\ref{connect}). After Barab\'asi, Albert, Jeong (1999).}
\label{fig_sfres}
\end{figure}

 As the power-law observed for real networks describes systems of rather different sizes, it is expected that a correct model should provide a time-independent degree distribution. Indeed, Eq. (\ref{cube}) predicts that asymptotically the degree 
distribution of the SF model is independent of time (and, subsequently, 
independent of the system size $N=m_0+t$), indicating that despite its 
continuous 
growth, the network reaches a stationary scale-free  state. 
Furthermore, Eq. (\ref{cube}) also indicates that the coefficient 
of the power-law distribution is proportional 
to $m^2$. All these predictions are confirmed by numerical simulations (see Fig. \ref{fig_sfres}).

{\it Master equation approach:} The method introduced by Dorogovtsev, Mendes and Samukhin (2000a, see also Kullmann and Kert\'esz 2000) studies the probability $p(k,t_i,t)$ that at time $t$ a node $i$ introduced at time $t_i$ has a degree $k$. In the SF model when a new node with $m$ edges enters the system, the degree of node $i$ increases with $1$ with a probability $m\Pi(k)=k/2t$, otherwise it stays the same. Consequently the master equation governing $p(k,t_i,t)$ for the SF model has the form
\begin{equation}
\label{sf_master}
p(k,t_i,t+1)=\frac{k-1}{2t}p(k-1,t_i,t)+\left(1-\frac{k}{2t}\right)p(k,t_i,t).
\end{equation} The degree distribution can be obtained as
\begin{equation} 
P(k)=\lim_{t\rightarrow \infty}\left(\sum_{t_i} p(k,t_i,t)\right)/t.
\end{equation}
Eq. (\ref{sf_master}) implies that $P(k)$ is the solution of the recursive equation

\begin{equation}
\label{recursive}
P(k)=\left\{\begin{array}{lcl}
\frac{k-1}{k+2}P(k-1) & \mbox{for} & k\geq m+1\\
2/(m+2) & \mbox{for} & k=m\\
\end{array}\right..
\end{equation}
giving
\begin{equation}
P(k)=\frac{2m(m+1)}{k(k+1)(k+2)},
\end{equation} very close to (\ref{beta_gamma}) obtained using the continuum theory.

{\it Rate equation approach:} The rate equation approach, introduced by Krapivsky, Redner and Leyvraz (2000), focuses on the average 
number $N_k(t)$ of nodes with $k$ edges at time $t$. When a new node enters the network in the scale-free model, $N_k(t)$ changes  as

\begin{equation}
\label{sf_rate}
\frac{dN_k}{dt}=m\frac{(k-1) N_{k-1}(t)-k N_k(t)}{\sum_k kN_k(t)}+\delta_{k,m}.
\end{equation}Here the first term accounts for the new edges that 
connect to nodes with $k-1$ edges, thus increasing their degree to $k$. The second term 
describes the new edges connecting to nodes with $k$ edges turning them into  
nodes with $k+1$ edges, decreasing the number of nodes with $k$ edges. The third term accounts for the new nodes with $m$ edges. In the asymptotic limit $N_k(t)=tP(k)$ and $\sum_k kN_k(t)=2mt$, leading to the same recursive equation, (\ref{recursive}), as predicted by the master equation approach. 

The master equation and rate equation approaches are completely equivalent, and offer the same asymptotic results as the continuum theory. Thus for calculating the scaling behavior of the degree distribution they can be used interchangeably. In addition, these methods, not using a continuum assumption, appear more suitable to obtain exact results in more challenging network models.
 
\subsection{Limiting cases of the SF model}
\label{sect_limit}

The the power-law scaling in the SF model indicates 
that growth and preferential attachment play an important 
role in network development. But are both of them necessary for the emergence 
of 
power-law scaling? To address this question, two limiting cases of the SF 
model 
have been investigated, which contain only one of these two mechanisms 
(Barab\'asi and Albert 1999, Barab\'asi, Albert and Jeong 1999).

{\it Model A} keeps the growing character of the network without preferential 
attachment. Starting with a small number of nodes ($m_0$),
at every time step we add a new node with $m(\le m_0)$ 
edges. We assume that the new node connects with equal probability to the 
nodes 
already present in the system,
i.e.  $\Pi(k_i)=1/(m_0+t-1)$, independent of $k_i$.

\begin{figure}[htb]
\vspace{1cm}
\centerline{\hspace{-1.9cm}\psfig{figure=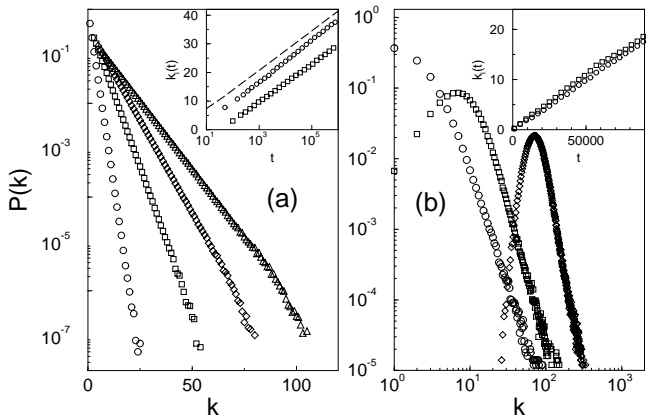,width=2.8in,angle=-90}}
\vspace{-2cm}
\caption{(a) Degree distribution for model A for $m_0=m=1$ (circles), 
$m_0=m=3$ 
(squares), $m_0=m=5$ (diamonds) and $m_0=m=7$ (triangles).  The size of the 
network is $N=800,000$. Inset: time evolution for the degree of two vertices 
added to the system at $t_1=7$ and $t_2=97$. Here $m_0=m=3$. The dashed line 
follows $k_i(t)=m\ln(m_0+t-1)$.(b) The degree distribution for model B for 
$N=10,000$ and $t=N$ (circles), $t=5N$ (squares), and $t=40N$ (diamonds). 
Inset: 
time dependence of the degrees of two vertices. The system size is 
$N=10,000$. After Barab\'asi, Albert and Jeong (1999).}
\label{fig_sfabres}
\end{figure}

The continuum theory predicts that $k_i(t)$ follows a logarithmic time 
dependence, and for $t\rightarrow \infty$ the degree distribution decays 
exponentially, following (Fig.$\,$\ref{fig_sfabres}a)
\begin{equation}
P(k)=\frac{e}{m}\exp\left(-\frac{k}{m}\right).
\end{equation}
 The exponential character of the distribution indicates that the absence of 
preferential attachment eliminates the scale-free character of the resulting 
network.

{\it Model B} starts with $N$ nodes and no edges. At each time step a node is 
selected randomly and connected with probability $\Pi(k_i)=k_i/\sum_j k_j$ to  
a 
node $i$ in the system. Consequently, model B eliminates the growth process, 
the 
number of nodes being kept constant during the network evolution.
Numerical simulations indicate that while at early times the model exhibits 
power-law scaling, $P(k)$ is not stationary (Fig \ref{fig_sfabres}). Since $N$ 
is 
constant, and the number of edges increases with time, after $T\simeq N^2$ 
timesteps the system reaches a  state in which all nodes are connected.

The time-evolution of the individual degrees can be calculated analytically 
using 
the continuum theory, indicating that
\begin{equation}
k_i(t)\simeq\frac{2}{N} t,
\label{evol_1}
\end{equation} assuming $N>>1$, in agreement with the numerical results ( 
Fig.$\,$\ref{fig_sfabres}b).

Since the continuum theory predicts that after a transient period the average 
degree of all nodes should have the same value given by Eq. (\ref{evol_1}), we 
expect that the degree distribution becomes a Gaussian around its mean value. 
Indeed, Fig.$\,$\ref{fig_sfabres}b shows that the shape of $P(k)$ changes from 
the initial power-law to a Gaussian. 

Motivated by correlations between stocks in finantial markets and airline route maps, a prior model incorporating preferential attachment, while keeping $N$ constant was independently proposed and studied by Amaral {\it et al.} (1999).

The failure of models A and B to lead to a scale-free distribution 
indicates that growth and preferential attachment are needed simultaneously 
to reproduce the stationary power-law distribution observed in
real networks.

\subsection{Properties of the SF model}

While the SF model captures the power-law tail of the degree distribution, it 
has 
other properties which may or may not agree with empirical 
results 
on real networks. As we discussed in Sect. \ref{sect_introd}, a characteristic 
feature of real networks is the 
coexistence of clustering and short path lengths. Thus we need to investigate 
if 
the network generated by the SF model has a small-world character.  

\subsubsection{Average path length}

\begin{figure}[htb]
\centerline{\hspace{-1.5cm}\psfig{figure=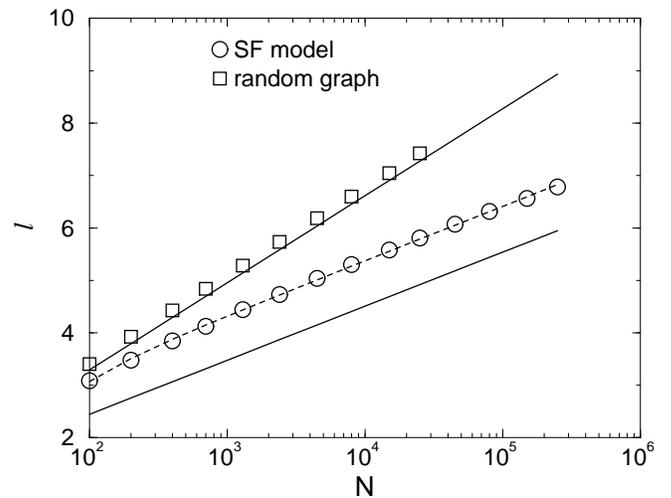,width=2.8in,angle=-90}}
\caption{Characteristic path length $\ell$ versus network size $N$ in a 
network 
with $\langle k\rangle=4$ generated by the SF model (circles), compared with a random 
graph 
with the same size and average degree generated with the algorithm described 
in 
Sect. \ref{sect_er} (squares). The dashed line follows Eq. 
(\ref{dist_sf}), and the solid lines 
represent 
Eq. [\ref{path}] with $z_1=\langle k\rangle$ and $z_2$ the numerically obtained number of second neighbors in the respective networks. }
\label{fig_avdist}
\end{figure}

Figure \ref{fig_avdist} shows the average path length of a network with 
average 
degree $\langle k\rangle=4$ generated by the SF model as a function of the 
network size, $N$, compared with the average path length of a random graph with 
the same size and average degree. The figure indicates that the average path length is smaller in the SF network  than 
in 
a random graph for any $N$, indicating that the heterogeneous 
scale-free topology is more efficient in bringing the nodes close than the 
homogeneous topology of random graphs. We find that the average path length of the SF network increases approximately 
logarithmically with $N$, the best fit following a generalized logarithmic 
form 
\begin{equation}
\label{dist_sf}
\ell=A \log(N-B)+C.
\end{equation}

In Figure \ref{fig_avdist} we also show the prediction of Eq. (\ref{path}) for 
these networks, using the numerically determined number of first and second 
neighbors. While the fit is good for the random graph, Eq. (\ref{path}) 
systematically 
underestimates the average path length of the SF network, as it does with the 
average path length of real networks (see Table \ref{table_gamma}, last three columns).

In summary, apart from the empirical fit (\ref{dist_sf}) there is no 
theoretical 
expression that would give a good approximation for the path length in the 
scale-free model (although encouraging first steps concerning the loop 
structure 
were made in Gleis {\it et al.} 2000). The failure of (\ref{path}) underlies the fact that the topology of the network generated by the 
SF 
model is different from the topology of a random network with power-law degree 
distribution (Sect. \ref{sect_sf_graph}). The dynamical process that generates the network introduces 
nontrivial correlations that affect all topological properties.

\subsubsection{Node degree correlations} 

In the random graph models with arbitrary degree distribution (see Abello {\it 
et 
al.} 2000 and Newman {\it et al.} 2000) the node degrees are uncorrelated. Krapivsky and Redner (2000) have shown that in the SF model correlations develop 
spontaneously between the degree of connected nodes.

 Let us consider all node pairs with degree $k$ and $l$ connected by an edge. 
Without loss of generality we assume that the node with degree $k$ was added 
later to the system, implying that $k<l$ since, according to Eq. 
(\ref{connect}), 
older nodes have higher degree than younger ones, and for simplicity we use 
$m=1$. Denoting by $N_{kl}(t)$ the number of connected pairs of nodes with 
degree 
$k$ and $l$, we have

\begin{eqnarray}
\frac{d N_{kl}}{dt} & =& \frac{(k-1)N_{k-1,l}-kN_{kl}}{\sum_k k 
N(k)}+\frac{(l-1)N_{k,l-1}-lN_{kl}}{\sum_k kN(k)}\nonumber\\
&+& (l-1)N_{l-1}\delta_{k1}.
\end{eqnarray}
 The first term on the r.h.s. accounts for the change in $N_{kl}$ due to the 
addition of an edge to a node of degree $k-1$ or $k$ which is connected to a 
node 
of degree $l$. Since the addition of a new edge increases the node's degree by 
$1$, the first term in the numerator corresponds to a gain in $N_{kl}$, while 
the 
second to a loss. The second term on the r.h.s. incorporates the same effects 
as 
the first applied to the other node. The last term takes into account the 
possibility that $k=1$, thus the edge that is added to the node with degree 
$l-1$ 
is the same edge that connects the two nodes. 

This equation can be transformed into a time-independent recursion relation 
using 
the hypotheses $\sum_k kN(k)\rightarrow 2t$ and $N_{kl}(t)\rightarrow 
tn_{kl}$. 
Solving for $n_{kl}$ we obtain

\begin{eqnarray}
\label{nkl}
n_{kl}&=&\frac{4(l-1)}{k(k+1)(k+l)(k+l+1)(k+l+2)}\nonumber\\
&+&\frac{12(l-1)}{k(k+l-1)(k+l)(k+l+1)(k+l+2)}.
\end{eqnarray}
For a network with an arbitrary degree distribution, if the edges are placed 
randomly, $n_{kl}=n_k n_l$. The most important feature of the 
result 
(\ref{nkl}) is that the joint distribution does not factorize, i.e. $n_{kl}\neq n_k 
n_l$. This indicates the spontaneous appearance of correlations between the 
degrees of the connected nodes. The only case when $n_{kl}$ can be simplified 
to 
a factorized expression is when $1\ll k\ll l$, and $n_{kl}$ becomes

\begin{equation}
n_{kl}\simeq k^{-2} l^{-2},
\end{equation}
but even then it is different from $n_{kl}=k^{-3}l^{-3}$, expected if 
correlations are absent from the network. This result offers the first 
explicit proof that the dynamical process that creates the scale-free network 
builds up nontrivial correlations between the nodes that are not present in 
the 
uncorrelated models discussed in Sect.\ref{sect_sf_graph}.

\subsubsection{Clustering coefficient}

\begin{figure}[htb]
\centerline{\hspace{-1.5cm}\psfig{figure=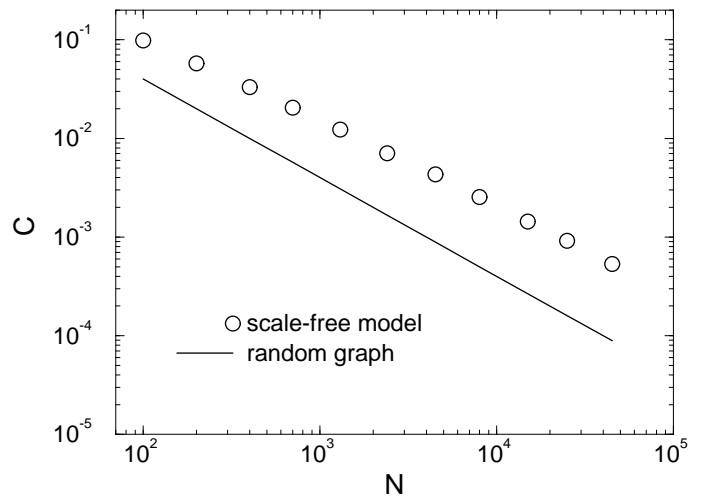,width=2.8in,angle=-90}}
\caption{Clustering coefficient versus size of the SF model with $\langle 
k\rangle=4$, compared with the clustering coefficient of a random graph, 
$C_{rand}\simeq \langle k\rangle/N$.}
\label{fig_sfclust}
\end{figure}

While the clustering coefficient has been much investigated for the WS model 
(Sect. \ref{sect_clust_coef_ws}), there is no analytical prediction for the SF 
model. Figure \ref{fig_sfclust} shows the clustering coefficient of the SF 
network 
with average degree $\langle k\rangle=4$ and different sizes, compared with 
the 
clustering coefficient $C_{rand}=\langle k\rangle/N$ of a random graph. We 
find 
that the clustering coefficient of the scale-free network is about $5$ times 
higher than that of the random graph, and this factor slowly increases with 
the 
number of nodes. However, the clustering coefficient of the SF model decreases with the network size following approximately a power-law $C\sim N^{-0.75}$, which, while a slower decay than the $C=\langle k\rangle N^{-1}$ decay observed for random graphs, is still different from the behavior of the small-world models, where $C$ is independent of $N$.

\subsubsection{Spectral properties}

\begin{figure}[htb]
\vspace{-2cm}
\centerline{\psfig{figure=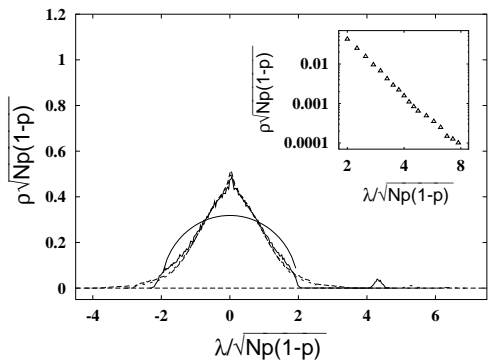,width=2.8in}}
\vspace{-1.5cm}
\caption{Rescaled spectral density of three networks generated by the 
scale-free 
model having $m=m_0=5$ and size $N=100$ (continuous line), $N=300$ (dashed 
line) 
and $N=1000$ (short-dashed line). The semi-circle law corresponding to random 
graphs is drawn for comparison. The isolated peak corresponds to the largest 
eigenvalue, which increases as $N^{1/4}$. Inset: the edge of the spectral 
density 
decays as a power-law. After Farkas {\it et al.} (2001).}
\label{fig_spect_ba}
\end{figure}

The spectral density of the scale-free model is continuous, but it has a markedly different shape than the semicircular spectral density of random graphs (Farkas {\it et al.} 
2001, Goh, Kahng and Kim 2001). Numerical simulations indicate that the bulk of 
$\rho(\lambda)$ has a triangle-like shape with top lying well above the 
semi-circle and edges decaying as a power-law (Fig. \ref{fig_spect_ba}). This  power-law decay is due to the eigenvectors localized on the highest 
degree 
nodes. As in the case of random graphs (and unlike small-world networks), the 
principal eigenvalue, $\lambda_1$, is clearly separated from the bulk of the 
spectrum. A lower bound for $\lambda_1$ can be given as the square-root of the 
network's largest degree $k_1$, and since the node degrees in the scale-free model increase as 
$N^{1/2}$, it results that $\lambda_1$ increases approximately as $N^{1/4}$. 
Numerical results indicate that $\lambda_1$ deviates from the expected 
behavior 
for small network sizes, reaching it asymptotically for $N\rightarrow\infty$. 
This crossover indicates the presence of correlations between the longest row 
vectors, offering additional evidence for correlations in the scale-free model. 

The principal eigenvalue plays an important role in the moments of 
$\rho(\lambda)$, determining the loop structure of the scale-free 
network. In contrast with the subcritical random graph (i.e. $p<1/N$), where 
the 
fraction of loops becomes negligible, in a scale-free network the fraction of 
loops with more than four edges increases with $N$, and their growth rate 
increases with the size of the loop. Note that the fraction of triangles 
decreases as $N\rightarrow\infty$ (Gleis {\it et al.} 2000, Bianconi 2000b). 

While for random graphs $\rho(\lambda)$ follows the semi-circle law (Wigner 
1955, 1957, 1958), deriving a similarly simple expression for small-world (see Sect. \ref{sect_swspectra}) and scale-free networks remains a considerable challenge.

\section{EVOLVING NETWORKS}
\label{sect_evol}

The scale-free model discussed in the previous section is a 
minimal model that captures the mechanisms responsible for the power-law degree distribution. Compared to real networks, it is easy to notice its limitations: it predicts a power-law degree 
distribution with a fixed exponent, while the exponents measured for real 
networks vary between $1$ and $3$ (see Table \ref{table_gamma}). 
Also, 
the degree distribution of real networks can have non-power-law features like 
exponential cutoffs (see  Amaral {\it et al.} 2000, Newman 2001b,c and Jeong 
{\it et al.} 2001) or a saturation for small $k$. The discrepancies between the model and real 
networks led to a surge of research aiming to answer several basic questions 
of 
network evolution: How can we change the scaling exponents? Are there 
universality classes similar to those seen in critical phenomena, 
characterized 
by unique exponents? How do various microscopic processes, known to be present 
in 
real networks, influence the network topology? Are there quantities, beyond 
the 
degree distribution, that could help classifying networks? While the community 
is 
still in the process of answering these questions, several robust results are already available, offering much insight into network evolution and 
topology. 

\subsection{Preferential attachment $\Pi (k)$ }
\label{sect_pref_attach}

A central ingredient of all models aiming to generate scale-free networks is preferential attachment, i.e. the 
assumption that the likelihood of receiving new edges increases with the 
node's 
degree. The SF model assumes that the 
probability $\Pi(k)$ that a node attaches to node $i$ is proportional to the 
degree $k$ of node $i$ (see Eq. (\ref{pi})). This assumption involves two 
hypotheses: the first that $\Pi(k)$ depends on $k$, in contrast with random 
graphs in which $\Pi(k)=p$. Second, the functional form of $\Pi(k)$ is linear in $k$. The 
precise form of $\Pi(k)$ is more than a purely academic question, as lately a 
series of studies have demonstrated that the degree distribution depends 
strongly 
on $\Pi(k)$. To review these developments we start by discussing the empirical 
results on the functional form of 
$\Pi(k)$, 
followed by the theoretical work predicting the effect of $\Pi(k)$ on the 
network 
topology.

\subsubsection{Measuring $\Pi(k)$ for real networks}

The functional form of $\Pi(k)$ can be determined for networks for which we know the time at which each node joined the 
network (Jeong, N\'eda, Barab\'asi 2001, Newman 2001d, Pastor-Satorras {\it et al.} 2001). Such dynamical data is available for the 
coauthorship network of researchers, the citation network of articles, the 
actor 
collaboration network and the Internet at the domain level (see Sect. \ref{sect_real_data}).

 Consider the state of the network at a given time, and record the number of 
"old" nodes present in the network and their degrees. Next measure the 
increase of the degree of the "old" nodes over a time interval 
$\Delta T$, much shorter than the age of the network. Then, according to 
(\ref{pi}), plotting the relative increase $\Delta k_i/\Delta k$ in function 
of 
the earlier degree $k_i$ for every node gives the $\Pi(k)$ function. Here 
$\Delta 
k$ is the number of edges added to the network in the time $\Delta T$. We can reduce the fluctuations in the data  by plotting the cumulative distribution
\begin{equation}
\kappa(k)=\sum_{k_i=0}^k \Pi (k_i).
\end{equation}

\begin{figure}[htb] 
\centerline{\psfig{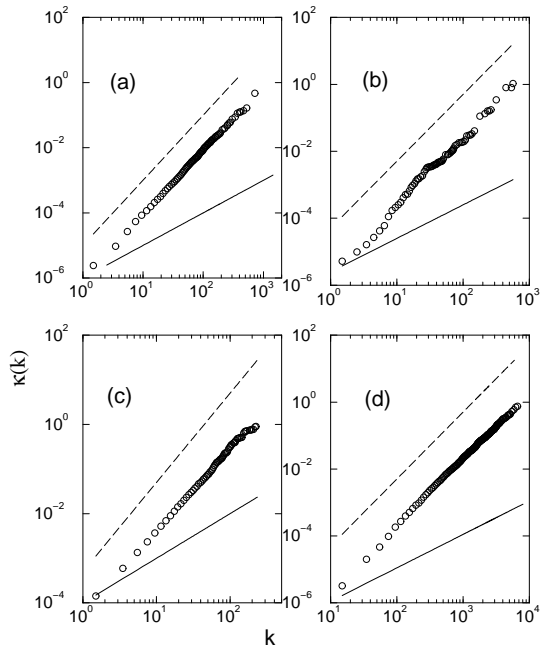}}
\caption{ Cumulated preferential attachment for (a) the citation network; (b) the Internet; (c) the neuroscience scientific 
collaboration network; (d) the actor collaboration network. In all panels the dashed line corresponds to linear preferential attachment, and the continuous line to no preferential attachment. After Jeong, N\'eda, Barab\'asi (2001).}
\label{fig_pref_attach}
\end{figure}

As Fig. \ref{fig_pref_attach} shows, the obtained $\Pi(k)$ supports the existence of preferential attachment. Furthermore, it 
appears that in each case $\Pi(k)$ follows a power-law, i.e. 
\begin{equation}
\label{nonlin}
\Pi(k)\sim k^{\alpha}.
\end{equation}
While in some cases, such as the Internet (Jeong, N\'eda, Barab\'asi 2001, Pastor-Satorras {\it et al.} 2001), citation network (Jeong, N\'eda , Barab\'asi 2001), Medline and  Los Alamos archive (Newman 2001d) we have $\alpha\simeq 1$, 
i.e. $\Pi(k)$ depends linearly on $k$ as assumed in the SF model, for 
some collaboration networks the dependence is sublinear, having $\alpha=0.8\pm 0.1$ for the neuroscience coauthorship and the 
actor collaboration network (Jeong {\it et al.} 2001).

\subsubsection{Nonlinear preferential attachment} 
 
The effect of a nonlinear $\Pi(k)$ on the network dynamics and topology was explained by Krapivsky, Redner and Leyvraz (2000). Replacing the linear preferential 
attachment (\ref{pi}) with (\ref{nonlin}) in a directed network model, 
Krapivsky, 
Redner and Leyvraz calculate the average 
number $N_k(t)$ of nodes with $k-1$ incoming edges at time $t$ by the rate equation approach (see Sect. \ref{sect_cont_theor}). The time evolution of $N_k(t)$ follows

\begin{equation}
\label{rate}
\frac{dN_k}{dt}=\frac{1}{M_\alpha}[(k-1)^\alpha N_{k-1}-k^\alpha 
N_k]+\delta_{k1},
\end{equation} where $M_\alpha(t)=\sum k^\alpha N_k(t)$ is the $\alpha$th 
moment 
of $N_k(t)$. In (\ref{rate}) the first term accounts for the new nodes that 
connect to nodes with $k-1$ edges, thus increasing their degree to $k$. The second term 
describes the new nodes connecting to nodes with $k$ edges turning them into  
nodes with $k+1$ edges, decreasing the number of nodes with $k$ edges. The third term accounts for the continuous introduction of new nodes with a single 
outgoing edge. 

Depending on the value of $\alpha$, several distinct phases have been 
identified:

{\it Sub-linear case} ($\alpha<1$): In this regime in the long-time limit 
$M_\alpha(t)$ satisfies 
$M_\alpha(t)=\mu t$,
with a prefactor $1\leq\mu=\mu(\alpha)\leq 2$. Substituting $M_\alpha (t)$ and $N_k$ into Eq. 
(\ref{rate}) we obtain the degree distribution
\begin{equation}
\label{nk}
P(k)=\frac{\mu}{k^\alpha}\prod_{j=1}^k\left(1+\frac{\mu}{j^\alpha}\right)^{-1}
.
\end{equation}
This product can be expanded in series, and the result is a stretch-exponential 
in 
which a new term arises whenever $\alpha$ decreases below $1/l$, where $l$ is 
an 
arbitrary positive integer.

{\it Superlinear preferential attachment}  ($\alpha>1$): In this regime  Eq. 
(\ref{rate}) has no analytical solution, but its discretized version can be 
used 
to determine recursively the leading behavior of each  $N_k$ as 
$t\rightarrow\infty$. For $\alpha >2$ a  
"winner-takes-all" phenomenon emerges, such that almost all  
nodes have a single edge, connecting them to a "gel" node which has the rest 
of 
edges of the network. For $3/2<\alpha<2$ the number of nodes with two edges 
grows 
as $t^{2-\alpha}$, while the number of nodes with more than two edges is again 
finite. Again, the rest of the edges belong to the gel node. In general for 
$(l+1)/l<\alpha<l/(l-1)$ the number of nodes with more than $l$ edges is 
finite 
even in infinite systems, while $N_k\sim t^{k-(k-1)\alpha}$ for $k\leq l$.

In conclusion, the analytical calculations of Krapivsky, Redner and Leyvraz 
demonstrate that the scale-free nature of the network is destroyed for 
nonlinear 
preferential attachment. The only case in which the topology of the network is 
scale-free is when the preferential attachment is asymptotically linear, i.e. 
$\Pi(k_i)\sim a_\infty k_i$ as $k_i\rightarrow\infty$. In this case the rate equation leads to 

\begin{equation}
P(k)\sim k^{-\gamma}\quad\quad \mbox{with}\quad \gamma=1+\frac{\mu}{a_\infty}.
\end{equation}
This way the 
exponent of the degree distribution can be tuned to any value between $2$ and 
$\infty$. 

\subsubsection{Initial attractiveness}
\label{sect_init_attr}

Another general feature of $\Pi(k)$ in real 
networks is that $\Pi(0)\neq 0$, i.e. there is a nonzero probability that a 
new 
node attaches to an isolated node (Jeong, N\'eda, Barab\'asi 2001). Thus in 
general 
$\Pi(k)$ has the form 

\begin{equation}
\Pi(k)=A +k^\alpha,
\end{equation}
where $A$ is the initial attractiveness of the node $i$ (Dorogovtsev, Mendes, 
Samukhin 2000a). Indeed, if $A=0$, a node that has $k=0$ can never increase 
its 
connectivity according to Eq. (\ref{pi}). However, in real networks every node 
has a finite chance to be "discovered" and linked to even if it has no edges 
to 
start with. Thus the parameter $A$ describes the likelihood that an isolated 
node 
is discovered, such as a new article is cited the first time. 

Dorogovtsev, Mendes and Samukhin (2000a) gave an exact solution for a 
class of growing network models using the master equation approach (see Sect. \ref{sect_cont_theor}). In their 
model at every timestep a new node is added to the network, followed by the 
addition of $m$ 
directed edges pointing from any node in the network to preferentially 
chosen 
nodes. The probability that a node receives an incoming edge is proportional 
to 
the sum of an initial attractiveness and the number of its incoming edges, 
i.e. 
$\Pi(k_{in})=A+k_{in}$. The calculations indicate that the degree distribution follows $P(k)\sim k^{-\gamma}$ with 
$\gamma=2+\frac{A}{m}$. 
Consequently, initial attractiveness does not destroy the scale-free nature of 
the 
degree distribution, only changes the degree exponent. These results agree 
with 
the conclusion of Krapivsky, Redner and Leyvraz (2000), who find that the power-law 
$P(k)$ is preserved for a shifted linear $\Pi(k)$, since the effect 
of 
the initial attractiveness diminishes as $k\rightarrow \infty$. A generalization of the Dorogovtsev, Mendes, Samukhin model (Dorogovtsev, 
Mendes 
and Samukhin 2000b) allows for the random distribution of $n_r$ edges and an 
initial degree $n$ of every new node. These changes do not modify the 
asymptotically linear scaling of the preferential attachment, thus this model 
also gives a power-law degree distribution with $\gamma=2+(n_r+n+A)/m$.

\subsection{Growth}
\label{sect_growth}

In the SF model the number of nodes and edges increases linearly in time, and 
consequently the average degree of the network is constant. In this section we discuss 
the effect of the nonlinear growth rates on the network dynamics and topology.

\subsubsection{Empirical results}

 The fact that networks can follow different growth patterns is supported by 
several recent measurements. For example, the average degree of the Internet 
in 
November of 1997 was $3.42$, but it increased to $3.96$ by December of 1998 
(Faloutsos {\it et al.} 1999). Similarly, the WWW has increased its average 
degree 
from $7.22$ to $7.86$ in the five months between the 
measurements 
of Broder {\it et al.} (2000). The average degree of the coauthorship network of scientists  has been found to 
continuously increase  over an eight year period (Barab\'asi {\it et al.} 
2000). 
Finally, comparison of the metabolic network of organisms of different sizes  
indicates that the average degree of the substrates increases approximately 
linearly with the number of substrates involved in the metabolism (Jeong {\it 
et 
al.} 2000). The increase of the average degree indicates that in many real 
systems the number of edges increases faster than the number of nodes, 
supporting 
the presence of a phenomenon called {\it accelerated growth}.

\subsubsection{Analytical results}

Dorogovtsev and Mendes (2000d) studied analytically the effect 
of 
accelerated growth on the degree distribution, generalizing the directed model with asymptotically linear 
preferential 
attachment of Dorogovtsev, Mendes and Samukhin 2000a (Sect 
\ref{sect_pref_attach}). In this model at every step a new node is added to 
the 
network which receives $n$ incoming edges from random nodes in the system. 
Additionally $c_0t^\theta$ new edges are distributed, each of 
them 
being directed from a randomly selected node to a node with high incoming 
degree 
with asymptotically linear preferential attachment $\Pi(k_{in})\propto 
A+k_{in}$. 
The authors show that accelerated growth, controlled by the exponent 
$\theta$, does not change the scale-free nature of the degree distribution, 
but 
it modifies the degree exponent, which now becomes 

\begin{equation}
\gamma=1+\frac{1}{1+\theta}.
\end{equation} 

While the model of Dorogovtsev and Mendes (2000d) is based on a directed network, Barab\'asi {\it et al.} 2001 discuss an undirected model motivated by measurements on the evolution of the 
coauthorship 
network. In the model  new nodes are added to the system with a constant rate, and these new nodes connect to $b$ nodes already in the system with 
preferential attachment 

\begin{equation}
P_i=b\frac{k_i}{\sum_j k_j}.
\end{equation}
Additionally, at every timestep a linearly increasing number of edges 
(constituting a fraction $a$ of the nodes which are present in the network) 
are 
distributed between the nodes, the probability that an edge is added between 
nodes 
$i$ and $j$ being

\begin{equation}
P_{ij}=\frac{k_i k_j}{\sum_{s,l}^{'} k_s k_l}N(t)a.
\label{prod_pref}
\end{equation}
 
Here $N(t)$ is the number of nodes in the system and the summation goes over 
all 
non-equal values of $s$ and $l$. As a result of these two processes the 
average 
degree of the network increases linearly in time, following $\langle 
k\rangle=at+2b$, in agreement with the measurements on the real coauthor 
network. 
The continuum theory predicts that the time-dependent degree distribution displays a crossover at a critical degree
\begin{equation}
k_c=\sqrt{b^2t}(2+2at/b)^{3/2},
\end{equation}
such that for $k\ll k_c$, $P(k)$ follows a power-law with exponent $\gamma=1.5$ 
and 
for $k \gg k_c$ the exponent is $\gamma=3$. This result explains the 
fast-decaying tail of the degree distributions measured by Newman (2001a), 
and it indicates that as time increases the scaling behavior with $\gamma=1.5$ 
becomes increasingly visible. An equivalent model, proposed by Dorogovtsev and Mendes (2001), was able to reproduce the two separate power-law regimes in the distribution of word coocurences (Ferrer i Cancho, Sol\'e 2001).

\subsection{Local events}
\label{sect_events}

The SF model incorporates only one mechanism for network growth: the 
addition of new nodes that connect to the nodes already in the system. In real 
systems, however, a series of microscopic events shape the network evolution, 
including the addition or rewiring of new edges or the removal of nodes or 
edges. Lately several models have been proposed  
to investigate the effect of selected processes on the scale-free nature of 
the 
degree distribution, offering a more realistic 
description 
of various real networks. Any local change in the network topology can be 
obtained through a combination of four elementary processes: addition and  
removal of a node and addition or removal of an edge. But in reality these 
events 
come jointly, such as the rewiring of an edge is a combination of an edge 
removal 
and addition. Next we briefly review several studies that address in general 
terms the effect of local events on the network topology.

\subsubsection{Internal edges and rewiring}

A model that incorporates new edges between existing nodes and the rewiring of 
edges was discussed by Albert and Barab\'asi (2000). Starting with $m_0$ 
isolated nodes, at each timestep we perform one of the following three 
operations:

(i)  With probability $p$ we add $m$($m\leq m_0$) new
edges. One end of a new edge is selected randomly, the other with probability

\begin{equation}
\label{prefer}
\Pi(k_i)=\frac{k_i+1}{\sum_j (k_j+1)}.
\end{equation}

(ii) With probability $q$ we rewire $m$ edges. For this
we randomly select a node $i$ and remove an edge $l_{ij}$ connected to
it, replacing it with a new edge
$l_{ij'}$ that connects $i$ with node $j'$ chosen with
probability $\Pi(k_j')$ given by (\ref{prefer}).

(iii) With probability $1-p-q$ we add a new node.
The new node has $m$ new edges that with probability
$\Pi(k_i)$ are connected to nodes $i$ already present in
the system.

In the continuum theory the growth rate of the degree of a node $i$ is given by: 

\begin{equation}
\label{partial}
\frac{\partial k_i}{\partial t}=(p-q)m\frac
1N+m\frac{k_i+1}{\sum_j (k_j+1)}.
\end{equation} The first term in the r.h.s. corresponds to the random 
selection of node $i$ as a starting point of a new edge (with probability $p$) 
or as the endpoint from which an edge is disconnected (with probability $q$). The second term corresponds to the selection of node $i$ as an endpoint of an edge with the preferential attachment present in all three of the possible 
processes.

The solution of (\ref{partial}) has the form
\begin{equation} 
\label{sol}
k_i(t)=\left(A(p,q,m)+m+1\right)\left(\frac{t}{t_i}\right)^{\frac{1}{B(p,q,m)}
}-A
(p,q,m)-1,
\end{equation}
where
\begin{eqnarray}
A(p,q,m)&=&(p-q)\left(\frac{2m(1-q)}{1-p-q}+1\right),\nonumber\\
B(p,q,m)&=&\frac{2m(1-q)+1-p-q}{m}.
\end{eqnarray}

The corresponding degree distribution has the generalized power-law form
\begin{equation}
\label{probab}
P(k)\propto(k+\kappa(p,q,m))^{-\gamma(p,q,m)},
\end{equation}
where $\kappa (p,q,m)=A(p,q,m)+1$ and $\gamma(p,q,m)=B(p,q,m)+1$.

Eq.$\,$(\ref {probab}) is valid only when $A(p,q,m) + m+1>0$, which, for fixed 
$p$ and $m$, translates into
$q<q_{max}={\rm min}\{1-p,(1-p+m)/(1+2m)\}$. Thus the
$(p,q)$ phase diagram separates into two regions: For
$q<q_{max}$ the degree distribution is given by (\ref{probab}), following a 
generalized power-law.  For $q>q_{max}$, however, Eq.(\ref{probab}) is not 
valid, but numerical simulations indicate that $P(k)$ approaches an exponential.

While a power-law tail is present in any
point of the scale-free regime, for small
$k$ the probability saturates at $P(\kappa(p,q,m))$, a feature seen in many 
real 
networks (Fig \ref{fig_comp_pk}b,d). Also, the exponent
$\gamma(p,q,m)$ characterizing the tail of $P(k)$ for
$k>>\kappa(p,q)$ changes continuously with $p$, $q$ and $m$, predicting a 
range 
of exponents between 2 and $\infty$. The realistic nature of $P(k)$ was 
confirmed 
by successfully fitting it to the degree distribution of the actor 
collaboration 
network (Albert and Barab\'asi 2000). 

\subsubsection{Internal edges and edge removal}

 Dorogovtsev and Mendes (2000c) consider a class of undirected models in which new edges 
are 
added between old sites and existing edges can be removed. In the first 
variant of the model 
called developing network, $c$ new edges are introduced at every timestep, 
which 
connect two unconnected nodes $i$ and $j$ with a probability proportional to 
the 
product of their degrees (as in Eq. \ref{prod_pref}), an assumption confirmed 
by 
the empirical measurements on the coauthorship network (Barab\'asi {\it et 
al.} 
2000). It is assumed that $c$ can be tuned continuously, such that $c>0$ for 
the 
developing and $c<0$ for the decaying network. The continuum theory predicts that the rate 
of 
change of the node degrees has the form

\begin{equation}
\label{develop}
\frac{\partial k_i}{\partial t}=\frac{k_i(t)}{\int_0^t 
k_j(t)dt_j}+2c\frac{k_i(t)\left[\int_0^t 
k_j(t)dt_j-k_i(t)\right]}{\left[\int_0^t 
k_j(t)dt_j\right]^2-\int_0^t k^2_j(t)dt_j},
\end{equation} 
where the summation over all nodes $\sum_j k_j$ has been approximated by an 
integral over all introduction times $t_j$. The first term in the right hand 
side 
incorporates linear preferential attachment, while the second term corresponds 
to 
the addition of $c$ new edges. Every node can be at either end of the new edge, 
and 
the probability of a node $i$ becoming an end of the new edge is proportional 
to 
the product of its degree $k_i$ and the sum of the degrees $k_j$ of all other 
nodes. The normalization factor is the sum of all products $k_i k_j$ 
with $i$ different from $j$.  

In the asymptotic limit the second term can be neglected compared with the 
first 
term in both the numerator and denominator, and (\ref{develop}) becomes
\begin{equation}
\label{simpler}
\frac{\partial k_i}{\partial t}=(1+2c)\frac{k_i(t)}{\int_0^t k_j(t)dt_j},
\end{equation}
which predicts the dynamic exponent (\ref{connect}) as
\begin{equation}
\label{beta_dev} 
\beta=\frac{1+2c}{2(1+c)},
\end{equation}
and the degree exponent as 
\begin{equation}
\label{gamma_dev}
\gamma=2+\frac{1}{1+2c}.
\end{equation}
The limiting cases of this developing network are $c=0$ when the familiar SF 
values $\beta=1/2$ and $\gamma=3$ are obtained, and $c\rightarrow\infty$, when 
$\beta\rightarrow 1$ and $\gamma\rightarrow 2$. 

In the decaying network at every timestep $|c|$ edges 
are 
removed randomly. The decrease of the node degrees due to this process is 
proportional to their current value, so equation (\ref{simpler}) applies here as well, the only difference being that now $c<0$. A more rigorous 
calculation accounting for the fact that only existing edges can be removed 
confirms that the end result is identical with Eqs. (\ref{beta_dev}) and (\ref{gamma_dev}), 
only 
with negative $c$. The limiting value of $c$ is $-1$, since the rate of 
removal 
of edges cannot be higher than the rate of addition of new nodes and edges, leading to the limit exponents $\beta\rightarrow -\infty$ and 
$\gamma\rightarrow\infty$.

\subsection{Growth constraints}
\label{sect_aging}

For many real networks the nodes have a finite lifetime (for example in social 
networks) or a finite edge capacity (Internet routers or nodes in the 
electrical 
power grid). Recently several groups have addressed the degree to which such 
constraints effect the degree distribution. 

\subsubsection{Aging and cost}

\begin{figure}[htb]

\centerline{\psfig{figure=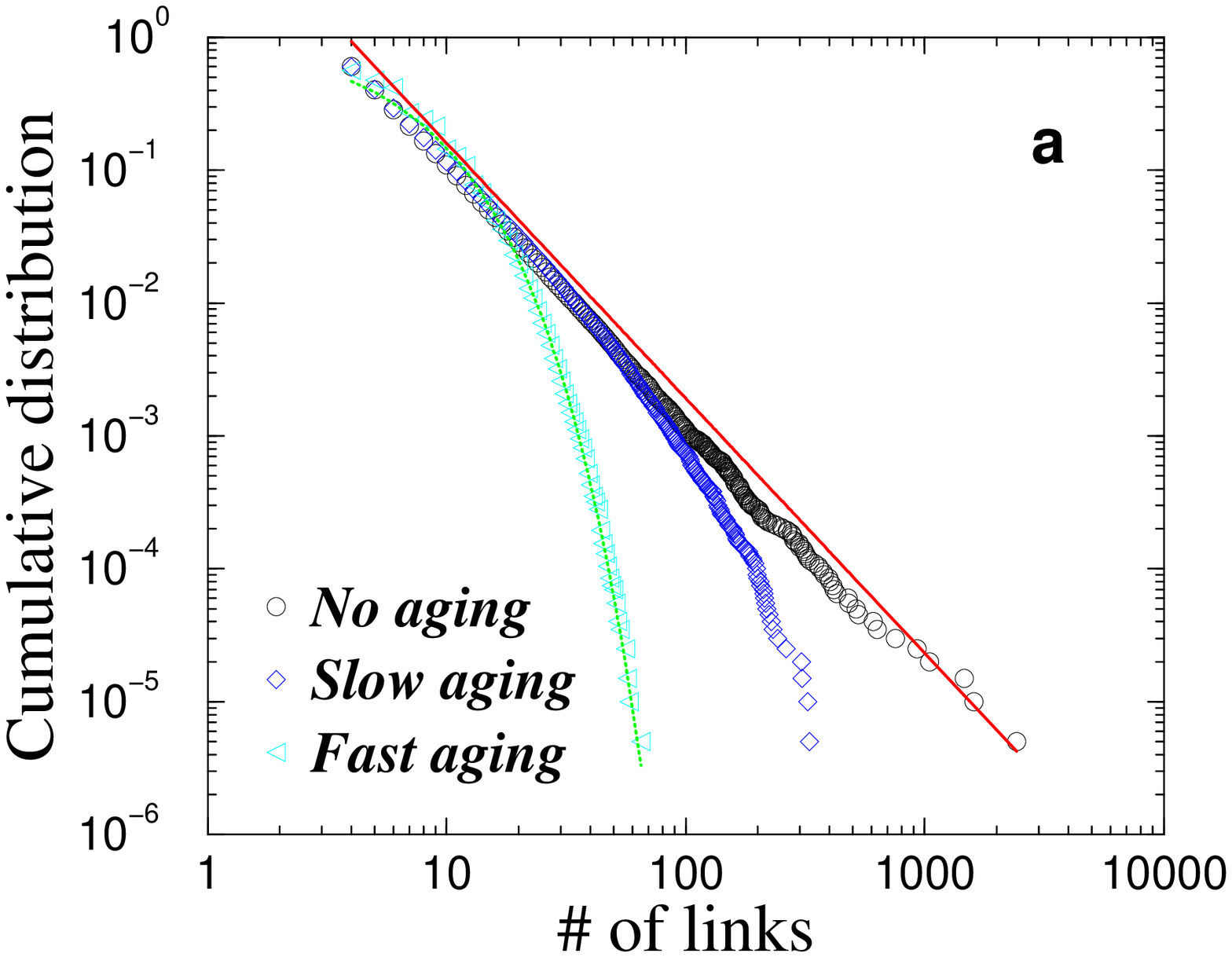,height=4.5cm,width=4.5cm}\psfig{figure=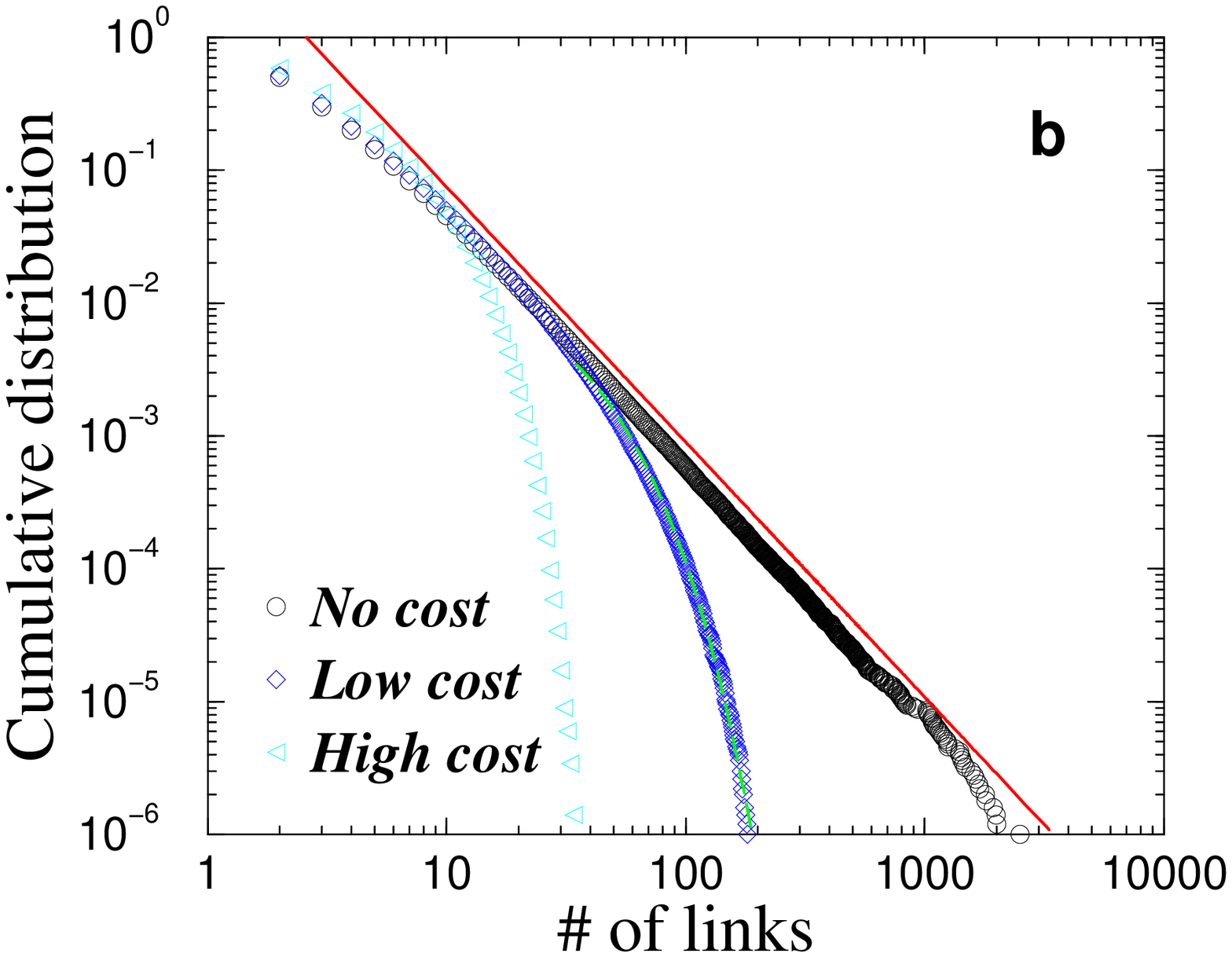,height=4.5cm,width=4.5cm}}
\caption{Deviation from a power-law of the degree distribution due to adding age (a) and 
capacity 
(b) constraints to the SF model. The constraints result in cutoffs of 
the 
power-law scaling. After Amaral {\it et al.} (2000).}
\label{fig_cutoff}
\end{figure}

Amaral {\it et al.} (2000) suggested that while several networks do show 
deviations 
from the power-law behavior, they are far from being random networks.  For 
example, the degree distribution of the electric powergrid of Southern 
California and of the neural network of the worm {\it C. elegans} is more 
consistent with a single-scale exponential distribution. Other networks, like 
the 
extended actor collaboration network, in which TV films and series are 
included, 
have a degree distribution in which the power-law scaling is followed by an 
exponential cutoff for large $k$. In all these examples there are 
constraints 
limiting the addition of new edges. For example, the actors have a finite 
active 
period during which they are able to collect new edges, while for the 
electrical powergrid or neural networks there are constraints on the total 
number 
of edges a particular node can have, driven by economic, physical or 
evolutionary 
reasons.  Amaral {\it et al.} propose that in order to explain these 
deviations 
from a pure power-law we need to incorporate aging and cost or capacity 
constraints. The model studied by them evolves following growth and 
preferential 
attachment, but when a node reaches a certain age (aging) or has more than a critical 
number of edges (capacity constraints), new edges cannot 
connect 
to it. In both cases numerical simulations indicate that while for 
small 
$k$ the degree distribution still follows a power-law, for large $k$ an 
exponential cutoff develops (Fig. \ref{fig_cutoff}).

\begin{figure}[htb]
\vspace{-1cm}
\psfig{figure=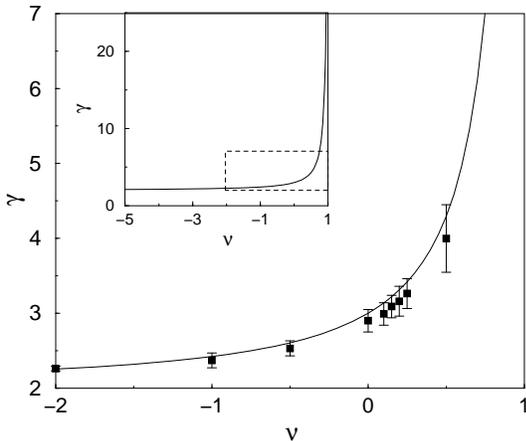,width=2.4in}
\caption{ The dependence of the degree exponent $\gamma$ on the aging exponent 
$\nu$ in the model of Dorogovtsev and Mendes (2000b). The points are obtained 
from simulations, while the solid line is the prediction of the continuum theory. After Dorogovtsev and Mendes (2000b).}
\label{fig_age}
\end{figure}

\subsubsection{Gradual aging}
  
Dorogovtsev and Mendes (2000b) propose that in some systems the probability that 
a 
new node connects to a node $i$ is proportional not only to the degree $k_i$ 
of 
node $i$, but it also depends on its age, decaying  as, $(t-t_i)^{-\nu}$, 
where 
$\nu$ is a tunable parameter. Papers or actors gradually lose their ability to attract more 
edges, the model assuming that this phase-out follows a power-law. The calculations predict 
that the degree distribution depends on the exponent $\nu$: the 
power-law scaling is present only for $\nu<1$, and the degree exponent depends on $\nu$ (Fig. \ref{fig_age}). Moreover, when $\nu>1$ the power-law 
scaling completely disappears, the degree distribution approaching an 
exponential.

\subsection{Competition in evolving networks}

The SF model assumes that all nodes increase their degree following a 
power-law time dependence with the same dynamic exponent $\beta=1/2$ (Eq. (\ref{connect})). As a consequence, the 
oldest nodes have the highest number of edges, since they had the longest 
lifetime to accumulate them. However, numerous examples indicate that in real 
networks a node's degree and growth rate do not depend on age alone. For 
example, on the World-Wide Web some documents acquire a large number of edges 
in 
a very short time through a combination of good content and marketing (Adamic 
and Huberman 2000), or some research papers acquire much more citations than 
their peers. Several studies offered models that address this shortcoming.

\subsubsection{Fitness model}
\label{sect_fit}

Bianconi and Barab\'asi (2000a) argue that real networks 
have 
a competitive aspect, as each node has an intrinsic ability to compete for 
edges 
at the expense of other nodes. They propose a model in which each node is 
assigned a fitness parameter $\eta_i$ which does not change in time. 
Thus at every timestep a new node $j$ with a fitness $\eta_j$ is added to the 
system, where $\eta_j$ is chosen from a distribution $\rho(\eta)$. Each new 
node 
connects with $m$ edges to the nodes already in the network, and the 
probability 
to connect to a node $i$ is proportional to the degree and the fitness of node 
$i$,

\begin{equation}
\Pi_i=\frac{\eta_i k_i}{\sum_j \eta_j k_j}.
\end{equation}
This generalized preferential attachment assures that even a relatively young 
node with a few edges can acquire edges at a high rate if it has a large 
fitness 
parameter. The continuum theory predicts that the rate of change of the degree of node $i$ 
is

\begin{equation}
\label{fitness_k}
\frac{\partial k_i}{\partial t}=m\frac{\eta_i k_i}{\sum_k \eta_j k_j}.
\end{equation} 

Assuming that the time-evolution of $k_i$ follows (\ref{connect}) with a fitness dependent $\beta(\eta)$,

\begin{equation}
\label{fit_assume}
k_{\eta_i}(t,t_i)=m\left(\frac{t}{t_i}\right)^{\beta(\eta_i)},
\end{equation} the dynamic exponent satisfies
\begin{equation}
\beta(\eta)=\frac{\eta}{C} \quad\quad \mbox{with}\quad C=\int 
\rho(\eta)\frac{\eta}{1-\beta(\eta)} d\eta.
\label{consist}
\end{equation}

\begin{figure}[htb]
\vspace{-0.5cm}
\psfig{figure=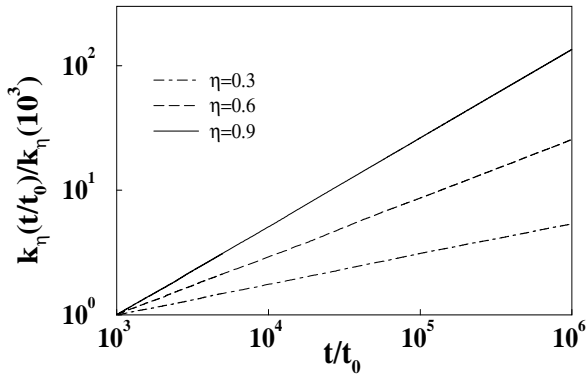,height=2in,width=3in}
\caption{Time dependence of the degree $k_{\eta}(t)$, for nodes with
fitness $ \eta =0.3,\ 0.6$ and $0.9$. Note that $k_{\eta}(t)$
follows a power-law in each case and  the dynamic exponent
$\beta(\eta)$, given by the slope of $k(t)$, increases with $\eta$. After Bianconi and Barab\'asi (2000a)}
\label{fig_fit}
\end{figure}

Thus $\beta$ is described by a spectrum 
of 
values governed by the fitness distribution (Fig. \ref{fig_fit}). Equation (\ref{fit_assume}) 
indicates that nodes with larger fitness increase their degree faster than 
those 
with smaller fitness. Thus the fitness model allows for late but fit nodes to take a 
central role in the network topology. The degree distribution of the model is 
a 
weighted sum of different power-laws
\begin{equation}
P(k)\sim \int 
\rho(\eta)\frac{C}{\eta}\left(\frac{m}{k}\right)^{\frac{C}{\eta}+1},
\end{equation}
which depends on the choice of the fitness distribution (see Sect. \ref{sect_bose}). For 
example, for a uniform fitness distribution Eq. (\ref{consist}) gives 
$C=1.255$ 
and $\beta(\eta)=\eta/1.255$, and the degree distribution is
\begin{equation}
\label{fit_pk}
P(k)\sim \frac{k^{-C-1}}{\log(k)},
\end{equation}
i.e, a power-law with a logarithmic correction. The fitness model can be extended to incorporate additional processes, such as internal edges, which affect the exponents, a problem studied by Erg\"{u}n and Rodgers (2001).

\subsubsection{Edge inheritance}

A different mechanism that gives individuality to the new nodes is proposed by 
Dorogovtsev, Mendes and Samukhin (2000c). They build on the evolving directed 
network algorithm introduced in their earlier paper (Dorogovtsev, Mendes and 
Samukhin 2000a), this time assuming that the degree of the new nodes is not 
constant, but it depends on the state of the network at the time the new node 
is 
added to the system. Specifically, every new node is assumed to be an "heir" 
of a 
randomly chosen old node, and it inherits a fraction $c$ of the old node's 
incoming edges (i.e. a fraction $c$ of the nodes which point to the parent 
node 
will point also to the heir). The parameter $c$ is assumed to be distributed with a probability density $h(c)$. 

The time-dependent degree distribution for uniformly distributed $c$ indicates that the fraction of nodes with no incoming edges 
increases and tends to $1$ asymptotically. The distribution of nonzero incoming 
edges  tends to 
distribution

\begin{equation}
P(k_{in}, k_{in}\neq 0)=\frac{d}{k_{in}^{\sqrt{2}}}\ln(ak_{in}), 
\end{equation}
where $d\simeq 0.174$ and $a\simeq 0.84$.

\subsection{Alternative mechanisms for preferential attachment}
\label{sect_altern}

 It is now established that highly connected nodes have better chances of 
acquiring new edges than their less connected counterparts. The SF model 
reflects this fact by incorporating it explicitly through the preferential 
attachment (\ref{pi}). But where does preferential attachment come from? We do 
not yet have a universal answer to this question, and there is a 
growing suspicion that the mechanisms responsible for preferential attachment 
are 
system-dependent. On the other hand, recently several papers have offered 
promising proposals and models that shed some light into this issue. The 
unifying 
theme of these models is that while preferential attachment is not explicitly 
introduced, the mechanisms used to place nodes and edges 
effectively induces one. The diversity of the proposals vividly illustrates the 
wide range of microscopic mechanisms that could effect the evolution of 
growing 
networks and still lead to the observed scale-free topologies. 

{\it Copying mechanism:} Motivated by the desire to explain the power-law degree distribution of the 
WWW, Kleinberg {\it et al.} (1999) and Kumar {\it et al.} (2000a,b) assume that new webpages 
dedicated to a certain topic copy links from existing pages on the same 
topic. In this model at each timestep a new node is added to the network, 
which 
connects to the rest of the nodes with a constant number of directed edges. In 
the same time a "prototype" node is chosen randomly from the nodes already in 
the 
system. The outgoing edges of the new node are distributed in the following 
way: 
with probability $p$ the destination of the $i$th edge is selected 
randomly, 
and with probability $1-p$ it is taken to be the destination of the $i$th 
edge of the prototype node. This second process increases the probability of 
high degree nodes to receive new incoming edges. In fact, since the 
prototype 
nodes are selected randomly, the probability that a webpage with degree $k$ 
will 
receive a new hyperlink is proportional 
with 
$(1-p)k$, indicating that the copying mechanism effectively amounts to a 
linear preferential attachment. Kumar {\it et al.} prove that the expectation of the incoming degree 
distribution 
is 
\begin{equation}
P(k_{in})=k^{-(2-p)/(1-p)},
\end{equation}
thus $P(k)$ follows a power-law with an exponent which varies between $2$ (for 
$p\rightarrow 0$) and $\infty$ (for $p\rightarrow 1$).

\end{multicols}
\begin{table}[htb]
\caption{Summary of the mechanisms behind the current evolving network models. For each model (beyond the SF model)  we list 
the concept or mechanism deviating from the linear growth and preferential attachment, the basic ingredients of the SF model, and 
the interval in which the exponent $\gamma$ of the degree distribution can vary.}
\label{table_evolv}
\begin{tabular}{|c|c|c|}
New concept or mechanism & Limits of $\gamma$ & Reference \\
\hline
Linear growth, linear pref. attachment & $\gamma=3$ & Barab\'asi and Albert 1999\\
\hline
Nonlinear preferential attachment & &\\
$\Pi(k_i)\sim k_i^\alpha$ & no scaling for $\alpha\neq 1$ & Krapivsky, Redner, 
Leyvraz 2000\\
\hline
Asymptotically linear pref. attachment &  $\gamma\rightarrow 2$ if 
$a_\infty\rightarrow \infty$ &\\
$\Pi(k_i)\sim a_{\infty}k_i$ as $k_i\rightarrow\infty$ & 
$\gamma\rightarrow\infty$ if $a_\infty\rightarrow 0$ & Krapivsky, Redner, 
Leyvraz 2000\\
\hline
Initial attractiveness & $\gamma=2$ if $A=0$ &\\
$\Pi(k_i)\sim A+k_i$ & $\gamma\rightarrow\infty$ if $A\rightarrow \infty$ 
& 
Dorogovtsev, Mendes, Samukhin 2000a,b\\ 
\hline
Accelerating growth $\langle k\rangle\sim t^{\theta}$ & $\gamma=1.5$ if 
$\theta\rightarrow 1$ &\\
constant initial attractiveness & $\gamma\rightarrow 2$ if $\theta\rightarrow 
0$& 
Dorogovtsev and Mendes 2000d\\
\hline
Accelerating growth  & $\gamma=1.5$ for $k\ll k_c(t)$ & Barab\'asi {\it et 
al.} 
2000\\
$\langle k\rangle=at+2b$ & $\gamma=3$ for $k\gg k_c(t)$ & Dorogovtsev, Mendes 2001\\
\hline
Internal edges with probab. $p$ & $\gamma=2$ if $q=\frac{1-p+m}{1+2m}$ &\\
Rewiring of edges with probab. $q$ & $\gamma\rightarrow\infty$ if $p, q, 
m\rightarrow 0$ & Albert and Barab\'asi 2000\\
\hline
$c$ internal edges  & $\gamma\rightarrow 2$ if $c\rightarrow\infty$ &\\
or removal of $c$ edges & $\gamma\rightarrow \infty$ if $c\rightarrow -1$ & 
Dorogovtsev and Mendes 2000c\\
\hline
Gradual aging & $\gamma\rightarrow 2$ if $\nu\rightarrow -\infty$ &\\
$\Pi(k_i)\sim k_i(t-t_i)^{-\nu}$ & $\gamma\rightarrow\infty$ if 
$\nu\rightarrow 
1$ & Dorogovtsev and Mendes 2000b\\
\hline
Multiplicative node fitness  & $P(k)\sim \frac{k^{-1-C}}{\log(k)}$ & \\
$\Pi_i \sim \eta_i k_i$ & & Bianconi and Barab\'asi 2000\\
\hline
Additive-multiplicative fitness & $P(k)\sim \frac{k^{-1-m}}{\ln(k)}$ &\\
$\Pi_i \sim \eta_i (k_i-1)+\zeta_i$ & $1\leq m\leq 2$ & Erg\"un, Rodgers 2001\\
\hline
Edge inheritance & $P(k_{in})=\frac{d}{k_{in}^{\sqrt{2}}}\ln(ak_{in})$ & 
Dorogovtsev, Mendes, Samukhin 2000c\\
\hline
Copying with probab. $p$ & $\gamma=(2-p)/(1-p)$ & Kumar {\it et al.} 2000a,b\\
\hline
Redirection with probab. $r$ & $\gamma=1+1/r$ & Krapivsky, Redner 2000\\
\hline
Walking with probab. $p$ & $\gamma\simeq 2$ for $p>p_c$ & V\'azquez 2000\\
\hline
Attaching to edges & $\gamma=3$ & Dorogovtsev, Mendes, Samukhin 2000d\\
\hline
$p$ directed internal edges  & $\gamma_{in}=2+p\lambda$ &\\
$\Pi(k_i,k_j)\propto (k^{in}_i+\lambda)(k^{out}_j+\mu)$ & $\gamma_{out}=1+(1-p)^{-1}+\mu p/(1-p)$ & Krapivsky, Rodgers, Redner 2001\\
\hline
$1-p$ directed internal edges & $\gamma_{in}=2+p$ &\\
Shifted linear pref. activity & $\gamma_{out}\simeq 2+3p$ & Tadi\'c 2001a\\
\end{tabular}
\end{table}
\begin{multicols}{2}

{\it Edge redirection:} Although inspired by a different mechanism, the growing network with redirection 
model of Krapivsky and Redner (2000) is mathematically equivalent with the 
model 
of Kumar {\it et al.} (2000a,b) discussed above. In this model at every timestep a new 
node 
is added to the system and an earlier node $i$ is selected uniformly as a 
possible target for attachment. With probability $1-r$ a directed edge from 
the 
new node to $i$ is created, however, with probability $r$ the edge is 
redirected 
to the ancestor node $j$ of node $i$ (i.e. the node at wich $i$ attached when 
it 
was first added to the network).

Applying the rate-equation approach (Sect. \ref{sect_cont_theor}), the number of nodes $N(k)$ with degree $k$ evolves as 

\begin{eqnarray}
\frac{dN(k)}{dt}&=&\delta_{k1}+\frac{1-r}{M_0}(N_{k-1}-N_k)\nonumber\\
	        &+&\frac{r}{M_0}[(k-2)N_{k-1}-(k-1)N_k].
\end{eqnarray}
The first term corresponds to the nodes which are just added to the network. 
The 
second term indicates the random selection of a node that the new node will 
attach to. This process affects $N(k)$ if this node has a degree $k-1$ (in 
which 
case its degree will become $k$, increasing $N(k)$) or $k$ (in which case 
$N(k)$ 
decreases by one). The normalization factor $M_0$ is the sum of all degrees. 
The 
third term corresponds to the rewiring process. Since the initial node is 
chosen 
uniformly, if redirection does occur, the probability that a node with $k-1$ 
pre-existing edges receives the redirected edge is proportional with $k-2$, 
the 
number of pre-existing incoming edges. Thus redirection also leads to a linear 
preferential attachment. 

This rate equation is equivalent with Eq. (\ref{rate}) with an asymptotically 
linear attachment $\Pi(k)\sim k-2+1/r$. Thus this model leads to a power-law 
degree distribution  with degree exponent $\gamma=1+1/r$, which can be tuned to any 
value larger than $2$.

{\it Walking on a network:} The walking mechanism proposed by V\'azquez (2000) was inspired by  
citation 
networks. Entering a new field, we are usually aware of a few important papers and follow the 
references included in these to find other relevant articles. This process is 
continued recursively, such that a manuscript will contain 
references 
to papers discovered this way. V\'azquez formulates the corresponding 
network 
algorithm in the following way: We start with an isolated node. At every 
timestep 
a new node is added which links with a directed edge to a randomly selected 
node, 
and then it follows the edges starting from this node and links to their endpoints with 
probability $p$. This last step is repeated starting from the nodes to which 
connections were established, until no new target node is found. In fact, this 
algorithm is similar to the breadth-first search used in determining the 
cluster 
structure of a network, with the exception that not all edges are followed, 
but only 
a fraction equal with $p$. In the special case of $p=1$ one can see 
that nodes with high in-degree will be more likely to acquire new incoming 
edges, 
leading to a preferential attachment $\Pi(k)=(1+k)/N$. Consequently, the degree 
distribution follows a power-law with $\gamma=2$. If $p$ varies between $0$ 
and 
$1$, numerical simulations indicate a phase-transition: for 
$p<p_c\simeq 0.4$ the degree distribution decays exponentially, while for 
$p>p_c$ 
it has a power-law tail with $\gamma$ very close to $2$, the value 
corresponding to $p=1$. Thus, while the model does not explicitly 
include 
preferential attachment, the mechanism behind creating the edges induces one.

{\it Attaching to edges:} Perhaps the simplest model of a scale-free network without explicit 
preferential 
attachment was proposed by Dorogovtsev, Mendes and Samukhin (2000d). In this 
model at every timestep a new node connects to both ends of 
a 
randomly selected edge. Consequently, the probability that a node receives a 
new 
edge is directly proportional with its degree, in other words, this model has 
exactly the same preferential attachment as the SF model. It readily follows 
that 
the degree distribution has the same asymptotic form as the SF model, i.e. 
$P(k)\sim k^{-3}$.

\medskip

The evolving network models presented in this section focus on capturing the mechanisms that govern the evolution of network topology (see Table \ref{table_evolv}), guided by the information contained in the degree distribution. Less is known, however, about the clustering coefficient of these models. Notable exceptions are the models of Barab\'asi {\it et al.} (2001) (Sect. \ref{sect_growth}) and Dorogovtsev, Mendes, Samukhin (2000c) (Sect. \ref{sect_altern}). The clustering coefficient of the model of Barab\'asi {\it et al.} displays a complex behavior as the network increases, first decreasing, going through a minimum, then increasing again, while the model of Dorogovtsev, Mendes and Samukhin (2000c) has a constant asymptotic clustering coefficient. These results suggest that evolving network models can capture the high clustering coefficient of real networks.

\subsection{Connection to other problems in statistical mechanics}

Modeling complex networks offered a fertile ground for statistical 
mechanics. Indeed, many advances in understanding the scaling properties of 
both small world and evolving 
networks have benefited from concepts ranging from critical phenomena to nucleation theory and gelation. On the 
other hand, there appears to be another close link between statistical 
mechanics 
and evolving networks: the continuum theories proposed to 
predict the degree distribution can be, often exactly, mapped into some well  
known problems investigated in statistical physics. In the following we will 
discuss two such mappings, relating evolving networks to the Simon model 
(1955) (Amaral {\it et al.} 2000, Bornholdt and Ebel 2000) and a Bose gas (Bianconi and Barab\'asi 2000b).
 
\subsubsection{The Simon model}

Aiming to account for the wide range of empirical distributions following a 
power-law, such as the frequency of word occurrences (Zipf 1949), the number 
of 
articles published by scientists (Lotka 1926), the city populations (Zipf 
1949) 
or incomes (Pareto 1898), Simon (1955) proposed a class 
of 
stochastic models that result in a power-law distribution 
function. The simplest variant of the Simon model, described in term of word 
frequencies, 
has the following algorithm: Consider a book that is being written, and 
has reached a length of $N$ words. Denote by $f_N(i)$ the number of different 
words 
that each occurred exactly $i$ times in the text. Thus 
$f_N(1)$ denotes the number of different words that have occurred only once. 
The text is continued by adding a new word. With probability $p$, this 
is a new word. However, with probability $1-p$, this 
word is already present. In this case Simon assumes that the probability that the $(N+1)$th word has 
already 
appeared $i$ times is proportional to $if_N(i)$, i.e. the total number of 
words 
that occurred $i$ times. 

As noticed by Bornholdt and Ebel (2000), the Simon model can be mapped exactly into the following network model: Starting from a small seed network, we record the 
number 
of nodes that have exactly $k$ incoming edges, $N_k$. At every timestep one of 
two processes can happen:

 (a) With probability $p$ a new node is added, 
and a 
randomly selected node will point to the new node.

 (b) With  probability $1-p$ a directed edge between two existing nodes is added. The starting 
point of this edge is selected randomly, while its endpoint is selected such 
that 
the probability that a node belonging to the $N_k$ nodes with $k$ incoming 
edges 
will be chosen is 
\begin{equation}
\label{eq_simon}
\Pi(class\,k)\propto kN_k.
\end{equation} 

To appreciate the nature of this mapping, we need to clarify several issues:

(1) While Eq. (\ref{eq_simon}) represents a form of "rich-gets-richer" phenomenon, it does not imply the preferential attachment (\ref{pi}) used in the scale-free model. On the other hand, (\ref{pi}) implies (\ref{eq_simon}). Thus the Simon model describes a general class of stochastic processes that can result in a power-law distribution, appropriate to capture Pareto and Zipf's laws.

(2) The interest in the scale-free model comes from its ability to describe the {\it topology} of complex networks. The Simon model does not have an underlying network structure, as it was designed to describe events whose frequency follows a power-law. Thus network measures going beyond the degree distribution such 
as the average path length, spectral properties, clustering coefficient cannot be obtained from this mapping. 

(3) The mapping described above leads to a directed network with internal edges, different from the scale-free model. On the other hand, it is close to the model proposed by Dorogovtsev, Mendes and Samukhin (2000a,b) discussed in Sect. 
\ref{sect_init_attr}, with the only difference that here the initial 
attractiveness is present only for the isolated nodes. Since (\ref{eq_simon}) corresponds to an asymptotically linear preferential attachment, a correspondence can be made with the model of Krapivsky, Redner and Leyvraz (2000) as well. 

\subsubsection{Bose-Einstein condensation}
\label{sect_bose}

\begin{figure}[htb]
\vspace{-0.5cm}
\centerline{\psfig{figure=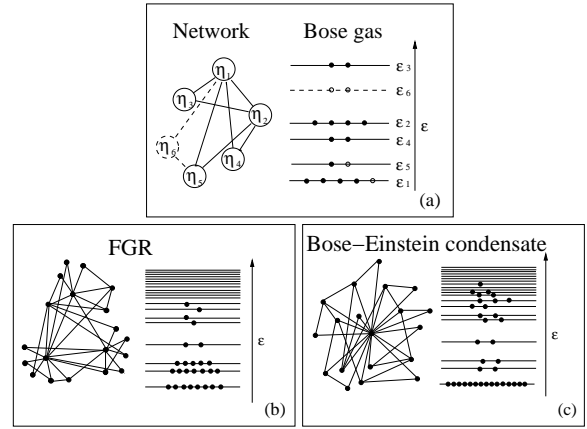,width=3in}}
\vspace{1cm}
\caption{Mapping between the network model and
the Bose gas. (a) On the left we have a network of five nodes, each
 characterized by a fitness $\eta_i$. Equation $(\ref{etaep})$ assigns
an energy $\epsilon_i$ to each $\eta_i$ (right). An edge from node $i$ to node 
$j$ corresponds to a particle at level $\epsilon_i$ and one at $\epsilon_j$. 
The 
network evolves by adding a new node (dashed circle, $\eta_6$) which connects 
to 
$m=2$ other nodes (dashed lines), chosen following $(\ref{pi})$. In the gas  
this 
results in the addition of a new energy
level ($\epsilon_6$, dashed) populated by $m=2$ new particles (open circles), 
and 
 the deposition of $m=2$ other particles to energy levels to which the new 
node 
is connected to ($\epsilon_2$ and $\epsilon_5$). (b) In the FGR phase we have a continuous degree distribution, the several high degree nodes linking the low degree nodes together. In the energy diagram this corresponds to a decreasing occupation number with increasing energy. (c) In the Bose-Einstein condensate the fittest node attracts a finite fraction of all edges, corresponding to a highly populated ground level, and sparsely populated higher energies. After Bianconi and Barab\'asi (2000b).}
\label{fig_bose}
\end{figure}

 Bianconi and Barab\'asi (2000b) show the existence of a close link between evolving networks and an 
equilibrium 
Bose gas. Starting with the fitness model introduced in Sect. \ref{sect_fit}, 
the 
mapping to a Bose gas can be done by assigning an energy $\epsilon_i$ to each 
node, determined by its fitness through the relation
\begin{equation}
\epsilon_i=-\frac{1}{\beta}\log \eta_i,
\label{etaep}
\end{equation} where $\beta=1/T$ plays the role of inverse temperature. An 
edge 
between two nodes $i$ and $j$, having energies $\epsilon_i$ and $\epsilon_j$, 
corresponds to two non-interacting particles, one on each energy level (see 
Fig. 
\ref{fig_bose}). Adding a new node, $l$, to the network corresponds to adding 
a 
new energy level $\epsilon_l$ and $2m$ new particles to the system. Half of 
these 
particles are deposited on the level $\epsilon_l$ (since all new edges start 
from 
the new node), while the other half are distributed between the energy levels 
of 
the endpoints of the new edges, the probability that a particle lands on level 
$i$ being given by
\begin{equation}
\Pi_i=\frac{e^{-\beta \epsilon_i} k_i}{\sum e^{-\beta\epsilon_i} k_i}.
\label{pi_bose}
\end{equation}

The continuum theory predicts that the rate at which particles accumulate on  
energy level $\epsilon_i$ is given by
\begin{equation}
\frac{\partial k_i(\epsilon_i, t, t_i)}{\partial 
t}=m\frac{e^{-\beta\epsilon_i}k_i(\epsilon_i, t, t_i)}{Z_t},
\label{k_bose}
\end{equation}
where $k_i(\epsilon_i, t, t_i)$ is the occupation number of level $i$ and 
$Z_t$ 
is the partition function, defined as $Z_t=\sum_{j=1}{t} e^{-\beta \epsilon_j} 
k_j(\epsilon_j, t, t_j)$. The solution of Eq. (\ref{k_bose}) is 

\begin{equation}
k_i(\epsilon_i, t, t_i)=m\left(\frac{t}{t_i}\right)^{f(\epsilon_i)},
\label{sol_bose}
\end{equation}
where the dynamic exponent $f(\epsilon)$ satisfies 
$f(\epsilon)=e^{-\beta(\epsilon-\mu)}$, $\mu$ plays the role of the chemical 
potential, satisfying the equation
\begin{equation}
\int\deg(\epsilon)\frac{1}{e^{\beta(\epsilon -\mu)} -1}=1,
\label{mu_bose}
\end{equation}
and $deg(\epsilon)$ is the degeneracy of the energy level $\epsilon$.
Eq. (\ref{mu_bose}) suggests that in the $t\rightarrow\infty$ limit the 
occupation number, giving the number of particles with energy $\epsilon$, 
follows 
the familiar Bose statistics
\begin{equation}
n(\epsilon)=\frac{1}{e^{\beta(\epsilon-\mu)}-1}.
\end{equation}

The existence of the solution (\ref{sol_bose}) depends on the functional form 
of 
the distribution $g(\epsilon)$ of the energy levels, determined 
by 
the $\rho(\eta)$ fitness distribution (see Sect. \ref{sect_fit}). 
Specifically, if Eq. (\ref{mu_bose}) has no nonnegative solution for a given 
$g(\epsilon)$ and $\beta$, we can observe a Bose-Einstein condensation, 
indicating that a finite fraction of the particles condensate on the lowest 
energy level (see Fig. \ref{fig_bose}c).

This mapping to a Bose gas predicts the existence of two distinct phases as a 
function of the energy distribution. In the fit-get-rich (FGR) phase, describing the 
case of uniform fitness discussed in Sect. \ref{sect_fit}, the fitter nodes  
acquire edges at a higher rate than older but less fit nodes. In the end the 
fittest node will have the most edges, but the richest node is not an absolute 
winner, since its share of the edges (i.e. the ratio of its edges and the 
total 
number of edges in the network) decays to zero for large system sizes (Fig. \ref{fig_bose}b). The 
unexpected outcome of this mapping is the possibility of a Bose-Einstein 
condensation for $T<T_{BE}$, when the fittest node acquires a finite fraction 
of 
the edges, and maintains this share of edges over time (Fig. \ref{fig_bose}c). A representative 
fitness 
distribution which leads to condensation is $\rho(\eta)=(1-\eta)^{\lambda}$ 
with 
$\lambda>1$. 

The temperature in (\ref{etaep}) plays the role of a dummy variable, 
since if we define a fixed distribution $\rho(\eta)$, the existence of a 
Bose-Einstein condensation or the fit-get-rich phase depends only on the 
functional form of $\rho(\eta)$ and is independent of $\beta$. Indeed, $\beta$ 
falls out at the end from all topologically relevant quantities. As 
Dorogovtsev 
and Mendes (2000d) have subsequently shown, the existence of Bose-Einstein 
condensation can be derived directly from the fitness model, without employing 
the 
mapping to a Bose gas. While the condensation phenomenon appears to be 
similar 
to the gelation process observed by Krapivsky, Redner and Leyvraz (2000a) superlinear preferential attachment, it is not clear at 
this point if this similarity is purely accidental or there is a deeper 
connection between the fitness model and the fitness-free superlinear model. 

\section{ERROR AND ATTACK TOLERANCE}
\label{sect_error_attack}

Many complex systems display a surprising degree of tolerance against errors. 
For 
example, relatively simple organisms grow, persist and reproduce despite 
drastic 
pharmaceutical or environmental interventions, an error tolerance attributed 
to 
the robustness of the underlying metabolic and genetic network (Jeong {\it et al.} 2000, 2001). Complex  communication networks display a high 
degree of robustness: while key components regularly malfunction, local 
failures 
rarely lead to the loss of the global information-carrying ability of the 
network. The stability of these and other complex systems is often attributed 
to 
the redundant wiring of their underlying network structure. But could the 
network 
topology, beyond redundancy, play a role in the error tolerance of such 
complex 
systems?

While error tolerance and robustness almost always has a dynamical component, here we will focus only on the topological aspects of 
robustness, caused by edge and/or node removal. The first results regarding 
network reliability under {\it edge removal} came from random graph theory (Moore 
and 
Shannon 1956a,b, Margulis 1974 and Bollob\'as 1985). Consider an arbitrary 
connected 
graph $H_N$ of $N$ nodes, and assume that a $p$ fraction  of the edges have 
been 
removed. What is the probability that the resulting subgraph is connected, and 
how does it depend on the removal probability $p$? For a broad class of 
starting 
graphs $H_N$ (Margulis 1974) there exists a threshold probability $p_c(N)$ 
such that if $p<p_c(N)$ the subgraph is connected, but if $p>p_c(N)$ it is 
disconnected. This phenomenon is in fact an inverse bond percolation problem 
defined on a graph, with the slight difference (already encountered in the 
evolution of a random graph) that the critical probability depends on $N$.

 As the removal of a node 
implies the malfunctioning of all of its edges as well, {\it node removal} inflicts more damage than edge removal. Does a threshold phenomenon appear for node 
removal too? And to what degree does the topology of the network determine the 
network's robustness? In the following we will 
call a network error tolerant (or robust) if it contains a giant cluster 
comprised of most of the nodes even after a fraction of its nodes are removed. The results indicate a strong correlation between robustness and network topology. In particular, scale-free networks are more robust than random 
networks 
against random node failures, but are more vulnerable when the most 
connected nodes are targeted.

\subsection{Numerical results}

 To compare the robustness of the Erd\H{o}s-R\'enyi random graph and the 
scale-free model, Albert, Jeong and Barab\'asi (2000) investigated networks 
that 
have the same number of nodes and edges, differing only in the degree 
distribution. Two types of node removal were considered. Random perturbations 
can 
cause the failure of some nodes, thus the first mechanism studied was the 
{\it removal 
of randomly selected nodes}. The second mechanism, in which {\it the most highly 
connected nodes are removed} at each step, was selected because it is the most 
damaging to the integrity of the system. This second choice emulates an 
intentional attack on the network. 

\begin{figure}[htb]
\vspace{-2cm}
\centerline{\psfig{figure=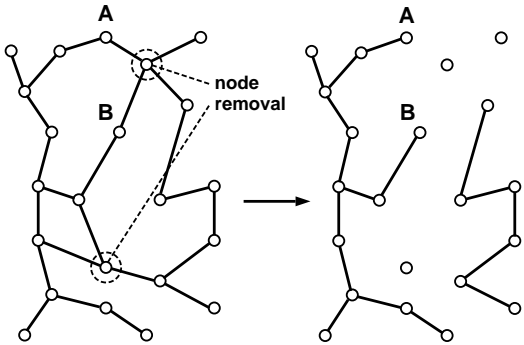,width=2.8in}}
\vspace{-2cm}
\caption{Illustration of the effects of node removal on an initially connected network. In the unperturbed state the distance between node A and B is $2$, but after two nodes are removed from the system, it increases to $6$. In the same time the network breaks into five isolated clusters.}
\label{fig_break}
\end{figure}

 Let us start from a connected network, and at each timestep remove a node. 
The 
disappearance of the node implies the removal of all edges that connect it to 
it, 
disrupting some of the paths between the remaining nodes. One way to monitor the disruption of an initially connected network is to 
study 
the relative size of the largest cluster that remains connected, $S$, and the average path 
length $\ell$ of this cluster, in function of the fraction $f$ of the nodes removed from the system. We expect that the size of the 
largest cluster decreases and its average path length increases as an 
increasing 
number of nodes are removed from the network.

\subsubsection{Random network, random node removal}

\begin{figure}[htb]
\vspace{1.8cm}
\centerline{\hspace{-1.7cm}\psfig{figure=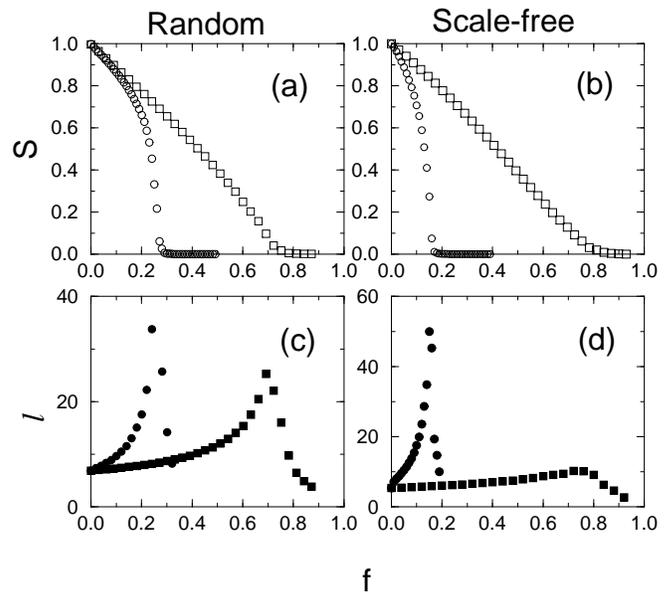,width=2.8in,angle=-90}}
\vspace{-0.5cm}
\caption{The relative size $S$ (a, b) and average path length $\ell$ (c, d) of the 
largest 
cluster in an initially connected network when a fraction $f$ of the nodes are 
removed.  (a, c) Erd\H{o}s-R\'enyi random network with $N=10,000$ and $\langle 
k\rangle=4$. (b, d) Scale-free network generated by the scale-free model with 
$N=10,000$ and $\langle k\rangle=4$. Squares indicate random node 
removal, while circles correspond to preferential removal of the most connected nodes. After Albert, Jeong, Barab\'asi (2000).}
\label{error_num}
\end{figure}

We start by investigating the response of a random network to the random 
removal 
of its nodes (see Fig. \ref{error_num}a, squares), looking at the changes in 
the 
relative size of the largest cluster, $S$ (i.e. the fraction of nodes contained in the largest cluster),  and its average path length, $\ell$, as an 
increasing number of nodes are randomly removed. 

{\it The size of the largest cluster:}  As expected, for a random network $S$ 
decreases from $S=1$ as $f$ increases. If only the removed nodes would 
be missing from the largest cluster, $S$ would follow the diagonal 
corresponding 
to $S=1$ for $f=0$ and $S=0$ for $f=1$. While for small $f$, $S$ follows this line, as $f$ increases the decrease becomes more rapid, indicating that
clusters of nodes become isolated from the main cluster. At a critical 
fraction 
$f_c$, $S$ drops to $0$, indicating that the network breaks into tiny 
isolated clusters. These numerical results indicate an inverse percolation transition. Indeed, percolation theory can 
be 
used to calculate the critical fraction $f_c$ (Sect. \ref{sect_failure}). 

{\it The average path length:} The behavior of $\ell$ also confirms this 
percolation-like transition: it starts from the value characteristic to an 
unperturbed random graph; increases with $f$ as paths are disrupted in the 
network, and it peaks at $f_c$ (Fig. \ref{error_num}c, filled squares). After 
the 
network breaks into isolated clusters $\ell$ decreases as well since in this regime the 
size of the largest cluster decreases very rapidly. 

When $f$ is small we can use the prediction of 
random graph theory, (\ref{diam_er}), indicating that $\ell$ scales as 
$\log(SN)/\log(\langle k\rangle)$, where $\langle k\rangle$ is the average 
degree 
of the largest cluster (Sect. \ref{sect_parall}). Since the number of edges decreases more rapidly than 
the 
number of nodes during node removal (the disruption of each node inducing the 
disruption of several edges), $\langle k\rangle$ decreases faster with 
increasing $f$ than $SN$, and consequently $\ell$ increases. On the other 
hand, 
for $f\simeq f_c$ the prediction of percolation theory becomes valid, and Eq. 
(\ref{diam_clust_inf}) indicates that $\ell$ does not depend on $\langle 
k\rangle$ any longer, and decreases with $S$. 

\subsubsection{Scale-free network, random node removal}

While a random 
network undergoes an inverse percolation transition when a critical fraction 
of 
its nodes are randomly removed, the situation is dramatically different for 
the 
SF network (Fig. \ref{error_num}b, d, squares). Simulations indicate that 
while the size of the largest cluster decreases, it reaches $0$ at a higher 
$f$. 
In the same time, $\ell$ increases much slower than in the random case, and 
its 
peak is much less prominent. The behavior of the system still suggests a 
percolation transition, but analytical calculations indicate that this is 
merely 
a finite size effect, and $f_c\rightarrow 1$ for a scale-free network as the 
size 
of the network increases (Sect. \ref{sect_failure}). In simple terms, 
scale-free networks display an exceptional robustness against random node 
failures.

\subsubsection{Preferential node removal}
  
In the case of an intentional attack, when the nodes with the highest number 
of 
edges are targeted, the network breaks down faster than in the case of random 
node removal. The general breakdown scenario follows again an inverse percolation transition, but now the critical fraction is much lower than in the random 
case. 
This is understandable, since at every step the highest possible number of 
edges 
are removed from the system. Again, the two network topologies respond 
differently to attacks (Fig. \ref{error_num}, circles): the scale-free network, due to 
its reliance on the highly connected nodes, breaks down earlier than the 
random 
network.

In conclusion, the numerical simulations indicate that scale-free networks display a topological robustness against random node failures. The origin of 
this 
error tolerance lies in their heterogeneous topology: low degree nodes are far 
more abundant than nodes with high degree, thus random node selection will 
more 
likely affect the nodes that play a marginal role in the overall network 
topology. But the same heterogeneity makes scale-free networks fragile to 
intentional attacks, since the removal of the highly connected nodes has a 
dramatic disruptive effect on the network.

\subsection{Error tolerance: analytical results}
\label{sect_failure}

The critical threshold for fragmentation, $f_c$, of a network under random node failures was first calculated by Cohen {\it et al.} (2000) and Callaway {\it et al.} (2000). 
Cohen {\it et al.} (2000a) argue that for a  random network with a given degree 
distribution $f_c$ can be determined using the following criterion: a giant 
cluster, with size proportional to the size of the original network, exists if 
an 
arbitrary node $i$, connected to a node $j$ in the giant cluster, is also 
connected to at least one other node. If $i$ is connected only to $j$, the 
network is fragmented. If we assume  that 
loops 
can be neglected (true  for large fragmented 
systems), this criterion can be written as
\begin{equation}
\label{criterion}
\frac{\langle k^2 \rangle}{\langle k\rangle}=2.
\end{equation}

Consider a node with initial degree $k_0$ chosen from an initial distribution 
$P(k_0)$. After the random removal of a fraction $f$ of the nodes the 
probability 
that the degree of that node becomes $k$ is $C_{k_0}^k (1-f)^k f^{k_0-k}$, and 
the new degree distribution is
\begin{equation}
P(k)=\sum_{k_0=k}^\infty P(k_0) C_{k_0}^k (1-f)^k f^{k_0-k}.
\end{equation} Thus the average degree and its second moment for the new 
system 
follows $\langle k\rangle=\langle k_0\rangle (1-f)$ and $\langle k^2\rangle= 
\langle k_0^2 \rangle (1-f)^2 +\langle k_0\rangle f(1-f)$, allowing us to 
rewrite 
the criterion (\ref{criterion}) for criticality as

\begin{equation}
\label{crit_frac}
f_c=1-\frac{1}{\frac{\langle k_0^2\rangle}{\langle k_0\rangle}-1},
\end{equation}
where $f_c$ is the critical fraction of removed nodes and $\langle 
k_0^2\rangle$, 
$\langle k_0\rangle$ are computed from the original distribution before the 
node 
removal.

{\it Random graphs:} As a test of the 
applicability of Eq. (\ref{criterion}), let us remove a fraction $f$ of the nodes from a random graph. Since in the original graph $k_0=pN$ and $k_0^2=(pN)^2+pN$ (see  Sect. \ref{sect_graph_evol}), Eq. (\ref{crit_frac}) implies that 
$
f_c=1-1/(pN).
$
If in the original system $\langle k_0^2 \rangle/\langle k_0\rangle=2$, 
meaning 
that $pN=\langle k\rangle=1$ (the familiar condition of the appearance of the giant cluster in a random graph), the above equation indicates that $f_c=0$, 
i.e. any amount of node removal leads to the network's fragmentation. The 
higher the original degree $\langle k_0\rangle$ of the network, the larger the damage it can survive without breaking apart. 

{\it Scale-free networks:} The critical probability is rather different if the  
degree distribution follows a power-law 
\begin{equation}
\label{cohen_sf}
P(k)=ck^{-\gamma},\quad\quad k=m,m+1,...K,
\end{equation}
where $m$ and $K\simeq mN^{1/\gamma-1}$ are the smallest and the largest degree values, 
respectively. Using a continuum approximation valid in the 
limit $K\gg m\gg 1$, we obtain that

\begin{equation}
\label{cases}
\frac{\langle k_0^2 \rangle}{\langle 
k_0\rangle}\rightarrow\frac{|2-\gamma|}{|3-\gamma|}\times\left\{
\begin{array}{lll}
m & \mbox{if} & \gamma>3;\\
m^{\gamma-2}K^{3-\gamma}, & \mbox{if}& 2<\gamma<3;\\
K & \mbox{if} & 1<\gamma<2.
\end{array}\right.
\end{equation}

We can see that for $\gamma>3$ the ratio is finite and there is a transition 
at 
\begin{equation}
\label{crit_cohen}
f_c=1-\frac{1}{\frac{\gamma-2}{\gamma-3}m-1}.
\end{equation}  

However, for $\gamma<3$ Eq. (\ref{cases}) indicates that the ratio diverges 
with 
$K$ and thus $f_c\rightarrow 1$ when $N\rightarrow \infty$. This result 
implies 
that infinite systems do not break down under random failures, as {\it a spanning 
cluster exists for arbitrarily large $f$}. In finite systems a transition is 
observed, although the transition threshold is very high. This result is in 
agreement with the numerical results discussed in the previous subsection 
(Albert, Jeong and Barab\'asi 2000) indicating a delayed and very small peak 
in 
the $\ell$ curve for the failure of the SF model (having $\gamma=3$).

\begin{figure}[htb]
\centerline{\psfig{figure=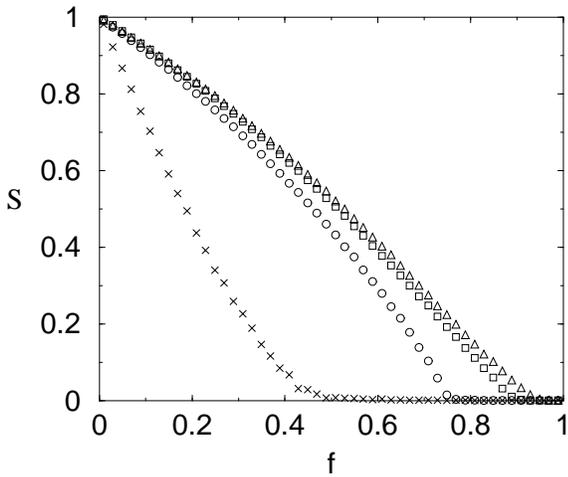,width=2.8in}}
\caption{The fraction of nodes in the giant cluster, $S$, as a function of the 
fraction of randomly removed nodes for scale-free random networks with 
$\gamma=3.5$ (crosses) and $\gamma=2.5$ (all other symbols). In the latter 
case 
three different system sizes were used, with corresponding largest degree 
values 
$K=25$ (circles), $K=100$ (squares) and $K=400$ (triangles). The different 
curves 
illustrate that the fragmentation transition exists only for finite networks, 
while $f_c\rightarrow 1$ as $N\rightarrow \infty$. After Cohen {\it et al.} 
2000}
\label{cohen_fail_large}
\end{figure}

 Callaway {\it et al.} (2000) investigate percolation on generalized random networks, considering that the 
occupation probability of nodes is coupled to the node degree. The 
authors use the method of generating functions discussed in Sect. 
\ref{sect_gen_funct} and generalize it to include the probability of occupancy 
of 
a certain node. The generating function for the degree distribution, 
corresponding to Eq. (\ref{gen0}) in Sect. \ref{sect_gen_funct}, becomes 
\begin{equation}
\label{error_gen_funct}
F_0(x)=\sum_{k=0}^\infty P(k)q_kx^k,
\end{equation}
where $q_k$ stands for the probability that a node with degree $k$ is present. 
The overall fraction of nodes which are present in the network is $q=F_0(1)$ 
which is also equal to $1-f$ where $f$ is the fraction of nodes missing from 
the 
system. This formulation includes the random occupancy (or conversely, 
random failure) case as the special case of uniform occupation probability 
$q_k=q$.

 The authors consider random networks with a truncated power-law degree distribution

\begin{equation}
P(k)=\left\{\begin{array}{lcl}
0 & \mbox{for} & k=0\\
Ck^{-\gamma}e^{-k/\kappa} & \mbox{for} & k\geq 1
\end{array}\right.
\end{equation}
The exponential cutoff of this distribution has the role of regularizing the 
calculations in the same way as the largest degree $K$ in the study of Cohen 
{\it 
et al.} (2000).

In the case of uniform occupation 
probability 
$q$ corresponding to the random breakdown of a fraction $f=1-q$ of the nodes,  the critical occupation probability follows

\begin{equation}
q_c=1-f_c= 
\frac{1}{\frac{Li_{\gamma-2}(e^{-1/\kappa})}{Li_{\gamma-1}(e^{-1/\kappa})}-1}.
\end{equation}
Here $Li_n(x)$ is the $n$th polylogarithm of $x$, defined as 
$Li_n(x)=\sum_{k=1}^\infty x^k/k^n$. This expression is similar with 
Eq. 
(\ref{crit_cohen}) derived by Cohen {\it et al.}, but we can notice that in 
contrast with Eq. (\ref{crit_cohen}), which is valid for $\gamma>3$, this 
equation 
gives nonzero values even for $2<\gamma<3$. The origin of the discrepancy is 
the 
cutoff $\kappa$ which captures the effects of size and capacity constrains 
(see Sect. \ref{sect_aging}). Indeed, if we consider $\kappa\rightarrow\infty$, 
the 
expression for the critical occupation probability becomes

\begin{equation}
q_c=\frac{1}{\frac{\zeta(\gamma-2)}{\zeta(\gamma-1)}-1},
\end{equation}
where $\zeta(x)$ is the Riemann $\zeta$ function defined in the region $x>1$, 
thus this expression is valid only for $\gamma>3$. Since 
$\zeta(x)\rightarrow\infty$ as $x\rightarrow 1$, $q_c$ becomes zero as 
$\gamma$ 
approaches $3$, indicating that for infinite scale-free networks even 
infinitesimal occupation probabilities can assure the presence of an infinite 
cluster.

\subsection{Attack tolerance: analytical results}
\label{sect_attack}

\begin{figure}[htb]
\centerline{\psfig{figure=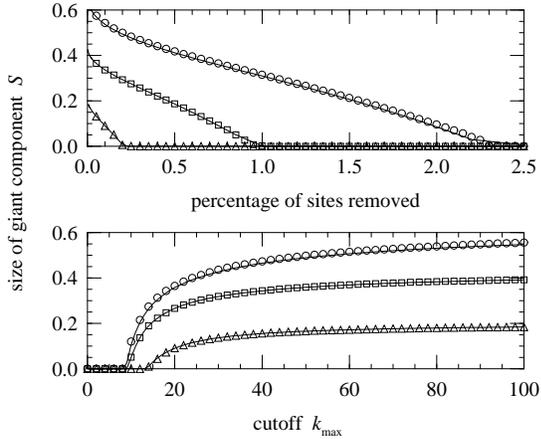,width=2.8in}}
\caption{Fraction of nodes in the spanning cluster in scale-free random 
networks 
with all nodes with degree greater than $k_{max}$ unoccupied, for $\gamma=2.4$ 
(circles), $\gamma=2.7$ (squares) and $\gamma=3.0$ (triangles). The solid 
lines 
are the analytical prediction. Upper frame: as a function of $f$. Lower frame: 
as 
a function of the cutoff $k_{max}$. After Callaway {\it et al.} 2000.}
\label{callaway_attack}
\end{figure}

In the general framework of Callaway {\it et al.} (2000), intentional attack targeted at the nodes with highest degree is equivalent with setting

\begin{equation}
q_k=\theta(k_{max}-k)=\left\{\begin{array}{lcl}
1 & \mbox{if} & k\leq k_{max}\\
0 &\mbox{if} & k>k_{max}\end{array}\right..
\end{equation}
This way only the nodes with degree $k\leq k_{max}$ are occupied, which is 
equivalent 
with removing all nodes with $k>k_{max}$. The number of removed nodes can be 
increased by lowering the value of $k_{max}$. Callaway {\it et al.} (2000) calculate the fraction of nodes in the infinite cluster $S$ as a function of $f$ and $k_{max}$ (Fig. 
\ref{callaway_attack}). This 
figure is in agreement with the results of Albert {\it et al.} (2000) indicating that scale-free networks become fragmented 
after a small fraction $f_c$ of highly connected nodes is removed. It also indicates 
that a small percentage of the highest connected nodes contains in fact nodes 
with surprisingly low degree, agreeing also with the finding of Broder {\it et 
al.} (2000) that the World-Wide Web is resilient to the removal of all nodes 
with degree higher than $5$.   

The theoretical framework of Cohen {\it et al.} (2000) can also be extended to the case of intentional attack on a scale-free network with degree 
distribution (\ref{cohen_sf}) (Cohen {\it et al.} 2001). Under attack two effects arise: (a) the cutoff 
degree $K$ is reduced to a new value $\tilde{K}<K$, and (b) the degree 
distribution of the remaining nodes is changed. The new cutoff can be 
estimated 
from the relation

\begin{equation}
\sum_{k=\tilde{K}}^{K} P(k)=\sum_{k=\tilde{K}}^\infty P(k)-\frac{1}{N}=f,
\end{equation}
which for large $N$ implies 

\begin{equation}
\label{cutoff}
\tilde{K}=mf^{1/(1-\gamma)}.
\end{equation}
The removal of a fraction $f$ of the most connected nodes results in a random 
removal of a fraction $\tilde{f}$ of edges from the remaining nodes. The  
probability that an edge leads to a deleted node equals the fraction of edges 
belonging to deleted nodes
 
\begin{equation}
\tilde{f}= \frac{\sum_{k=\tilde{K}}^K kP(k)}{\langle 
k_0\rangle}=f^{(2-\gamma)/(1-\gamma)},
\end{equation}
 for $\gamma>2$. In the limit $\gamma=2$, $\tilde{f}=\ln(Nf/m)$, thus very 
small 
$f$ values can lead to the destruction of a large fraction of the edges as 
$N\rightarrow \infty$.

Since for random node deletion the probability of an edge leading to a deleted 
node equals the fraction of deleted nodes, Cohen {\it et al.} (2001) argue that the network after undergoing an attack is equivalent with a scale-free network 
with 
cutoff $\tilde{K}$ that has undergone random removal of a fraction $\tilde{f}$ 
of 
its nodes. Replacing $f$ with $\tilde{f}$ and $K$ with $\tilde{K}$ in Eq. 
(\ref{crit_frac}), we obtain the following equation for $\tilde{K}$:
\begin{equation}
\label{cohen_attack_crit}
\left(\frac{\tilde{K}}{m}\right)^{2-\gamma}-2=
\frac{2-\gamma}{3-\gamma}m\left[\left(\frac{\tilde{K}}
{m}\right)^{3-\gamma}-1\right].
\end{equation}

This equation can be solved numerically to obtain $\tilde{K}$ as a function of 
$m$ and $\gamma$, then $f_c(m,\gamma)$ can be determined from (\ref{cutoff}). The 
results indicate that a breakdown phase transition exists for $\gamma>2$, 
and 
$f_c$ is very small for all $\gamma$ values, on the order of a few percent. An 
interesting feature of the $f_c(\gamma)$ curve is that it has a maximum around 
$\gamma=2.25$. It is not surprising that smaller $\gamma$ values lead to 
increased vulnerability to attacks due to the special role the highly 
connected 
nodes play in connecting the system. On the other hand, Cohen {\it et al.} (2001) 
argue 
that the cause of the increased susceptibility of high $\gamma$ networks is 
that 
for these even the original network is formed by several independent clusters, 
and the size of the largest cluster decreases with increasing $\gamma$. 
Indeed, 
the results of Aiello, Chung and Lu (2000) (see Sect.\ref{sect_sf_graph}) indicate that for  $2<\gamma<3.478$ the original network contains an infinite cluster and several smaller clusters of size at most $\log N$ and for $\gamma>3.478$ the original network has no infinite cluster.

\subsection{The robustness of real networks}

 Systematic studies on error and attack tolerance of real networks are available for 
four systems highly relevant to science an technology.

\begin{figure}[htb]
\vspace{1.8cm}
\centerline{\hspace{-1.7cm}\psfig{figure=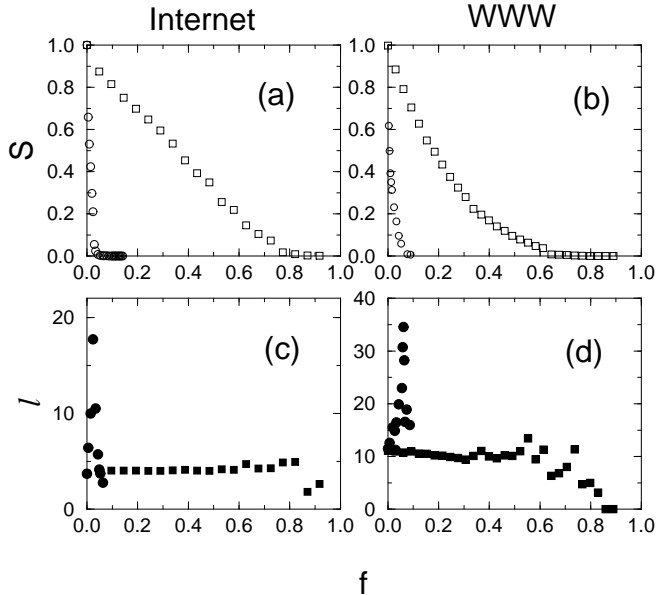,width=2.8in,angle=-90}}
\vspace{-0.5cm}
\caption{The relative size $S$ (a, b) and average path length $\ell$ (c, d) of the 
largest 
cluster in two communication networks when a fraction $f$ of the nodes are 
removed. (a, c), Internet at the domain level, $N=6,209$, $\langle 
k\rangle=3.93$. (b, d) Subset of the WWW with $N=325,729$ and $\langle 
k\rangle=4.59$. Squares indicate random node removal, circles mean 
preferential 
removal of the most connected nodes. After Albert, Jeong, Barab\'asi 2000.}
\label{error_data}
\end{figure}

{\it Communication networks:} The error and attack tolerance of the Internet and the WWW was investigated by Albert, Jeong  and Barab\'asi (2000). Of the two networks, the Internet's robustness has more practical 
significance, 
as about $0.3\%$ of the routers regularly malfunction (random errors), and the 
Internet is occasionally subject to hacker attacks targeting some of the most 
connected nodes. The results, based on the latest map of the Internet  topology at the inter-domain (autonomous system) level, indicate that the average 
path length on the Internet is unaffected by the random removal of as many as 
$60\%$ of the nodes, while if the most connected nodes are eliminated 
(attack), 
$\ell$ peaks at a very small $f$ (Fig. \ref{error_data}a). Similarly, the 
large 
connected cluster persists for high rates of random node removal, but if nodes 
are removed in the attack mode, the size of the fragments that break off 
increases rapidly, the critical point appearing at a very small threshold, 
$f_c^I\simeq 0.03$ (Fig. \ref{error_data}c). 

The WWW study was limited to a subset of the web containing 
$325,729$ 
nodes, the sample investigated in Albert, Jeong and 
Barab\'asi (1999). As the WWW is directed, not all nodes can be reached from 
all 
nodes even for the starting network. To resolve this problem, only distances 
between nodes that have a  path between them are included in the average 
distance 
between nodes. Second, directed networks cannot be separated into clusters 
unambiguously: two nodes can be seen as part of the same cluster when starting 
from a certain node, and appear to be in separate clusters when starting from 
another. This way the number of independent clusters is not unambiguous, but 
the 
largest cluster can be determined. Third, when simulating an attack on the 
WWW, the nodes with the highest 
number of outgoing edges were removed, since $k_{out}$ can be readily 
obtained by looking on a web document, while $k_{in}$ can 
be 
only determined when the map of the whole web is available. Despite these 
methodological differences, the response of the WWW is similar to the undirected networks: 
after 
a slight initial increase, $\ell$ remains constant in the case of random 
failures 
(Fig. \ref{error_data}b), while it increases for attacks. The network survives 
as 
a large cluster under high rates of failure, but under attack the system 
abruptly 
falls apart at $f_c^w=0.067$ (Fig. \ref{error_data}d).

{\it Cellular networks:} Cellular networks can be subject to random errors as a result of mutations or 
protein misfolding, as well as harsh external conditions eliminating essential 
metabolites. Jeong {\it et al.} (2000) study the response of the metabolic 
networks 
of several organisms  to random and preferential node removal. Removing up to 
$8\%$ of the substrates they obtain that the average path length does not 
increase when the nodes are removed randomly, but it increases rapidly under 
the 
removal of the most connected nodes, attaining a $500\%$ increase with the 
removal of only $8\%$ of the nodes. Similar results have been obtained for the 
protein network of yeast as well (Jeong {\it et al.} 2001, see also Vogelstein, Lane and Levine 2000).

{\it Ecological networks:} As a result of human actions or environmental changes, species are deleted 
from 
food webs, an issue of major concern for ecology and environmental 
science. Sol\'e 
and Montoya (2000) study the response of the food webs discussed in Sect. \ref{sect_real_data} to the removal of species 
(nodes) (Montoya 
and Sol\'e 2000). The authors measure the relative size $S$ of the largest cluster, the 
average size $\langle s\rangle$ of the rest of the species clusters, and the fraction of species becoming 
isolated 
due to removal of other species on whom their survival depends (secondary extinctions). The results indicate 
that under random species removal the fraction of species contained in the 
largest 
cluster decreases linearly. In the same time the values of $\langle s\rangle$ remain $0$ or $1$, 
and 
the secondary extinction rates remain very low (smaller than $0.1$) even when 
a 
high fraction of the nodes is removed. The estimate of Eq. (\ref{crit_frac}) 
for 
the critical fraction at which the network fragments  gives $f_c^{fail}$ 
values 
around $0.95$ for all networks, indicating that these networks are error 
tolerant. However, when the most connected (keystone) species are successively 
removed, $S$ decays quickly and becomes zero at $f_c^{attack}\simeq 0.2$, 
while 
$\langle s\rangle$ peaks. The secondary extinctions increase 
dramatically, 
reaching $1$ at relatively low values of $f$ ($f\simeq 0.16$ for the Silwood 
Park web).

\medskip

The results presented in this section offer a simple but compelling
picture: scale-free networks display a high degree of robustness against
random errors, coupled with a susceptibility to attacks. This double
feature is the result of the heterogeneity of the network topology,
encoded by the power-law degree distribution. While we focused on two
measures only, $S$ and $\ell$, it is likely that most network measures
will show distinct behavior for scale-free and random networks. 

The type of disturbances we considered were static, that is, the removal
of a node affected other nodes only in the topological sense. On the
other hand, in many networks there is a dynamical aspect to error
tolerance: the removal of a node could affect the functionality of other
nodes as well. For example, the removal of a highly connected router on
the Internet will redirect traffic to other routers that may not have
the capacity to handle the increased traffic, creating an effective
denial of service attack. Thus in many systems errors lead to cascading
failures, affecting a large fraction of the network. While little is
known about such events, Watts (2001) has recently shown that the
network topology makes a big difference under cascading failures as
well. He investigated a binary model in which the state of a node
changes from off to on if a threshold fraction of its neighbors are on.
In this model the probability that a perturbation in an initially
all-off state spreads to the entire network can be connected to the
existence of a giant cluster of vulnerable nodes. Using the method of
generating functions, Watts (2001) showed that scale-free random graphs
are much less vulnerable to random perturbations than Erd\H{o}s-R\'enyi
random graphs with the same average degree. 

It is often assumed that the robustness of many complex systems is
rooted in their redundancy, which for networks represents the existence
of many alternative paths that can preserve communication between nodes
even if some nodes are absent. We are not aware of any research that
would attempt to address this issue in quantitative terms, uncovering to
which degree does redundancy play a role. 

\section{OUTLOOK}

Just like its main subject, the field of complex networks is rapidly
evolving. While the potential for new and important discoveries is high,
it has attained a degree of coherence that made a review necessary and
appropriate. The fact that the obtained results have reached a critical
mass is best illustrated by the amount of work that we had no space to
review above. Being forced to make a choice, we focused on the
mechanisms and models that describe network topology. In the following
we briefly discuss the results that could not fit in this approach, but
are important for the field. In many ways the list is as prominent as
the works covered so far. 

\subsubsection{Dynamics on networks}

Most networks offer a support for various dynamical processes, as often
the topology plays a crucial role in determining the system's dynamical
features. The range of possible dynamical processes is wide. Watts
(1999) studied the impact of clustering on several processes, including
games, cooperation, the Prisoner's Dilemma, cellular automata and
synchronization (see also Lago-Fern\'andez {\it et al.} 2001). Wang and
Chen (2001) have shown that inhomogeneous scale-free topology plays an
important role in determining synchronization on a complex network, but
search and random walks on complex networks is also a much investigated
topic (Huberman {\it et al.} 1998, Kleinberg 2000, Adamic {\it et al.}
2001, Burda {\it et al.} 2001, Walsh 2001, Bilke and Peterson 2001).
Modeling dynamics on a fixed topology is legitimate when the timescales
describing the network topology and the dynamics on the network differ
widely. A good example is Internet traffic, whose modeling requires time
resolutions from milliseconds up to a day (Willinger {\it et al.} 1997,
Crovella and Bestavros 1997, Sol\'e and Valverde 2001), compared with
the months required for significant topological changes. Similarly,
within the cell the concentrations of different chemicals change much
faster than the cellular network topology (Schilling and Palsson 1998,
Savageau 1998, Gardner {\it et al.} 2000, Elowitz and Leibler 2000),
which is shaped by evolution over many generations. 

The network structure plays a crucial role in determining the spread of
ideas, innovations or computer viruses (Coleman, Menzel and Katz 1957,
Valente 1995). In this light, spreading and diffusion has been studied
both on regular (Kauffman 1992, Keeling 1999), random (Solomonoff,
Rapoport 1951, Rapoport 1957, Weigt and Hartmann 2001), small-world
(Abramson and Kuperman 2000, Newman and Watts 1999, Moukarzel 1999,
Newman, Moore, Watts 2000, Moore and Newman 2000a,b) and scale-free
(Johansen and Sornette 2000, Watts 2001, Bilke and Peterson 2001
,Tadi\'c 2001b) networks. A particularly surprising result was offered
recently by Pastor-Satorras and Vespignani (2001a,b), who studied the
effect of the network topology on disease spreading. They show that
while for random networks a local infection spreads to the whole network
only if the spreading rate is larger than a critical value $\lambda_c$,
for scale-free networks any spreading rate leads to the infection of the
whole network. That is, for scale-free networks the critical spreading
rate reduces to zero, a highly unexpected result that goes against
volumes of results written on this topic. 

When the timescales governing the dynamics on the network are comparable
to that characterizing the network assembly, the dynamical processes can
influence the topological evolution. This appears to be the case in
various biological models inspired by the evolution of communities or
the emergence of the cellular topology (Slanina and Kotrla 1999, 2000,
Bornholdt and Sneppen 2000, Jain and Krishna 2001, L\"assig {\it et al.}
2001). In the current models often these systems are not allowed to
"grow", but they exist in a stationary state that gives room for diverse
network topologies (Slanina and Kotrla 1999, 2000). Interestingly, these
models do not lead to scale-free networks in the stationary state, while
it is known that cellular networks are scale-free (Jeong {\it et al.}
2000, 2001, Wagner and Fell 2000). Thus it is an open challenge to
design evolutionary models that, based on selection or optimization
mechanisms, could produce topologies similar to those seen in the real
world. 

In general, when it comes to understanding dynamics on networks, as well
as the coupling between the dynamics and network assembly, we are only
at the beginning of a promising journey (Strogatz 2001). So far we are
missing simple organizing principles that would match the coherence and
universality characterizing network topology. Due to the importance of
the problem, and the rapid advances we witnessed in descibing network
topology, we foresee it as being a rapidly growing area. 

\subsubsection{Directed networks}

Many important networks, including the WWW or metabolic networks, have
directed edges. In directed networks, however, not all nodes can be
reached from a given node. This leads to a fragmented cluster structure,
where the clusters are not unique, but they depend on the starting point
of the inquiry. Beyond some general aspects, little is known about such
directed networks, but important insights could emerge in the near
future. A promising step in this direction is the empirical study of the
cluster structure of the World-Wide Web (Broder {\it et al.} 2000),
finding that the WWW can be partitioned into several qualitatively
different domains. The results indicate that $28\%$ of the nodes are
part of the strongly connected component (SCC), in which any pair of
nodes is connected by paths in either direction. Another $23\%$ of the
nodes can be reached from the SCC, but cannot connect to it in the other
direction, while a roughly equal fraction of the nodes have paths
leading to the SCC but cannot be reached from it. As several groups have
pointed out, this structure is not specific to the WWW, but common to
all directed networks, ranging from the cell metabolism to citation
networks (Newman, Strogatz and Watts 2000, Dorogovtsev, Mendes and
Samukhin 2001). 

Most network models (including small world and evolving networks) ignore
the network's directedness. On the other hand, as the WWW measurements
have shown, the incoming and outgoing edges could follow different
scaling laws. In this respect, the scale-free model (Barab\'asi and
Albert 1999) explains only the incoming degree distribution, as, due to
its construction, each node has exactly $m$ outgoing edges, thus the
outgoing degree distribution is a delta function (Sect.
\ref{sect_sf_mod}). While recently several models have investigated
directed evolving networks, obtaining a power-law for both the outgoing
and incoming edges (Tadi\'c 2001a,b, Krapivsky, Rodgers, Redner 2001),
the generic features of such complex directed models could hold further
surprises. 

\subsubsection{Weighted networks, optimization, allometric scaling}

Many real networks are weighted networks, in contrast with the binary
networks investigated so far, where the edge weights can have only two
values $0$ and $1$ (absent or present). Indeed, in social networks it is
often important to assign a strength to each acquaintance edge,
indicating how well the two individuals know each other (Newman
2001b,c). Similarly, cellular networks are characterized by reaction
rates, and the edges on the Internet by bandwidth. What are the
mechanisms that determine these weights? Do they obey nontrivial scaling
behavior? To what degree are they determined by the network topology?
Most answers to these questions come from two directions: theoretical
biology and ecology, concerned with issues related to allometric
scaling, and random resistor networks (Derrida and Vannimenus 1982,
Duxbury, Beale and Moukarzel 1995), a topic much studied in statistical
mechanics. Allometric scaling describes the transport of material
through the underlying network characterizing various biological
systems. Most of these systems have a branching tree-like topology. The
combination of the tree topology with the desire to minimize the cost of
transportation leads to nontrivial scaling in the weights of the edges
(West {\it et al.} 1997, Enquist {\it et al.} 1998, 1999). 

In a more general context, Banavar and collaborators have shown that
when the aim is to minimize the cost of transportation, the optimal
network topology can vary widely, ranging from tree-like structures to
spirals or loop-dominated highly interconnected networks (Banavar {\it
at al.} 1999, 2000). Beyond giving systematic methods and principles to
predict the topology of transportation networks, these studies raise
some important questions that need to be addressed in the future. For
example, to which degree is the network topology shaped by global
optimization, or the local processes seen in scale-free networks? There
are fundamental differences between transportation and evolving
networks. In transportation models the network topology is determined by
a global optimization process, in which edges are positioned to
minimize, over the whole network, some predefined quantity, such as cost
or energy of transportation. In contrast, for evolving networks such
global optimization is absent, as the decision where to link is
delegated to the node level. However, this decision is not entirely
local in scale-free networks either, as the node has information about
the degree of all nodes in the network, from which it chooses one
following (\ref{pi}), the normalization factor in making the system
fully coupled. The interplay between such local and global optimization
processes is far from being fully understood (Carlson and Doyle 1999,
2000, Doyle and Carlson 2000). 

While edge weights are well understood for trees and some much studied
physical networks, ranging from river networks (Rodr\'iguez-Iturbe and
Rinaldo 1997, Banavar {\it et al.} 1997) to random resistor networks
(Derrida and Vannimenus 1982, Duxbury, Beale and Moukarzel 1995), little
work is done on these problems in the case of small world or scale-free
networks. Recently Yook {\it et al.} (2001) have investigated an
evolving network model in which the weights were added dynamically,
resulting in unexpected scaling behavior. Newman (2001b) has also
assigned weights to characterize the collaboration strength between
scientists. These studies make an important point, however: despite the
practical relevance and potential phenomenological richness, the
understanding of weighted networks is still in its infancy. 

\subsubsection{Internet and World-Wide Web}

A few real networks, with high technological or intellectual importance,
have received special attention. In these studies the goal is to develop
models that go beyond the basic growth mechanisms and incorporate the
specific and often unique details of a given system. Along these lines
much attention has focused on developing realistic World-Wide Web models
that explain everything from the average path length to incoming and
outgoing degree distribution (Adamic, Huberman 1999, Flake {\it et al.}
2000, Tadi\'c 2001a, Krapivsky, Rodgers, Redner 2001). Many studies
focus on the identification of web communities as well, representing
clusters of nodes that are highly connected to each other (Gibson {\it
et al.} 1998, Flake {\it et al.} 2000, Pennock {\it et al.} 2000, Adamic
and Adar 2000). 

There is a race in computer science to create good Internet topology
generators (Paxson, Floyd 1997, Comellas {\it et al.} 2000). New
Internet protocols are tested on model networks before their
implementation, and protocol optimization is sensitive to the underlying
network topology (Labovitz {\it et al.} 2000). Prompted by the discovery
that the Internet is a scale-free network, all topology generators are
being reviewed and redesigned. These studies have resulted in careful
investigations into what processes could contribute to the correct
topology, reaffirming that growth and preferential attachment are
necessary conditions for realistic Internet models (Medina {\it et al.}
2000, Palmer, Steffan 2000, Yook {\it et al.} 2001b, Pastor-Satorras
{\it et al} 2001, Jeong, N\'eda, Barab\'asi 2001). In addition, an
interesting link has been recently found (Caldarelli {\it et al.} 2000)
to river networks, a much studied topic in statistical mechanics (see
Banavar {\it et al.} 1999, Dodds and Rothman 2000, 2001a,b,c). 

\subsubsection{General questions} 

The high interest in scale-free networks might give the impression that
all complex networks in nature have power-law degree distributions. As
we discussed in Sect. \ref{sect_real_data}, that is far from being the
case. It is true that several complex networks of high interest for the
scientific community, such as the WWW, cellular networks, Internet, some
social networks and the citation network are scale-free. However, others
like the power grid or the neural network of {\it C. elegans} appear to
be exponential. Does that mean that they are random? Far from it. These
systems are also best described by evolving networks. As we have seen in
many examples in Sect. \ref{sect_evol}, evolving networks can develop
both power-law and exponential degree distributions. While the power-law
regime appears to be robust, sublinear preferential attachment, aging
effects, growth constraints lead to crossovers to exponential decay.
Thus, while evolving networks are rather successful at describing a wide
range of networks, the functional form of $P(k)$ cannot be guessed until
the microscopic details of the network evolution are fully understood.
If all processes shaping the topology of a certain network are properly
incorporated, the resulting $P(k)$ often has a rather complex form,
described by a combination of power-laws and exponentials.

In critical phenomena we are accustomed to unique scaling exponents that
characterize complex systems. Indeed, the critical exponents are
uniquely determined by robust factors, such as the dimension of the
space or conservation laws (Stanley 1971, Ma 1976, Hohenberg and
Halperin 1977). The most studied exponents in terms of evolving networks
are the dynamic exponent, $\beta$ and the degree exponent $\gamma$.
While the former characterizes the network dynamics, the latter is a
measure of the network topology. The inseparability of the topology and
the dynamics of evolving networks is shown by the fact that these
exponents are related by the scaling relation (\ref{beta_gamma})
(Dorogovtsev, Mendes, Samukhin 2000a), underlying the fact that the
network assembly uniquely determines the network topology. However, in
no case are these exponents unique. They can be tuned continuously by
such parameters as the frequency of internal edges, rewiring rates,
initial node attractiveness and so on. While it is difficult to search
for universality in the value of the exponents, this does not imply that
the exponents are not uniquely defined. Indeed, if all processes
contributing to the network assembly and evolution are known, the
exponents can be calculated exactly. But they do not take up the
discrete values we are accustomed to in critical phenomena. 

Some real networks have an underlying bipartite structure (Sect.
\ref{sect_ccoeff_nws}). For example, the actor network can be
represented as a graph consisting of two types of nodes: actors and
movies, the edges always connecting two nodes of different type. These
networks can be described as generalized random graphs (Newman, Strogatz
and Watts 2000). It is important to note, however, that both subsets of
these bipartite graphs are growing in time. While it has not been
attempted yet, the theoretical methods developed for evolving networks
can be generalized for bipartite networks as well, leading to coupled
continuum equations. We expect that extending these methods, whenever
appropriate, would lead to a much more realistic description of several
real systems. 

The classical thinking of complex networks, rooted in percolation and
random graph theory (see Aldous 1999), is that they appear as a result
of a percolation process in which isolated nodes eventually join a giant
cluster as the number of edges increases between them. Thus a much
studied question concerns the threshold at which the giant cluster
appears. With a few exceptions (Callaway {\it et al.} 2001), evolving
networks do not follow this percolation picture, since they are
connected from their construction. Naturally, if node or edge removal is
allowed, percolation-type questions do emerge (Sect.
\ref{sect_error_attack}). 

\subsubsection{Conclusions}

The shift that we experienced in the past three years in our
understanding networks was rather swift and unexpected. We have learned
through empirical studies, models and analytic approaches that real
networks are far from being random, but display generic organizing
principles shared by rather different systems. These advances have
created a prolific branch of statistical mechanics, followed with equal
interest by sociologists, biologists and computer scientists. Our goal
here was to summarize, in a coherent fashion, what is known so far. Yet,
we believe that these results are only the tip of the iceberg. We have
uncovered some generic topological and dynamical principles, but the
answers to the open questions could hide new concepts and ideas that
might turn out to be equally exciting as those we encountered so far.
The future could bring a new infusion of tools as well, as the recent
import of ideas from field theory (Burda {\it et al.} 2001) and quantum
statistics (Bianconi 2000a, Bianconi and Barab\'asi 2000, Zizzi 2001)
indicate. Consequently, this article is intended to be as much a review
a catalyst for further advances. We hope that the latter aspect will
dominate. 

\subsubsection{Acknowledgments}

We wish to thank Altavista, Istv\'an Albert, Alain Barrat, Ginestra
Bianconi, Duncan Callaway, Reuven Cohen, Ramesh Govindan, Jos\'e Mendes,
Christian Moukarzel, Zolt\'an N\'eda, Mark Newman, Steven Strogatz,
Andrew Tomkins and Duncan Watts for allowing us to reproduce their
figures. We are grateful to Luis N. Amaral, Alain Barrat, Duncan
Callaway, Reuven Cohen, Imre Der\'enyi, Sergei Dorogovtsev, Ill\'es
Farkas, Shlomo Havlin, Miroslav Kotrla, Paul Krapivsky, Jos\'e Mendes,
Christian Moukarzel, Mark Newman, Sidney Redner, Frantisek Slanina,
Ricard Sol\'e, Steven Strogatz, Bosiljka Tadi\'c, Tam\'as Vicsek and
Duncan Watts for reading our manuscript and helpful suggestions. We have
benefited from discussions with Istv\'an Albert, Ginestra Bianconi,
Zolt\'an Dezs\H{o}, Hawoong Jeong, Zolt\'an N\'eda, Erzs\'ebet Ravasz
and Soon-Hyung Yook. This research was supported by NSF-DMR-9710998 and
NSF-PHYS-9988674.

\end{multicols}
\end{document}